\documentclass[11pt, arial,useAMS,aps,amsmath,amssymb,twocolumn]{article}
\usepackage[T1]{fontenc}
\usepackage[math]{iwona}

\pdfoutput=1
\usepackage[usenames,dvipsnames]{color}
\usepackage[colorlinks=true, linkcolor=BrickRed, citecolor=Blue, urlcolor=Blue, filecolor=Blue, draft]{hyperref}
\usepackage{longtable,epsfig,verbatim,multirow,multicol,hhline}
\usepackage{soul}
\usepackage{epstopdf,pslatex, psfrag, pdfpages,pdftricks}
\usepackage{graphicx}
\usepackage[scaled]{helvet}
\usepackage[small]{caption}
\usepackage{subfigure}
\usepackage{eurosym}
\usepackage{sectsty}
\usepackage{natbib}
\usepackage{url}
\usepackage{xspace}
\usepackage[left=2cm,right=2cm,top=2cm,bottom=2cm]{geometry}
\usepackage{fancyhdr}
\extrafloats{200}
\usepackage{longtable}
\usepackage[table]{xcolor}
\usepackage{floatrow}

\definecolor{FireBrick}{rgb}{0.70,0.13,0.13}

\makeatletter

\newcommand{\uas}{\ensuremath{\mu\mbox{as}}\xspace}

\newcommand{\microns}{\ensuremath{\mu\mbox{m}}\xspace}

\newcommand{\Msun}{\ensuremath{\mbox{M}_\odot}\xspace}

\newcommand{\gaia}{\textit{Gaia}\xspace}
\newcommand{\gaiap}{\textit{Gaia}'s~}

\newcommand{\theia}{\textit{Theia}\xspace}
\newcommand{\theiax}{\textit{Theia}}
\newcommand{\theiap}{\textit{Theia}'s~}

\newcommand{\hips}{\textit{Hipparcos~}}
\newcommand{\hst}{\textit{HST}}
\newcommand{\NEATx}{\textit{NEAT}}

\newcommand{\athena}{\textit{ATHENA}} 
\newcommand{\plato}{\textit {PLATO} } 
\newcommand{\euclidx}{\textit {Euclid}} 
\newcommand{\euclid}{\textit {Euclid} } 
 
\newcommand{\jwst}{\textit{JWST}} 
\newcommand{\wfirst}{\textit{WFIRST}}
\newcommand{\hn}{H_0}

\newcommand{\CB}[1]{\textcolor{Red}{{ [CB: \bf #1]}}}

\newcommand{\remove}[1]{}

\renewcommand{\figurename}{Fig.}
\renewcommand{\tablename}{Tab.}
\renewcommand{\thefigure}{\thesection.\arabic{figure}}
\renewcommand{\thetable}{\thesection.\arabic{table}}

\DeclareCaptionFont{NavyBlue}{\color{NavyBlue}}
\DeclareCaptionFont{Blue}{\color{Blue}}

\urldef{\theiaurl}\url{http://theia.osug.fr}

\title{ Theia: Faint objects in motion or the new astrometry frontier }

\onecolumn
\author{The Theia Collaboration  \thanks{c.m.boehm@durham.ac.uk. To join the collaboration, see \url{http://theia.phyip3.dur.ac.uk}}}

\begin{document}

\maketitle

\begin{abstract}
In the context of the ESA M5 (medium mission) call we proposed a new satellite mission, \textit{Theia}, based on relative astrometry and extreme precision to study the motion of very faint objects in the Universe. 
\textit{Theia} is primarily designed to study the local dark matter properties, the existence of Earth-like exoplanets in our nearest star systems and the physics of compact objects. Furthermore, about 15 $\%$ of the mission time was dedicated to an open observatory for the wider community to propose complementary science cases.  With its unique metrology system and ``point and stare'' strategy, \textit{Theia}'s precision would have reached the sub micro-arcsecond level. This is about 1000 times better than ESA/\textit{Gaia}'s accuracy for the brightest objects and represents a factor 10-30 improvement for the faintest stars (depending on the exact observational program). In the version submitted to ESA, we proposed an optical (350-1000nm) on-axis TMA telescope. Due to ESA Technology readiness level, the camera's focal plane would have been made of CCD detectors but we anticipated an upgrade with CMOS detectors. Photometric measurements would have been performed during slew time and stabilisation phases needed for reaching the required astrometric precision. 
\end{abstract}

\section*{Authors}

\subsection*{Management team: }
C\'eline Boehm (PI, \textit{Durham University - Ogden Centre, UK}), 
Alberto Krone-Martins (co-PI, \textit{Universidade de Lisboa - CENTRA/SIM, Portugal}) and, \textbf{in alphabetical order}, 
Ant\'onio Amorim  (\textit{FCUL,CENTRA/SIM, Portugal}),
Guillem Anglada-Escud\'e (\textit{Queen Mary University of London, UK}),
Alexis Brandeker (\textit{Stockholm University, Sweden}),
Frederic Courbin (\textit{EPFL, Switzerland}),
Torsten En{\ss}lin (\textit{MPA Garching, Germany}), 
Ant\'onio Falc\~ao (\textit{Uninova, Portugal}),
Katherine Freese (\textit{University of Michigan, USA $\&$ Stockholm University, Sweden}), 
Berry Holl  (\textit{Universit\'e de Gen\`eve, Switzerland}),
Lucas Labadie (\textit{Universit\"at zu K\"oln, Germany}),
Alain Leger (\textit{IAS-CNRS, France}),
Fabien Malbet (\textit{Universit\'e de Grenoble, France}),
Gary Mamon (\textit{IAP, France}),
Barbara McArthur (\textit{University of Texas at Austin, USA}),
Alcione Mora (\textit{Aurora Technology BV, Spain}), 
Michael Shao (\textit{JPL/NASA, USA}),
Alessandro Sozzetti  (\textit{INAF - Osservatorio Astrofisico di Torino, Italy}),
Douglas Spolyar (\textit{Stockholm University, Sweden}),
Eva Villaver  (\textit{Universidad Aut\'onoma de Madrid, Spain}),
\\
\subsection*{Science team (alphabetical order)}
Conrado Albertus (\textit{Universidad de Granada, Spain}),
Stefano Bertone (\textit{University of Bern, Switzerland }), 
Herv\'e Bouy (\textit{CAB INTA CSIC, Spain}),
Michael Boylan-Kolchin (\textit{University of Texas, USA }), 
Anthony Brown (\textit{Durham University - Ippp, UK}),
Warren Brown (\textit{Harvard-Smithsonian Centre for Astrophysics, USA}),
Vitor Cardoso (\textit{CENTRA, IST - Universidade de Lisboa, Portugal}),
Laurent Chemin (\textit{INPE, Brasil}),
Riccardo Claudi (\textit{INAF Astronomical Observatoy of Padova, Italy}),
Alexandre C. M. Correia (\textit{CIDMA, University of Aveiro, Portugal}),
Mariateresa Crosta (\textit{INAF, Italy}),
Antoine Crouzier (\textit{Observatoire de Paris, France}),
Francis-Yan Cyr-Racine (\textit{Harvard University, USA}),
Mario Damasso (\textit{INAF - Osservatorio Astrofisico di Torino, Italy}),
Ant\'onio da Silva (\textit{IA - Universidade de Lisboa, Portugal}),
Melvyn Davies (\textit{Lund University, Sweden}),
Payel Das (\textit{University of Oxford, UK}),
Pratika Dayal (\textit{Kapteyn, Netherlands}),
Miguel de Val-Borro (\textit{Princeton University, USA}),
Antonaldo Diaferio (\textit{University of Torino - Dept. of Physics, Italy}), 
Adrienne Erickcek (\textit{University of North Carolina, USA}),
Malcolm Fairbairn (\textit{King's College London, UK}),
Morgane Fortin (\textit{Nicolaus Copernicus Astronomical Center, Polish Academy of Sciences, Poland}),
Malcolm Fridlund (\textit{Leiden/Onsala, Netherlands/Sweden}),
Paulo Garcia (\textit{Universidade do Porto - CENTRA/SIM, Portugal}),
Oleg Gnedin (\textit{University of Michigan, USA}),
Ariel Goobar (\textit{Stockholm University, Sweden}),
Paulo  Gordo (\textit{Universidade de Lisboa - CENTRA/SIM, Portugal}),
Renaud Goullioud (\textit{JPL/NASA, USA}),
Nigel Hambly (\textit{University of Edinburgh, UK}),
Nathan Hara (\textit{IMCCE, Observatoire de Paris, France}), 
David Hobbs (\textit{Lund University, Sweden}),
Erik Hog (\textit{Niels Bohr Institute, Denmark}),
Andrew Holland (\textit{Open University, UK}),
Rodrigo Ibata (\textit{Universit\'e de Strasbourg, France}),
Carme Jordi (\textit{University of Barcelona, ICCUB-IEEC, Spain}),
Sergei Klioner (\textit{Lohrmann Observatory, Technische Universit\"at Dresden, Germany}),
Sergei Kopeikin (\textit{University of Missouri, USA}),
Thomas Lacroix (\textit{Institut Astrophysique de Paris, France}),
Jacques Laskar (\textit{IMCCE, Observatoire de Paris, France}),
Christophe Le Poncin-Lafitte (\textit{Observatoire de Paris Meudon, France}),
Xavier Luri (\textit{University of Barcelona, ICCUB-IEEC, Spain}),
Subhabrata Majumdar (\textit{Tata Institute of Fundamental Research, India}),
Valeri Makarov (\textit{US Naval Observatory, USA}),
Richard Massey (\textit{Durham University, UK}),
Bertrand Mennesson  (\textit{NASA JPL, USA}), 
Daniel Michalik (\textit{Lund University, Sweden}),
Andr\'e Moitinho de Almeida (\textit{Universidade de Lisboa - CENTRA/SIM, Portugal}),
Ana Mour\~ao (\textit{CENTRA, Instituto Superior T\'ecnico - Universidade de Lisboa, Portugal}),
Leonidas Moustakas (\textit{JPL/Caltech, USA}),
Neil Murray (\textit{The Open University, UK}),
Matthew  Muterspaugh  (\textit{Tennessee State University, USA}),
Micaela Oertel (\textit{LUTH, CNRS/Observatoire de Paris, France}),
Luisa Ostorero (\textit{Department of Physics - University of Torino, Italy}),
Angeles Perez-Garcia (\textit{USAL, Spain}),
Imants Platais  (\textit{Johns-Hopkins University, USA}), 
Jordi Portell i de Mora (\textit{DAPCOM Data Services S.L., Spain}),
Andreas Quirrenbach (\textit{Universit\"at Heidelberg, Germany}),
Lisa Randall  (\textit{Harvard University, USA}),
Justin Read (\textit{University of Surrey, UK}),
Eniko Regos (\textit{Wigner Research Institute for Physics, Hungary}),
Barnes Rory (\textit{University of Washington, USA}),
Krzysztof Rybicki (\textit{Warsaw University Astronomical Observatory, Poland}),
Pat Scott (\textit{Imperial College London, UK}),
Jean Schneider (\textit{Observatoire de Paris Meudon, France}),
Jakub Scholtz (\textit{Harvard University, USA}), 
Arnaud Siebert (\textit{Universit\'e de Strasbourg, France}),
Ismael Tereno (\textit{IA - Universidade de Lisboa, Portugal}),
John Tomsick (\textit{University of California Berkeley, USA}),
Wesley Traub (\textit{Jet Propulsion Laboratory, USA}),
Monica Valluri (\textit{University of Michigan, USA}),
Matt Walker (\textit{Carnegie Mellon University, USA}),
Nicholas Walton (\textit{University of Cambridge, UK}),
Laura Watkins (\textit{Space Telescope Science Institute, USA}),
Glenn White (\textit{Open University \& The Rutherford Appleton Laboratory, UK})
Dafydd Wyn Evans (\textit{Institute of Astronomy, Cambridge, UK}),
Lukasz Wyrzykowski (\textit{Warsaw University, Poland}),
Rosemary Wyse (\textit{Johns Hopkins University, USA}),

\subsection*{Additional members of the collaboration (alphabetical order)}
Ummi Abbas (\textit{Osservatorio Astrofisico di Torino, Italy}),
Jean-Michel Alimi (\textit{Observatoire de Paris Meudon, France}),
Martin Altmann (\textit{Astronomisches Recheninstitut, Germany}),
Jo\~ao Alves (\textit{University of Vienna, Austria}),
Richard Anderson (\textit{Johns Hopkins University, USA}),
Fr\'ed\'eric Arenou (\textit{CNRS/GEPI, Observatoire de Paris, France}),
Coryn Bailer-Jones  (\textit{Max Planck Institute for Astronomy, Heidelberg, Germany}), 
Carlton	Baugh 	(\textit{Durham University, UK}), 
Michael Biermann (\textit{ARI, Germany})
Sergi Blanco-Cuaresma (\textit{Harvard-Smithsonian Center for Astrophysics, USA}),
Aldo Stefano Bonomo (\textit{INAF - Osservatorio Astrofisico di Torino, Italy}),   
Avery Broderick (\textit{Univeristy of Waterloo, Canada}),
Giorgia Busso (\textit{Institute of Astronomy, University of Cambridge, UK}),
Juan Cabrera (\textit{DLR, Germany}),
Josep Manel Carrasco (\textit{University of Barcelona, ICCUB-IEEC, Spain}),
Carla Sofia Carvalho (\textit{IA - Universidade de Lisboa, Portugal}),
Marco Castellani (\textit{INAF - Rome Astronomical Observatory, Italy}),
Marina Cerme\~no-Gavil\'an (\textit{University of Salamanca, Spain}),
Paula Chadwick (\textit{Durham University - Ippp, UK}),
Jeremy Darling (\textit{University of Colorado, USA}),
Michael Davidson (\textit{Institute for Astronomy, University of Edinburgh, UK}),
Francesca De Angeli (\textit{Institute of Astronomy, University of Cambridge, UK}),
Reinaldo de Carvalho (\textit{National Institute for Space Research, Brazil}),
Mario Damasso 	(\textit{INAF - Osservatorio Astrofisico di Torino, Italy}), 
Silvano Desidera (\textit{INAF - Osservatorio Astronomico di Padova, Italy}),
Roland Diehl (Max Planck Institut f\"ur extraterrestrische Physik, Germany), 
Chris Done (\textit{Durham University - Ippp, UK}),
Christine Ducourant (\textit{LAB - Bordeaux Observatory, France}),
Denis	Erkal (\textit{Institute of Astronomy, Cambridge University, UK},
Laurent Eyer (\textit{Geneva Observatory, University of Geneva, Switzerland, Switzerland}),
Benoit Famey (\textit{Universit\'e de Strasbourg, France}),
Sofia Feltzing (\textit{Lund Observatory, Sweden}),
Emilio Fraile Garcia (\textit{European Space and Astronomy Centre, Spain}),
Facundo Ariel Gomez  (\textit{MPA Garching, Germany}), 
Carlos Frenk (\textit{Durham University, ICC, UK}),
Mario Gai (\textit{INAF - Osservatorio Astrofisico di Torino, Italy}),
Phillip Galli (\textit{Universidade de S\~ao Paulo, Brazil}),
Laurent Galluccio (\textit{Observatoire de la C\^ote d'Azur, France}),
Paulo Garcia (\textit{Universidade do Porto - CENTRA/SIM, Portugal}),
Panagiotis Gavras (\textit{National observatory of Athens, Greece}),
Paolo Giacobbe (\textit{INAF - Osservatorio Astrofisico di Torino, Italy}), 
Facundo Ariel Gomez (\textit{Max Planck Institute for Astrophysics, Germany}),
Ariel Goobar (\textit{Stockholm University, Sweden}),  (\textit{Universidade de Lisboa - CENTRA/SIM, Portugal}),
Raffaele Gratton (\textit{INAF - Osservatorio Astronomico di Padova, Italy}),
Fabrizia Guglielmetti	 (\textit{Max-Planck Institute for Astrophysics, Germany}),
Eike Gunther (\textit{TLS, DE}),
David Hall (\textit{Open University, UK}),
Diana Harrison (\textit{IoA, Cambridge, UK}),
Artie Hatzes (\textit{TLS, Germany})
Daniel Hestroffer (\textit{IMCCE, France}),
Emille Ishida (\textit{Universit\'e Blaise-Pascal, France}),
Pascale Jablonka (\textit{EPFL, Switzerland}),
Christopher Jacobs (\textit{JPL, USA}),
Markus Janson (\textit{Stockholm University, Sweden}),
Jens Jasche (\textit{Excellence Cluster Universe / Technical University of Munich, Germany}), 
Mathilde Jauzac (\textit{Durham University, UK}),
Hugh Jones (\textit{University of Hertfordshire, UK}),
Peter Jonker (\textit{SRON, Netherlands Institute for Space Research, The Netherlands}),
Francesc Julbe (\textit{Dapcom Data Services S.L., Spain}),
Jean-Paul Kneib (\textit{EPFL, Switzerland}),
Georges Kordopatis (\textit{Leibniz institute fur Astrophysik, Germany}),
Arianne Lancon (\textit{Universit\'e de Strasbourg, France}),
Mario Gilberto Lattanzi (\textit{INAF - Osservatorio Astrofisico di Torino, Italy}),
Jean-Michel Leguidou  (\textit{CNES, France}),
Matt Lehnert (\textit{IAP, France}),
Harry Lehto (\textit{Tuorla Observatory, University of Turku, Finland}),
Ilidio Lopes (\textit{CENTRA, IST - Universidade de Lisboa, Portugal }),
Jesus Maldonado (\textit{INAF -  Osservatorio Astronomico di Palermo, Italy}),
Marcella Marconi (\textit{INAF-Osservatorio Astronomico di Capodimonte, Italy}),
Nicolas Martin (\textit{Universit\'e de Strasbourg, France}),
Marie-Elisabeth	Maury (\textit{French Air Force, France}), 
Anupam Mazumdar (\textit{Lancaster, UK}),
Tatiana Michtchenko (\textit{Universidade de S\~ao Paulo, Brasil}),
Stefano Minardi (\textit{AIP, Germany}), 
Carlos Munoz (\textit{UAM \& IFT, Madrid, Spain}),
Giuseppe Murante (\textit{INAF - Osservatorio Astronomico di Trieste, Italy}),
Neil Murray (\textit{The Open University, UK}),
Ilaria Musella (\textit{INAF-Osservatorio Astronomico di Capodimonte, Italy}),
Gerhard	Ortwin	(\textit{Max-Planck-Inst. for Ex. Physics, Germany}), 
Isabella Pagano (\textit{INAF - Osservatorio Astrofisico di Catania, Italy}),
Paolo Pani (\textit{Sapienza U. of Rome \& CENTRA-IST Lisbon, Italy}),
Martin Paetzold (\textit{Universität zu Köln, Germany}), 
Daniel	Pfenniger	(\textit{Observatory of Geneva, Switzerland}), 
Giampaolo Piotto (\textit{Universita' di Padova, Italy}),
Olivier Preis (\textit{Laboratoire Lagrange - OCA - Nice, France}),
Nicolas Produit (\textit{University of Geneva, Switzerland}),
Jean-Pierre Prost (\textit{Thales Alenia Space, France}),
Heike Rauer (\textit{DLR, Germany}),
Sean Raymond (\textit{Laboratoire d'Astrophysique de Bordeaux, France}),
Rosa Reinaldo (\textit{National Institute for Space Research (INPE), Brazil}), 
Yves Revaz (\textit{Observatoire de Gen\`eve, Switzerland}), 
Rita	Ribeiro (\textit{UNINOVA, Portugal}), 
Lorenzo Rimoldini (\textit{University of Geneva, Dept. of Astronomy, Switzerland}),
Arnau Rios Huguet (\textit{University of Surrey, UK}),
Vincenzo Ripepi (\textit{INAF-Capodimonte Observatory, Italy}),
Pier-Francesco Rocci (\textit{Laboratoire Lagrange - CNRS/INSU, France}),
Maria D. Rodriguez Frias (\textit{UAH, Spain}),
Reinaldo R. Rosa (\textit{National Institute of Space Research, Brazil}), 
Johannes Sahlmann (\textit{Research Fellow within the ESA Science Operations Department, N/A}),
Ippocratis Saltas (\textit{IA - Universidade de Lisboa, Portugal}),
Jos\'e Pizarro Sande e Lemos (\textit{CENTRA, Instituto Superior Tecnico - Universidade de Lisboa, Portugal}),
Luis M. Sarro (\textit{UNED, Spain}),
Bjoern Malte Schaefer  (\textit{Heidelberg University, Germany}),
Jascha Schewtschenko (\textit{Durham, UK}),
Jean Schneider (\textit{Observatoire de Paris Meudon, France}),
Damien Segransan (\textit{University of Geneva, Switzerland}),
Franck Selsis (\textit{Laboratoire d'Astrophysique de Bordeaux, France}),
Joe Silk (\textit{IAP, France}),
Manuel Silva (\textit{CENTRA/SIM - FEUP - Universidade do Porto, Portugal}),
Filomena Solitro (\textit{ALTEC, Italy}),
Alessandro Spagna (\textit{INAF - Osservatorio Astrofisico di Torino, Italy}),
Volker Springel (\textit{Heidelberg University, Germany}),
Maria S\"uveges (\textit{University of Geneva, Switzerland}),
Ramachrisna Teixeira (\textit{Universidade de S\~ao Paulo, Brazil}),
Shindou	Tetsuo	(\textit{Kogakuin University, Japan}), 
Philippe Thebault (\textit{Observatoire de Paris, France}),
Feng Tian (\textit{Tsinghua University, China}),
Catherine Turon (\textit{GEPI, Observatoire de Paris, France}),
Jos\'e W. F. Valle (\textit{IFIC (\textit{UV-CSIC}), Spain}),
Eugene Vasiliev	(\textit{Oxford University, UK}), 
Juan Vladilo (\textit{INAF - Osservatorio Astronomico di Trieste, Italy}),
Martin	Vollmann (\textit{Technical University of Munich, Germany}), 
Martin Ward (\textit{Durham University - Ippp, UK}),
Jochen Weller	(\textit{Ludwig-Maximilians University Munich, Germany}), 
Mark Wilkinson (\textit{University of Leicester, UK}), 
Sebastian Wolf (\textit{Uni Kiel, Germany}), 
Fu Xiaoting (\textit{SISSA, Italy}),
Yoshiyuki Yamada (\textit{Kyoto University, Japan}), Mei Yu (Texas A\&M University, USA),
Sven Zschocke (\textit{Lohrmann Observatory at Dresden Technical University, Germany}),
Shay Zucker (\textit{Tel Aviv University, Israel})

\onecolumn

\thispagestyle{empty}

\onecolumn
\tableofcontents
\newpage

\section{Executive summary}

\subsection{\theia's aims}
What is the nature of dark matter? Are there habitable exo-Earths nearby? What is the equation of state of matter in extreme environments? These are the fundamental questions the \theia astrometric space observatory is designed to answer. Through its ultra-precise micro-arcsecond relative astrometry, \theia will address a large number of prime open questions in three themes of ESA's cosmic vision:

\vspace{-0.3cm} 
\subsection*{$\bullet$ Dark matter (the main focus of the mission)}
\vspace{-0.05cm} 
\theia will dramatically advance cosmology by determining the small-scale properties of the dark matter (DM) component in the local Universe. It is the first space observatory designed to test for signatures of models beyond the Standard Model of particle physics, and it will either confirm or invalidate Cold Dark Matter (CDM) and various theories of primordial inflation.
\theia will:
\vspace{-0.1cm} 
\begin{itemize}
\setlength\itemsep{0em} 
\item examine whether DM in the inner part of faint dwarf spheroidal galaxies is cuspy or more homogeneously distributed;
\item determine whether the outer halo of the Milky Way is prolate;
\item detect small DM halos by finding the gravitational perturbations they have left on the Milky Way disc; and
\item test inflationary models by detecting ultra-compact mini-halos of DM.
\end{itemize} 
This will help us understand the origin and composition of the Universe (theme 4 of ESA's Cosmic Vision). 

\vspace{-0.3cm} 
\subsection*{$\bullet$  Exoplanets}
\vspace{-0.05cm} 
\theia will provide the first direct measurements of the masses and inclinations of a significant sample of Earth and super-Earth planets orbiting our nearest star neighbours. This census of habitable exoplanets will be crucial for future exobiology missions. Spectroscopic follow-ups to \theia will enable the detection of possible signatures of complex life and the chemical pathways to it.
This will help us understand the conditions for planet formation and the emergence of life, and how the Solar System works (themes 1-2 of ESA's Cosmic Vision).
 
\vspace{-0.3cm} 
\subsection*{$\bullet$  Neutron stars and black holes}
\theia will determine the masses of more than 15 neutron stars by measuring binary orbital motion. In conjunction with X-ray measurements from other missions (e.g., Athena), \theia will improve neutron star radius measurements for a dozen systems, which will constrain their composition and equation of state. For black hole binaries, \theia will also make proper motion measurements to understand their formation, and orbital measurements to determine if their accretion discs are warped.
This will help us understand the fundamental physical laws of the Universe (theme 3 of ESA's Cosmic Vision).

\vspace{-0.3cm} 
\subsection{Scientific instruments}
The payload is deliberately simple: it includes a single three-mirror anastigmat telescope, with metrology subsystems and a camera. The telescope is an Korsch on-axis three-mirror anastigmat telescope (TMA) with an 80 cm primary mirror. The camera focal plane consists of 24 detectors, leading to a Nyquist sampled field of view $\simeq 0.5^\circ$, and four wavefront sensors. Its metrology subsystems ensure that \theia can achieve the sub-microarcsecond astrometric precision that is required to detect habitable exoplanets near us.

\vspace{-0.3cm} 
\subsection{Significant additional benefits}
\theiap main purpose is to observe the targets set by our science cases, but it will use its repointing and stabilization phases to perform photometric observations to infer the age of the Universe to a unique precision.
In addition, \theia will benefit the community by reserving 15$\%$ of the observing time for open call proposals, and allowing the public to "crowd-select" four astronomic objects to be scrutinised.
\theiap measurements will significantly improve the knowledge we gain from other key ground and space research programs. \theiap ultra-precise astrometry will serve as a new reference standard, and benefit the broader astronomical community, as the natural astrometric successor to ESA/\hips and \gaia. It will open  promising new avenues for scientific breakthroughs in astronomy, astrophysics and cosmology.

\newpage

\begin{table}
\begin{center}
\begin{tabular}{||l|l||}
\hline 
Science case  & Dark Matter, Exoplanets, Neutron stars and Binary Black Holes. \\
\hline 
Science objectives & $\bullet$ To discover the nature of dark matter; \\
&$\bullet$ To find nearby habitable Earths; \\
&$\bullet$ To probe Nature's densest environments.  \\ 
\hline 
Overview & $\bullet$ Spacecraft at L2 for 4.5 years; \\ 
&$\bullet$ Optical telescope (350nm-1000nm); \\ 
& $\bullet$  Micro-arcsecond astrometry, sub-percent photometry;\\ 
&$\bullet$ Point and stare strategy, to enable relative (differential) astrometry;\\
&$\bullet$ Built on \gaia's "absolute" reference frame. \\ 
\hline 
What makes \theia unique?  
& $\bullet$   Ultra-high-precision astrometry, only reachable from space: \\ 
&  \hspace{0.2cm} from 10 $\mu \rm{as}$ (dark matter) down to 0.15 $\mu \rm{as}$ (exoplanets);\\
&$\bullet$ Dedicated payload design to achieve the required astrometric precision;\\ 
& $\bullet$ Unprecedented sensitivity to DM targets, enabling particle physics tests; \\ 
& $\bullet$ True masses and orbital architecture of habitable-zone terrestrial planets,\\ 
& \hspace{0.2cm} and complete orbital characterization of planetary systems;\\
& $\bullet$ Measurements of orbits and distances to probe the interiors of neutron stars \\
& \hspace{0.2cm} and the structure of black hole accretion discs.\\
\hline
& $\bullet$ dwarf spheroidals $\&$  ultra-faint dwarf galaxies, hyper-velocity stars; \\
Main observational targets  &$\bullet$ nearby A, F, G, K, M stellar systems; \\ 
& $\bullet$ neutron stars in X-ray binaries;\\
& $\bullet$ Milky Way disc $+$ open observatory targets.\\
 \hline 
Payload 
    & $\bullet$ Korsch on-axis TMA telescope with controlled optical aberrations; \\
    & $\bullet$  Primary mirror: $D=0.8\,$m diameter; \\
    & $\bullet$ Long focal length, $f=32$\,m; \\ 
    &  $\bullet$ FoV $\sim $0.5\,deg, with 4 to 6 reference stars with magnitude $R\leq10.8$\,mag;\\
    & $\bullet$ Focal plane with 24 CCD detectors ($\sim$402 Mpixels, 350nm-1000nm); \\
    & $\bullet$ Nyquist sampling of the point-spread-function; \\
\\
    & $\bullet$ Metrology calibration of the focal plane array: relative positions of pixels \\
    &\hspace{0.2cm}   at the micropixel level using Young's interferometric fringes; \\ 
    & $\bullet$ Interferometric monitoring of the telescope: picometer level determination; \\ 
    &\hspace{0.2cm} of the telescope geometry using laser interferometric hexapods.\\
\hline 
Spacecraft & 
$\bullet$ Spacecraft dry mass with margin: 1063\,kg. Total launch Mass: 1325\,kg; \\
& 
$\bullet$ Attitude Control System: synergistic system with hydrazine, reaction \\ 
& \hspace{0.2cm} wheels and cold-gas thrusters. RPE: 20 mas rms in a few minutes ($1\sigma$);\\
& $\bullet$ Thermal Control System: active thermal control of telescope;\\ 
&\hspace{0.2cm}  dedicated radiator for the payload; \\
& 
$\bullet$ Telecommand, Telemetry and Communication: Ka-band, $\sim$95 GBytes of \\ 
& \hspace{0.2cm}  science data per day. High Gain Antenna and 35m stations.\\
\hline 
Launcher and operations & $\bullet$ Ariane 6.02. Lissajous orbit at L2. Launch in 2029;\\ 
& $\bullet$ Nominal mission: 4 yrs $+$ 6 months transit, outgassing $\&$ commissioning;\\
& $\bullet$ MOC at ESOC, SOC at ESAC. \\
\hline 
Data policy & $\bullet$ Instrument Science Data Centers at consortium member states;\\
& $\bullet$ Short proprietary period and 2 data releases.\\
\hline 
Consortium &$\bullet$ {> 180} participants from 22 countries; \\
& \hspace{0.2cm} UK, France, Germany, Italy, Spain, Switzerland, Poland, Portugal, \\ 
& \hspace{0.2cm} Sweden, The Netherlands, Hungary, Greece, Denmark,  Austria, Finland, \\
& \hspace{0.3cm}USA, Brazil,  China, Canada, India, Israel, Japan. \\ 
\hline 
Estimated cost & $\bullet$ 536\,M\euro\ for the spacecraft and telescope, including launcher (70),  \\ 
& \hspace{0.2cm} ground segment (85), project (53) and payload contribution (56).\\
  &  $\bullet$  51.3\,M\euro \ for the payload (consortium member states only)\\
\hline 
\end{tabular}
\end{center}
\end{table}


\twocolumn

\section{Science case }
Europe has always been a pioneer of astrometry, from the time of ancient Greece to Tycho Brahe, Johannes Kepler, the Copernican revolution and Friedrich Bessel. ESA's \hips and \gaia satellites continued this tradition,  revolutionizing our view of the Solar Neighborhood and Milky Way, and providing  a crucial foundation for many disciplines of astronomy.

\theia's unprecedented microarcsecond relative precision will advance European astrometry still further, setting the stage for breakthroughs on the most critical questions of cosmology, astronomy and particle physics.

\subsection{Dark Matter }

The current hypothesis of cold dark matter (CDM) urgently needs verification. Dark matter (DM) is essential to the $\Lambda$ + CDM cosmological model ($\Lambda$CDM), which successfully describes  the large-scale distribution of galaxies and  the angular fluctuations of the
  Cosmic Microwave Background, as confirmed by the ESA/Planck mission. Dark matter is the dominant form of matter ($\sim 85\%$) in the Universe, and ensures the formation and stability of enmeshed galaxies and clusters of galaxies. The current paradigm is that dark matter is made of heavy, hence cold, particles; otherwise galaxies would not form. However, the nature of dark matter is still unknown.

There are a number of open issues regarding $\Lambda$CDM on small-scales. 
Simulations based on DM-only predict a 1) large number of small objects orbiting the Milky Way, 2) a steep DM distribution in their centre and 3) a prolate Milky Way halo. However, hydrodynamical simulations, which include dissipative gas and violent astrophysical phenomena (such as supernovae explosions and jets from galactic nuclei) can change this picture. Quantitative predictions are based on very poorly understood sub-grid physics and there is no consensus yet on the results. Answers are buried at small-scales, which are extremely difficult to probe. A new astrometric mission such as \theia appears to be the best way to settle the nature of DM. 
\theia will allow us to validate or refute key predictions of $\Lambda$CDM, such as

\begin{itemize}
\item The DM distribution in dwarf spheroidal galaxies  
\item The outer shape of the Milky Way DM halo 
\item The lowest masses of the Milky Way satellites 
\item The power spectrum of density perturbations
These observations will significantly advance research into DM. \theia's observations may indicate that DM is warmer than $\Lambda$CDM predicts. Or we may find that DM is prone to self-interactions that reduce its density in the central part of the satellites of the Milky Way. We may discover that DM has small interactions that reduce the number of satellite companions. Alternatively, \theia's measurement of the Milky Way DM halo could reveal that DM is a sophisticated manifestation of a modification of Einstein's gravity. Astrometric microlensing (see Sec.~\ref{sec:Compact objects in the GC}) could even reveal that DM is made of primordial black holes rather than particles.
\end{itemize}

\subsubsection{ The Dark Matter distribution in dwarf spheroidal galaxies}

\begin{figure}[h]
\centering
\includegraphics[width=0.9\textwidth]{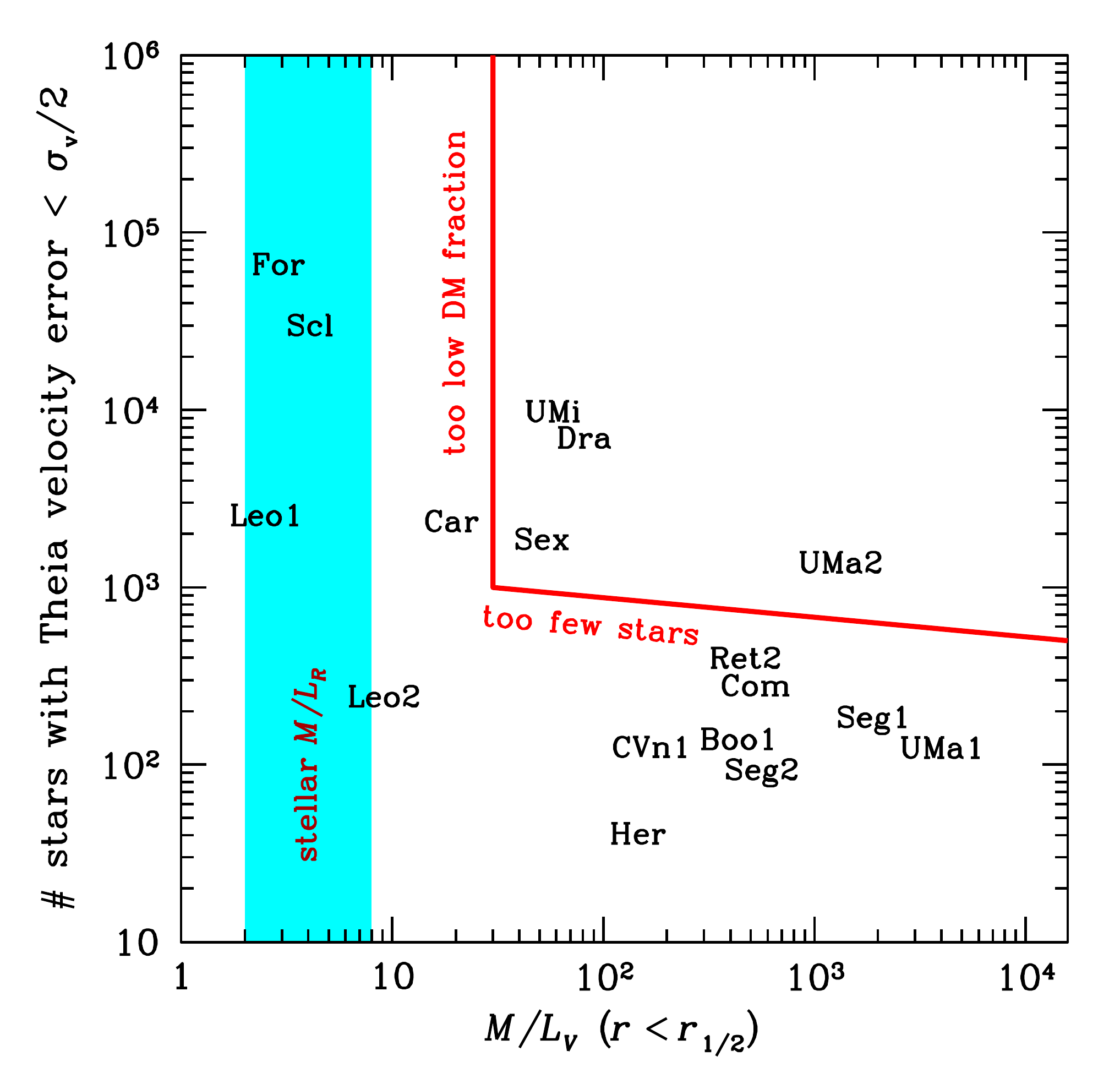}
\caption{Number of dwarf spheroidal galaxy stars  within the \theia field with expected plane-of-sky errors lower than half the galaxy's velocity dispersion as a function of the galaxy's estimated mass-to-light ratio within the effective (half-projected-light) radius of the galaxy. 
Luminosities and total masses within the half-light radii are mainly from \cite{Walker+09}.}
\label{fig:NvsMoverL}
\end{figure}

\begin{figure*}[t]
\centering
\includegraphics[width=0.35\textwidth]{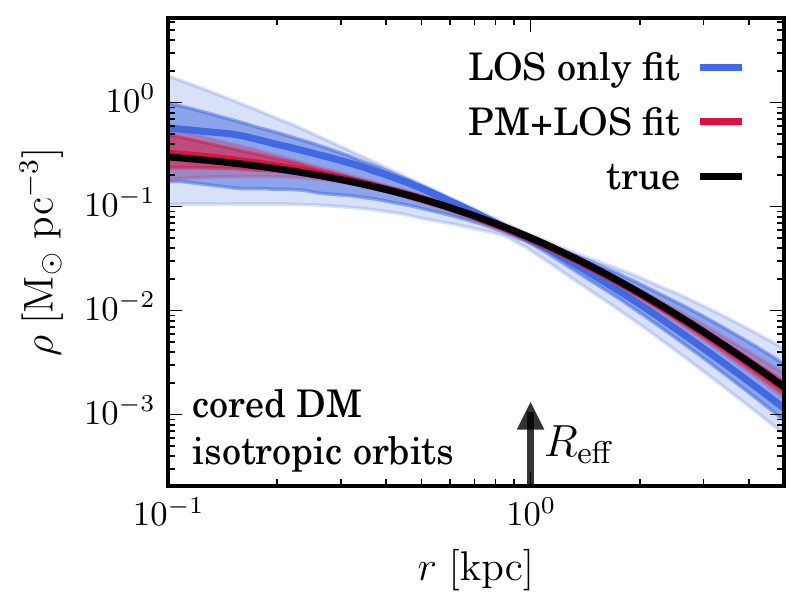}
\includegraphics[width=0.35\textwidth]{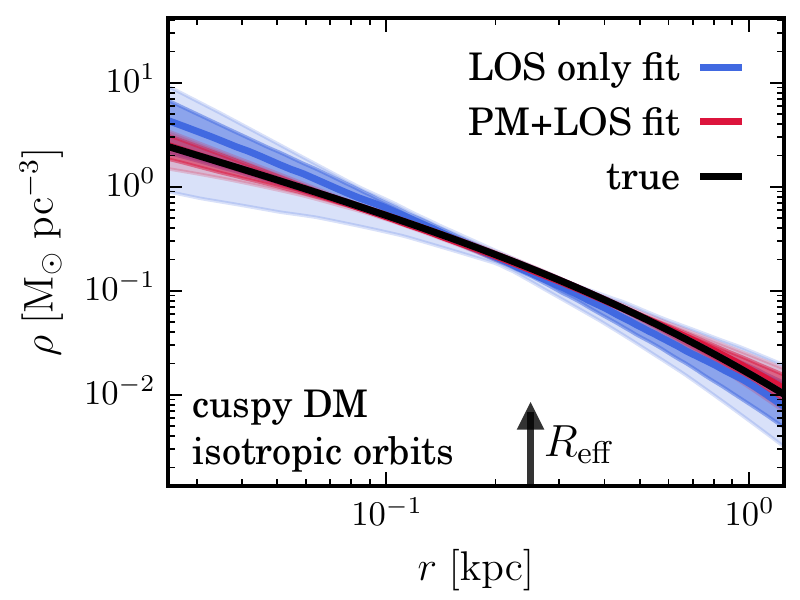}
\includegraphics[width=0.35\textwidth]{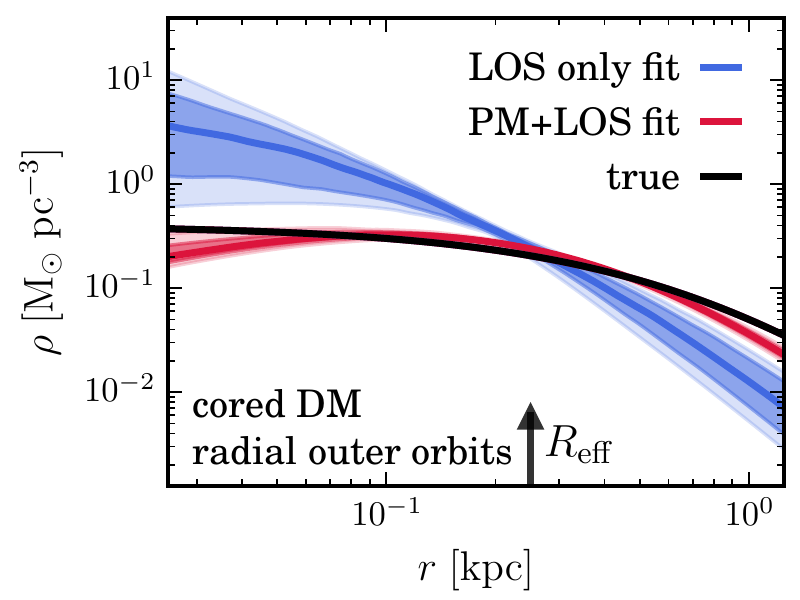}
\includegraphics[width=0.35\textwidth]{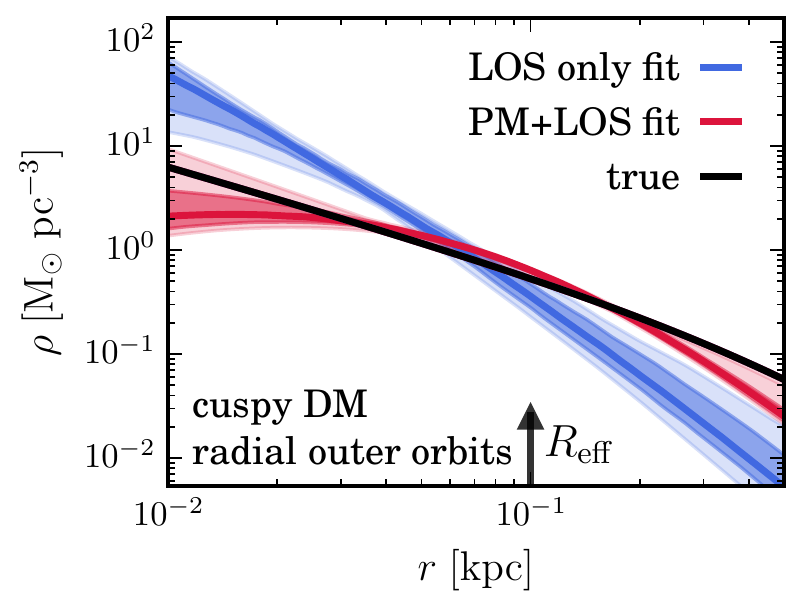}
\caption{Reconstruction of the DM halo profile of the Draco dSph without (\textit{blue}) and with (\textit{red}) proper motions using the mass-orbit modeling algorithm of \citet{Watkins+13}. Four mocks of Draco were used, with cored (\textit{left}) and cuspy (\textit{right}) DM halos, and with isotropic velocities everywhere (\textit{top}) or only in the inner regions with increasingly radial motions in the outer regions (\textit{bottom}). The effective (half-projected light) radii of each mock is shown with the \textit{arrows}. The stellar proper motions in the mocks were given errors, function of apparent magnitude, as expected with 1000 hours of observations spread over 4 years.
Only with proper motions can the DM density profile be accurately reconstructed, properly recovering its cuspy or cored nature.}
\label{fig:rhoofrPM}
\end{figure*}

Because they are DM-dominated (see Fig.~\ref{fig:NvsMoverL}, where the number of stars versus the mass-to-light ratio is displayed),  dwarf Spheroidal galaxies (dSphs) are excellent laboratories to test the distribution of DM within the central part of small galaxies and disentangle the influence of complex baryonic processes from that of dark matter at these scales.

Simulations (e.g. \citealt{Onorbe+15,Read+16}), show that the DM distribution (referred to as DM profile) in dSphs strongly depends on their star formation history. More specifically, these simulations find that CDM can be heated by bursty star formation inside the stellar half light radius $R_{1/2}$, if star formation proceeds for long enough. As a result, some dSphs like Fornax have formed stars for almost a Hubble time and so should have large central dark matter cores, while others, like Draco and Ursa Minor, had their star formation truncated after just $\sim 1-2$\,Gyrs and should retain their steep central dark matter cusp.

Large DM cores could also be  attributed however to strong self-interactions. 
Hence finding evidence for such cores in the faintest dSphs (which are even more DM dominated \citep{Wolf+10} than the classical ones), would bring tremendous insights about the history of baryonic processes in these objects and could even dramatically change our understanding of the nature of dark matter. 
Indeed, self-interacting DM \citep{Spergel&Steinhardt00} is expected to scatter in the dense inner regions of dSphs, and thus leads to homogeneous cores. Finding such a core DM distribution in dSphs could then reveal a new type of particle forces in the dark matter sector and provide us with new directions to  build extensions of Standard Model of particle physics. On the other hand, finding cuspy DM profiles in all dSphs (including the faintest ones) would confirm $\Lambda$CDM and place strong constraints on galaxy formation. As shown in Figs.~\ref{fig:vposerrvsDist} and \ref{fig:astroComparions}, with its micro-arcsecond astrometric precision, \theia has the ability to determine whether the DM distribution in dSphs is cuspy or has a core and therefore bring possible very significant breakthrough regarding the nature of DM.

To determine the inner DM distribution in dSphs, one needs to remove the  degeneracy between the radial DM profile and orbital anisotropy that quantifies whether stellar orbits are more radial or more tangential in the Jeans equation \citep{Binney&Mamon82}.   
This can be done by adding the proper motions of stars in dSphs.  Fig.~\ref{fig:rhoofrPM} shows that for the Draco dSph (which was obtained using  single-component spherical mock datasets from the \gaia Challenge Spherical and Triaxial Systems working group,\footnote{ \url{http://astrowiki.ph.surrey.ac.uk/dokuwiki/doku.php?id=tests:sphtri}}  and the number of stars expected to be observed by \theia), the inclusion of proper motions lifts the cusp/core degeneracy that line-of-sight-only kinematics cannot disentangle.

We remark in addition that \theia will be able to perform  follow-ups of \gaiap observations of dSphs streams of stars if needed. Not only will \theia provide the missing tangential velocities for stars with existing radial velocities, but it will also provide crucial membership information - and tangential velocities - for stars in the outer regions of the satellite galaxies that are tidally disrupted by the Milky Way.

\subsubsection{ The triaxiality of the Milky Way dark matter halo \label{triaxial} }

For over two decades cosmological simulations have shown that Milky Way\--like DM halos have triaxial shapes, with the degree of triaxiality varying with radius \citep[e.g.][]{dubins_94, kkzanm04}: halos are more round or oblate at the center, become triaxial at intermediate radii, and prolate at large radii \citep{zemp_etal_11}. 

These departures from spherical symmetry can be tested by precise measurement of the velocity of Hyper Velocity Stars (HVS), entirely independently of any other technique attempted so far (such as the tidal streams). 
HVS were first discovered  serendipitously  
\citep{Brown+05,Hirsch+05,Edelmann+05},
and later discovered in a targeted survey of blue main-sequence stars 
(\citealt{Brown15_ARAA} and references therein).
They are located between 20 and 100 kpc from the Galactic Center and have radial velocities that significantly exceed the Galactic escape velocity.

\begin{figure}[thb]
\centering
\includegraphics[width=0.9\textwidth,clip]{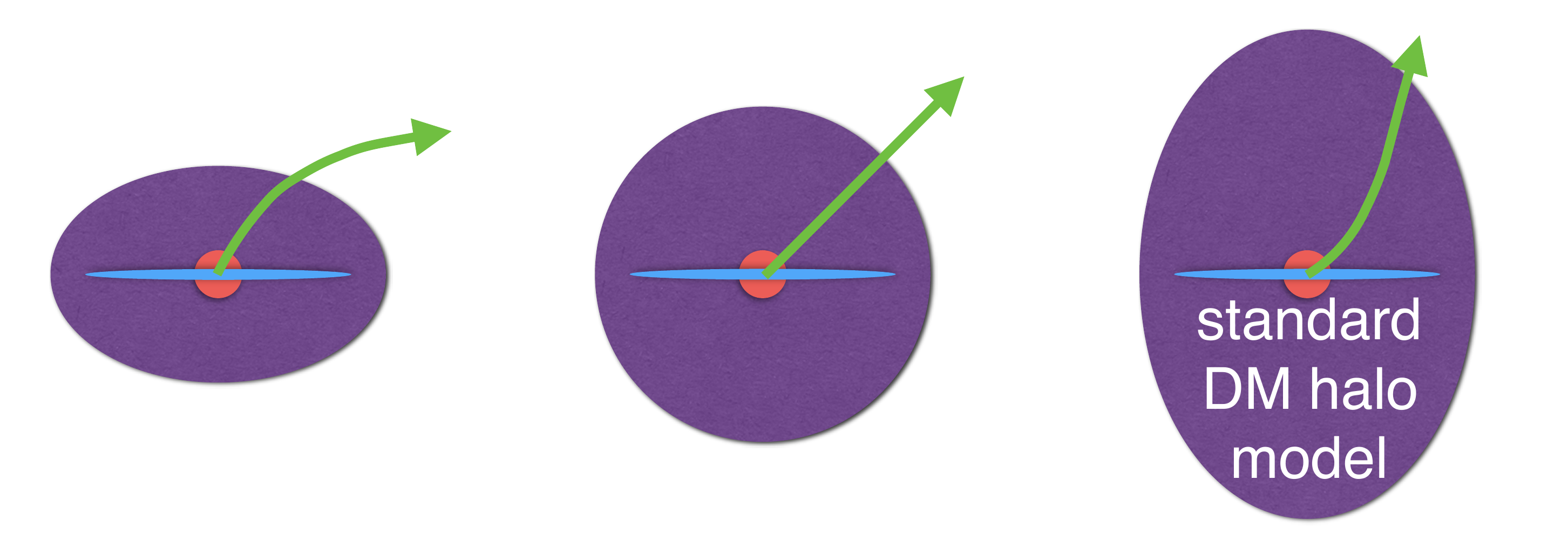}
\caption{Illustration of the trajectories of hyper velocity stars ejected from Galactic Centre for 3 different outer dark matter halo shapes: oblate (\textit{left}), spherical (\textit{middle}), and prolate (\textit{right}).} 
\label{Fig:triaxial_halos}
\end{figure}

Because these velocities exceed the plausible limit for a runaway star ejected from a binary, in which one component has undergone a supernova explosion, the primary mechanism for a star to obtain such an extreme velocity is assumed to be a three-body interaction and ejection from the deep potential well of the supermassive black hole at the Galactic center \citep{Hills88,Yu&Tremaine03}.
\begin{figure}[hbt]
    \includegraphics[width=0.9\textwidth,trim=0cm 0cm 0cm 1.5cm,]{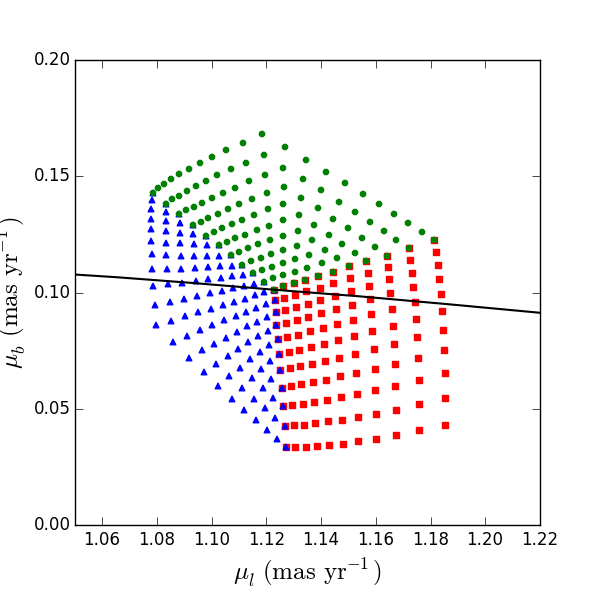}
\caption{Expected proper motions of HVS5 under different assumptions about the shape and orientation of the DM halo. The families of models are shown with the halo major axis along the Galactic X- (\textit{red squares}), Y- (\textit{blue triangles}), and Z-   (\textit{green circles}) coordinates.  The \textit{solid line} shows how the centroid of the proper motions would shift with a different distance to HVS5.}
\label{fig:pm_hvs5} 
\end{figure}

By measuring the three-dimensional velocity of these stars, we will reconstruct the triaxiality of the Galactic potential. In a spherical potential, unbound HVS ejected from the Galactic center should travel in nearly a straight line, as depicted in Fig.\ref{Fig:triaxial_halos}.  However, for triaxial halos, the present velocity vector should not point exactly from the Galactic Center because of the small curvature of the orbit caused by non-spherically symmetric part of the potential 
\citep{Gnedin+05,Yu&madau07}.
While both the halo and stellar disc induce transverse motions, the effect is dominated by halo triaxiality at the typical distance of HVS. The deflection contributed by the disc peaks around 10 kpc but quickly declines at larger distances,
while the deflection due to the triaxial halo continues to accumulate along the whole trajectory. Fig.~\ref{fig:pm_hvs5} actually shows the spread of proper motion for one star, HVS5, for different halo shapes (different halo axis ratios and different orientations of the major axis).

Proper motions of several HVSs were measured with the Hubble Space Telescope (\textit{HST}) by \cite{Brown+15},
using an astrometric frame based on background galaxies (the FOV was too small to include any quasars).  However, these measurements were not sufficiently accurate to constrain the halo shape or the origin of HVS. \theia has a sufficiently large FOV to include about 10 known quasars from the SDSS catalog around most HVSs. This will provide a much more stable and accurate astrometric frame, and will allow us to constrain the halo axis ratios to about 5$\%$. 

\begin{figure}[bht]
\centering
\includegraphics[width=0.8\textwidth, trim=0.cm 0.3cm 0.8cm 1cm ,clip]{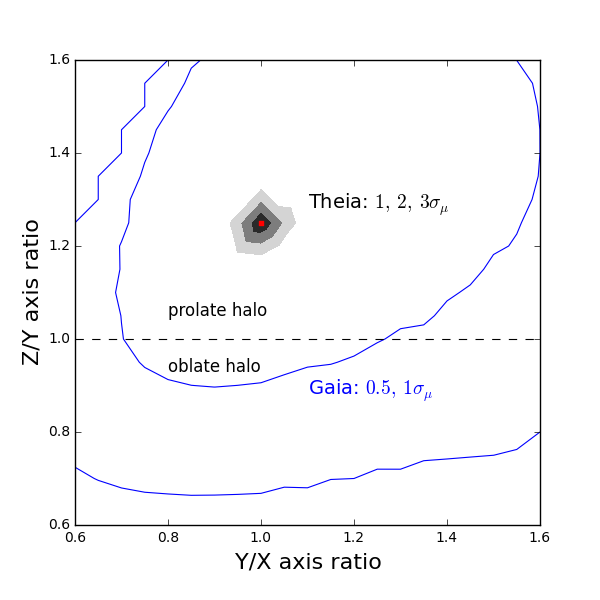}
  \caption{Example of a reconstruction of the Galactic halo shape from \theia's measurement of proper motion of HVS5.  The assumed proper motions correspond to a prolate model with $q_X = q_Y = 0.8\, q_Z$, marked by a \textit{red square}. \textit{Shaded contours} represent confidence limits corresponding to the expected 1, 2, and $3\, \sigma_\mu$ proper motion errors. The \textit{outer blue contours} show the accuracy that would be achieved by \gaia at the end of its mission, even if its expected error was reduced by a factor of 2.} \label{Fig:axishvs5d20theia} 
\end{figure}

Fig.\ref{Fig:axishvs5d20theia} shows indeed that with a precision of $4 \, \mu$as/yr (see Sec.~\ref{sec:requirements_HVS}) we can constrain the orientation of the halo major axis and measure the axis ratios to an accuracy of $\delta (q_Z/q_X) < 0.05$ for the typical HVS distance of 50 kpc.  For comparison, \gaia at the end of its mission would achieve only $40 - 150 \, \mu$as/yr, which is highly insufficient to provide useful constraints on the axis ratios.

Finally, an accurate measurement of HVS velocities may lead to improved understanding of the black hole(s) at the Galactic center. Indeed, theoretical models show that HVSs will have a different spectrum of ejection velocities from a binary black hole versus a single massive black hole.

\subsubsection{Orbital distribution of Dark Matter from the orbits of halo stars}

The orbits of DM particles in halos\footnote{For an analysis of orbital content of DM halos see \cite{valluri_etal_10, valluri_etal_12,bryan_etal_12,valluri_etal_13}.} cannot be detected directly since DM particles interact only weakly with normal matter. However, in a triaxial potential such as described above, it is expected that a large fraction of the DM orbits do have any net angular momentum. Hence these particles should get arbitrarily close to the center of the cusp, regardless of how far from the center they were originally. This allows dark matter particles, which annihilate within the cusp to be replenished for a timescale $10^4$ longer than in a spherical halo \citep[analogous to loss cone filling in the case of binary black holes][]{merritt_poon_04}.

\begin{figure*}[tbh]
 \includegraphics[width=0.4\textwidth]{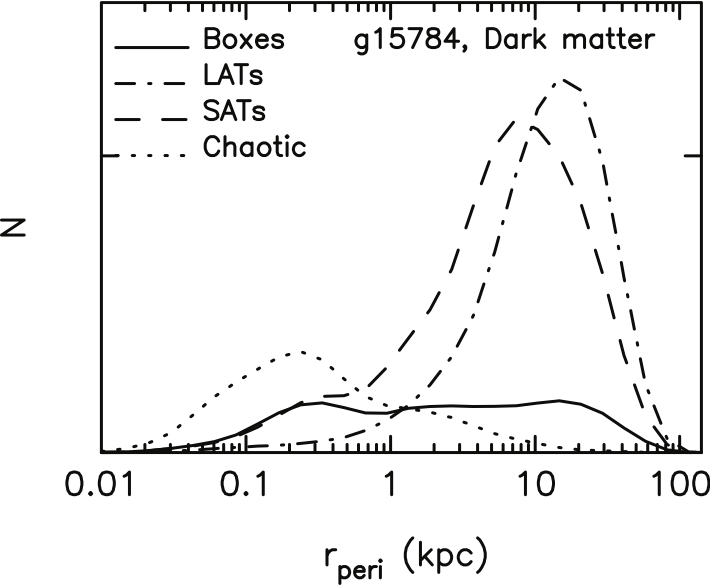}
 \includegraphics[width=0.4\textwidth]{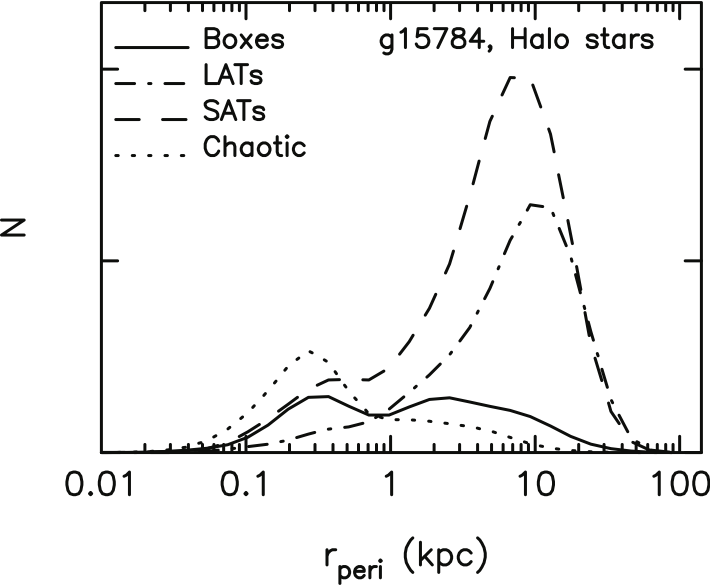}
 \vspace {-0.1in}
\caption{Kernel density distributions of orbit types as a function of orbital pericenter distance $r_{\rm peri}$ for $10^4$ halo dark matter particles (\textit{left}) and halo star particles (\textit{right}) in a cosmological simulation from the MaGICC suite (Stinson et al. 2012). In each panel the vertical axis is proportional to the number of orbits with a particular value of $r_{\rm peri}$, i.e. the sum of integrals over all curves in a panel is equal to the total number of orbits. Line styles indicate the 4 major orbit families (boxes, short-axis tubes, long-axis tubes and chaotic orbits). The orbits of dark matter and halo star particles at small radii are predominantly on centrophilic box and chaotic orbits [reproduced from Valluri et al. 2013.]}
\label{fig:dmOrbits}
\end{figure*}

Recent cosmological simulations show that the orbital distributions of halo stars are similar to those of DM particles \citep[][see Fig~\ref{fig:dmOrbits}]{valluri_etal_13}. The orbits reflect both the accretion/formation history and the current shape of the potential because DM halos are dynamically young (i.e. they are still growing and have not attained a long term equilibrium configuration where all orbits are fully phase mixed). This opens up the very exciting possibility that one can infer the orbital properties of DM particles by assuming that they are represented by the orbits of halo stars.

By combining the high accuracy determination of the shape (see Sec.\ref{triaxial}), the radial scale length, and density normalization of the dark matter halo of the Milky Way with accurate positions and velocities for halo field stars (which are obtained for free in by targeting HVS), we estimate that it will be possible to derive the orbits of 5000-10,000 field stars and thereby to infer the orbital distribution of dark matter particles.

\begin{figure*}[t]
\centering
\includegraphics[width=0.28 \textwidth]{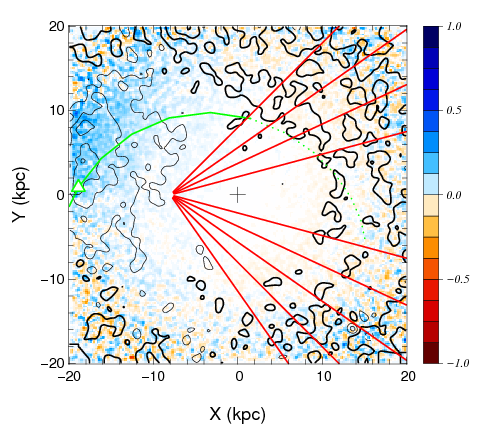}
\includegraphics[width=0.28 \textwidth]{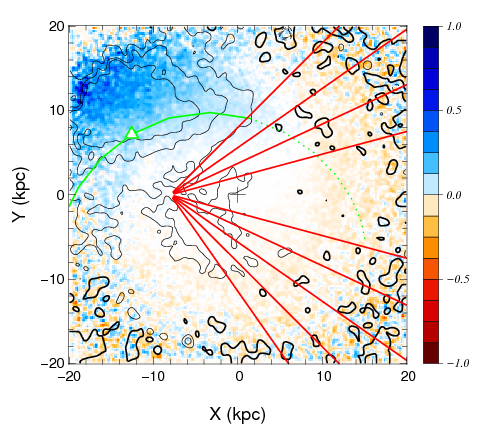}
\includegraphics[width=0.28 \textwidth]{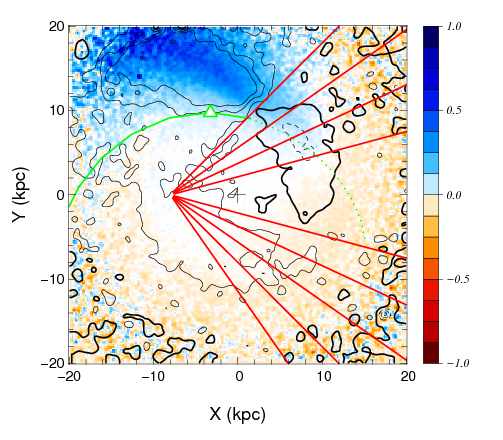}
\includegraphics[width=0.28 \textwidth]{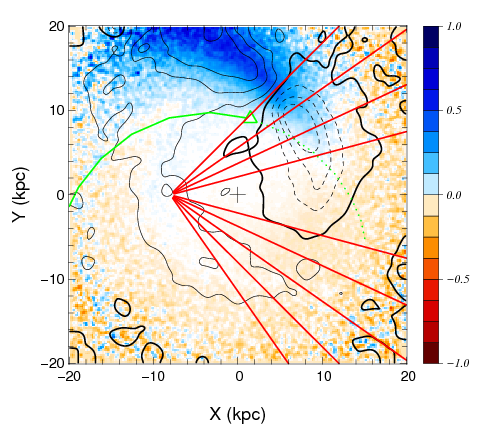}
\includegraphics[width=0.28 \textwidth]{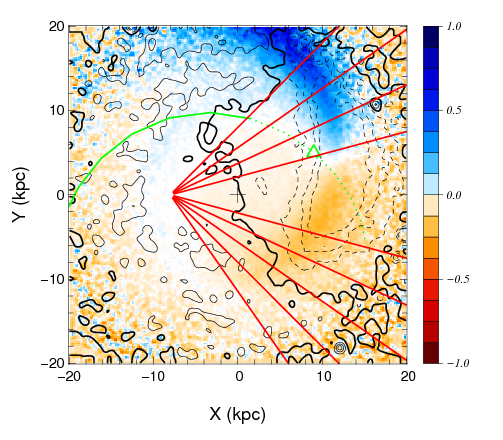}
\includegraphics[width=0.28 \textwidth]{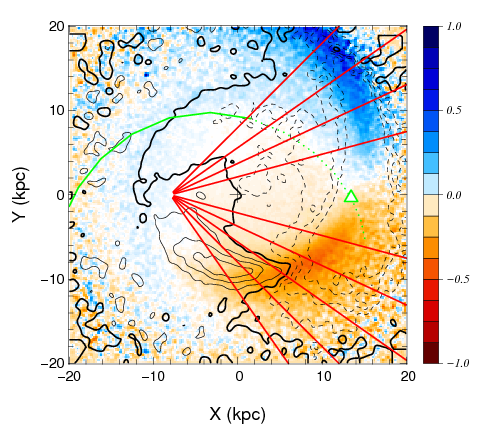}
\caption{Face-on view of the evolution of the perturbation of a Galactic Disc due to a DM subhalo of mass 3$\%$ of the mass of the disc crossing the disc from above. The \textit{upper} and \textit{lower} panels are before and after the crossing, respectively, for different times 125, 75 and 25 Myr before the crossing and 25,75,125 Myr after (from \textit{left} to \textit{right}).  The mean displacement amplitude is indicated in the color bar, while the \textit{contours} indicate the amplitude of the bending mode in velocity space,  using plain lines for positive values and dashed lines for negative values. The \textit{green line} shows the projected orbit of the subhalo (dashed line after the impact with the disc). The \textit{green triangle} shows the current location of the subhalo on its orbit.  The \textit{red lines} are our potential lines of sight for \theia, spaced by 10$^\circ$ in longitude with one pointing above the plane and one below the plane, that would allow us to map the disc perturbation behind the Galactic Center. }
\label{fig:dm:1}
\end{figure*}

\subsubsection{ Perturbations of the Galactic Disc by Dark Matter subhalos}
A central prediction of $\Lambda$CDM in contrast to many alternatives of DM (such as warm DM, e.g.  \citealt{Schaeffer:1984bt} or interacting DM, e.g. \citealt{Boehm:2014vja}) is the existence of numerous $10^6$ to $10^8$ M$_\odot$ DM \emph{subhalos} in the Milky Way halo. Their detection is extremely challenging, as they are very faint and lighter than dSphs. However, N-body simulations of the Galactic Disc show that such a DM halo  passing through the Milky Way disc would warp the disc and produce a motion (bending mode), as shown in Fig.~\ref{fig:dm:1}. 
This opens new avenues for detection as such perturbations of the disc would result in anomalous motions of the stars in the disc (e.g.~\citealt{2015MNRAS.446.1000F} for recent analysis), that could give rise to an astrometric signal.

\begin{figure*}[ht]
\centering
\includegraphics[width=0.45\textwidth, clip]{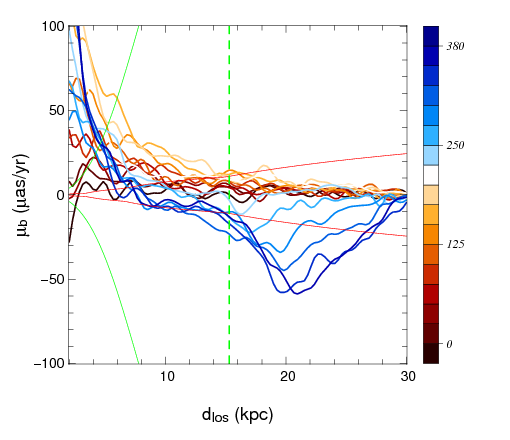}
\includegraphics[width=0.45 \textwidth,clip]{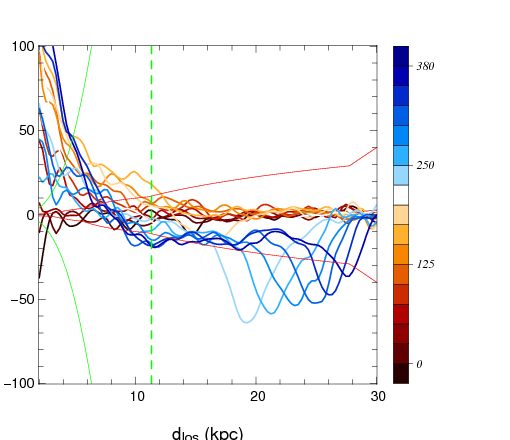}
\caption{Astrometric signatures in the proper motion along Galactic latitude of the perturbation of disc stars by a subhalo. The \textit{left} and \textit{right} panels show lines of sight as a function of distance along the line of sight and time,  for $\ell=-25^\circ$  and $\ell=+25^\circ$ 
respectively for \remove{a Galactic latitude of} $b=+2^\circ$. The color codes the time in Myr, \textit{red} for times prior to the crossing of the plane by the satellite, \textit{blue} for later times. The \textit{green line} is \gaia's expected end of mission performance for a population of red clump stars along these lines of sight. The \textit{vertical dashed line} is \gaia's detection limit ($G$=20) for the same population. The \textit{red lines} are \theia's expected 1$\sigma$ accuracy  for the same stars and for a 400~h exposure of the field over the course of the mission.
}
\label{fig:dm:2}
\end{figure*}

These anomalous bulk motions develop both in the solar vicinity \citep{2012ApJ...750L..41W} and on larger scales 
\citep{2015MNRAS.446.1000F}, see Fig.\ref{fig:dm:2}.
Therefore, measuring very small proper motions of individual faint stars in different directions towards the Galactic disc could prove the existence of these subhalos and confirm the CDM scenario. Alternatively, in case they are not found,  \theia's observations would provide tantalizing evidence for alternative DM scenarios, the most popular today being a warmer form of DM particle, though these results could also indicate dark matter interactions \citep{Boehm:2014vja}.   

\remove{The detection of such halos through the discovery of anomalous bulk motions would definitely rule out DM alternatives, such as light DM candidates with a mass smaller than 1 keV, thus providing a direct measurement of the lower limit on the DM mass. This would also improve the limits on the interaction cross section of DM with Standard model particles \citep{Boehm:2014vja}. }

A field of view of $1^\circ\times 1^\circ$ in the direction of the Galactic disc has $\sim 10^6$ stars with an apparent magnitude of $R \leq 20$ (given by the confusion limit). Given \theia's astrometric precision per field of view,  \theia 
could detect up to 3 impacts on the disc from sub-halos 
as small as a few $10^6\,\Msun$.

\subsubsection{ Ultra-compact minihalos of dark matter in the Milky Way} 
\label{sec:science_minihalos}
In the $\Lambda$CDM model, galaxies and other large-scale structures formed from tiny fluctuations in the distribution of matter in the early Universe. Inflation predicts a spectrum of primordial fluctuations in the curvature of spacetime, which directly leads to the power spectrum of initial density fluctuations.  This spectrum is observed on large scales in the cosmic microwave background and the large scale structure of galaxies, but is very poorly constrained on scales smaller than 2\,Mpc.  This severely restricts our ability to probe the physics of the early Universe.  \theia can provide a new window on these small scales by searching for astrometric microlensing events caused by \emph{ultra-compact minihalos} (UCMHs) of DM.

UCMHs form shortly after matter domination (at $z\sim1000$), in regions that are initially overdense ($\delta\rho/\rho > 0.001$; \citealt{2009ApJ...707..979R}).  UCMHs only form from fluctuations about a factor of 100 larger than their regular cosmological counterparts, so their discovery would indicate that the primordial power spectrum is not scale invariant.  This would rule out the single-field models of inflation that have dominated the theoretical landscape for the past thirty years.  Conversely, the absence of UCMHs can be used to establish upper bounds on the amplitude of the primordial power spectrum on small scales \citep{Bringmann11}, which would rule out inflationary models that predict enhanced small-scale structure \citep{Aslanyan16}.

\begin{figure}[h!tb]
  \includegraphics[width=0.8\textwidth]{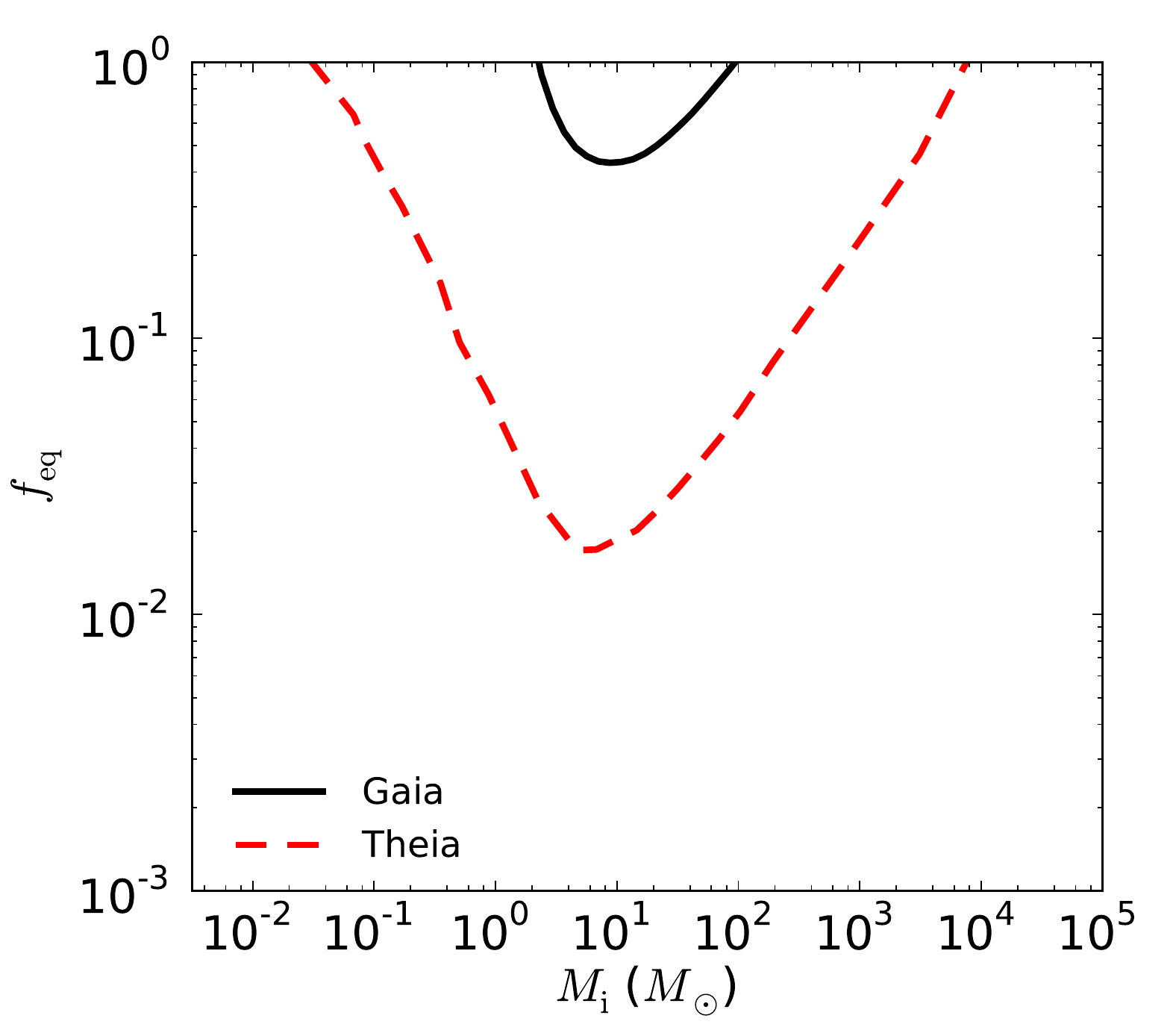}
  \caption{Projected sensitivity of \theia to the fraction of dark matter in the form of ultracompact minihalos (UCMHs) of mass $M_i$ at the time of matter-radiation equality.  Smaller masses probe smaller scales, which correspond to earlier formation times (and therefore to \textit{later} stages of inflation). A UCMH mass of 0.1\,M$_\odot$ corresponds to a scale of just 700\,pc.  Expected constraints from \gaia are given for comparison, showing that \theia will provide much stronger sensitivity, as well as probe smaller scales and earlier formation times than ever reached before.}
  \label{fig:ucmh}
\end{figure}

\begin{figure*}[t]
  \includegraphics[width=0.8\textwidth,trim = 0cm 6.3cm 0cm 7.5cm,clip]{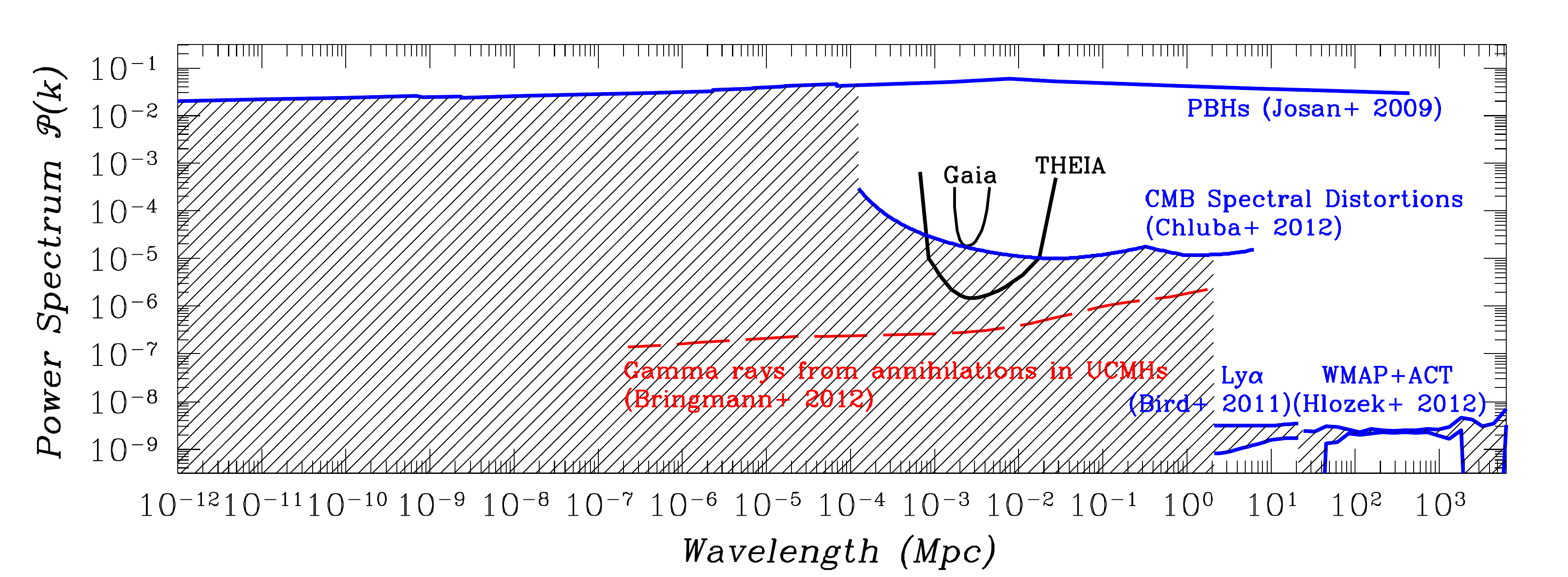}
  \caption{Limits on the power of primordial cosmological perturbations at all scales, from a range of different sources.  \theia will provide far stronger sensitivity to primordial fluctuations on small scales than \gaia, spectral distortions or primordial black holes (PBHs).  Unlike gamma-ray UCMH limits, \theia's sensitivity to cosmological perturbations will also be independent of the specific particle nature of dark matter.}
  \label{fig:ucmh2}
\end{figure*}

Like standard DM halos, UCMHs are too diffuse to be detected by regular photometric microlensing searches for MAssive Compact Halo Objects (MACHOs).  Because they are far more compact than standard dark matter halos, they however produce much stronger \textit{astrometric} microlensing signatures \citep{2012PhRvD..86d3519L}.  By searching for microlensing events due to UCMHs in the Milky Way, \theia will provide a new probe of inflation.  A search for astrometric signatures of UCMHs in the \gaia dataset could constrain the amplitude of the primordial power spectrum to be less than about $10^{-5}$ on scales around 2\,kpc \citep{2012PhRvD..86d3519L}.  Fig.\ \ref{fig:ucmh} shows that with its higher astrometric precision, \theia would provide more than \textit{an order of magnitude higher sensitivity} to UCMHs, and around \textit{four orders of magnitude greater mass coverage} than \gaia.  These projections are based on 8000\,hr of observations of 10 fields in the Milky Way disc, observed three times a year, assuming that the first year of data is reserved for calibrating stellar proper motions against which to look for lensing perturbations.  Fig.\ \ref{fig:ucmh2} shows that \theia would test the primordial spectrum of perturbations down to scales as small as 700\,pc, and improve on the expected limits from \gaia by over an order of magnitude at larger scales.

The results will be independent of the DM nature, as astrometric microlensing depends on gravity only, unlike other constraints at similar scales based on dark matter annihilation, from the \textit{Fermi} Gamma Ray Space Telescope \citep{Bringmann11}.  \theia's sensitivity will be four orders of magnitude stronger than constraints from the absence of primordial black holes (PBHs), and more than an order of magnitude better than CMB spectral distortions \citep{2012ApJ...758...76C}, which give the current best model-independent limit on the primordial power spectrum at similar scales.

\nocite{2011MNRAS.413.1717B, 2012ApJ...749...90H}

\subsection{Exoplanets}

\subsubsection{The Frontier of Exoplanet Astrophysics}

The ultimate exoplanetary science goal is to answer the enigmatic and ancient question, ``Are we alone?'' via unambiguous detection of biogenic gases and molecules in the atmosphere of an Earth twin around a Sun-like star 
\citep{Schwieterman+16}.
Directly addressing the age-old questions related to the uniqueness of the Earth as a habitat for complex biology constitutes today the 
vanguard of the field, and it is clearly recognized as one unprecedented, 
cross-technique, interdisciplinary 
endeavor.

Since the discovery of the first Jupiter-mass companion to a solar-type star 
 \citep{Mayor&Queloz95},
tremendous progress has been made in the field of exoplanets. Our knowledge is expanding ever so quickly due to the discovery of thousands of planets, and the skillful combination of high-sensitivity space-borne and ground-based programs that have unveiled the variety of planetary systems architectures that exist in the Galaxy (e.g., \citealt{2013Sci...340..572H}; \citealt{2011arXiv1109.2497M}). 
Preliminary estimates 
(e.g., \citealt{Winn&Fabrycky15})
are now also available for the occurrence rate $\eta_\uplus$ of terrestrial-type planets in the Habitable Zone (HZ) of stars more like the Sun ($\eta_\uplus\sim10\%$) and low-mass M dwarfs ($\eta_\uplus\sim50\%$). 

However, transiting or Doppler-detected HZ terrestrial planet candidates (including the recent discovery of the $m_{\rm p}\sin i=1.3$ $M_\oplus$ HZ-planet orbiting Proxima Centauri) 
lack determinations of their bulk densities $\varrho_{\rm p}$. Thus, the HZ terrestrial planets known to-date are not amenable to make clear statements on their habitability. 
The \textit{K2}, \textit{TESS}, and \textit{PLATO} missions 
are bound to provide tens of HZ Earths and Super Earths around bright M dwarfs and solar-type stars for which $\varrho_{\rm p}$ estimates might be obtained in principle, but atmospheric characterization for the latter sample 
might be beyond the capabilities of JWST and the Extremely Large Telescopes (ELTs). The nearest stars to the Sun are thus the most natural reservoir for the identification of potentially habitable rocky planets that might be characterized via a combination of high-dispersion spectroscopy and high-contrast imaging with the ELTs 
\citep{Snellen+15}
or via coronagraphic or interferometric  observations in space 
\citep{Leger15}.

 Unlike the Doppler and transit methods, astrometry alone can determine reliably and precisely the true mass and  three-dimensional orbital geometry of an exoplanet, which are fundamental inputs to models of planetary evolution, biosignature identification, and habitability. 
By determining the times, angular separation and position angle at periastron and apoastron passage, \theia's  exquisitely precise position measurements will allow the prediction of where and when a planet will be at its brightest (and even the likelihood of a transit event), thus (a) crucially helping in the optimization of direct imaging observations and (b) relaxing important model degeneracies in predictions of the planetary phase function
in terms of orbit geometry, companion mass, system age, orbital phase, cloud cover, scattering mechanisms and degree of polarization 
(e.g., \citealt{Madhusudhan&Burrows12}).
{\it Only \theia observations have the potential to 1) discover most of the potentially habitable planets around the nearest stars to the Sun, 2) directly measure their masses and system architectures, and 3) provide the most complete target list and vastly improve the efficiency of detection of potential habitats for complex exo-life with the next generation of space telescopes and ELTs.}

\subsubsection{Core Program} 

\begin{figure*}[htb]
\centering
\includegraphics[width=0.65 \textwidth, trim = 0cm 0cm 0cm 1cm, clip]
{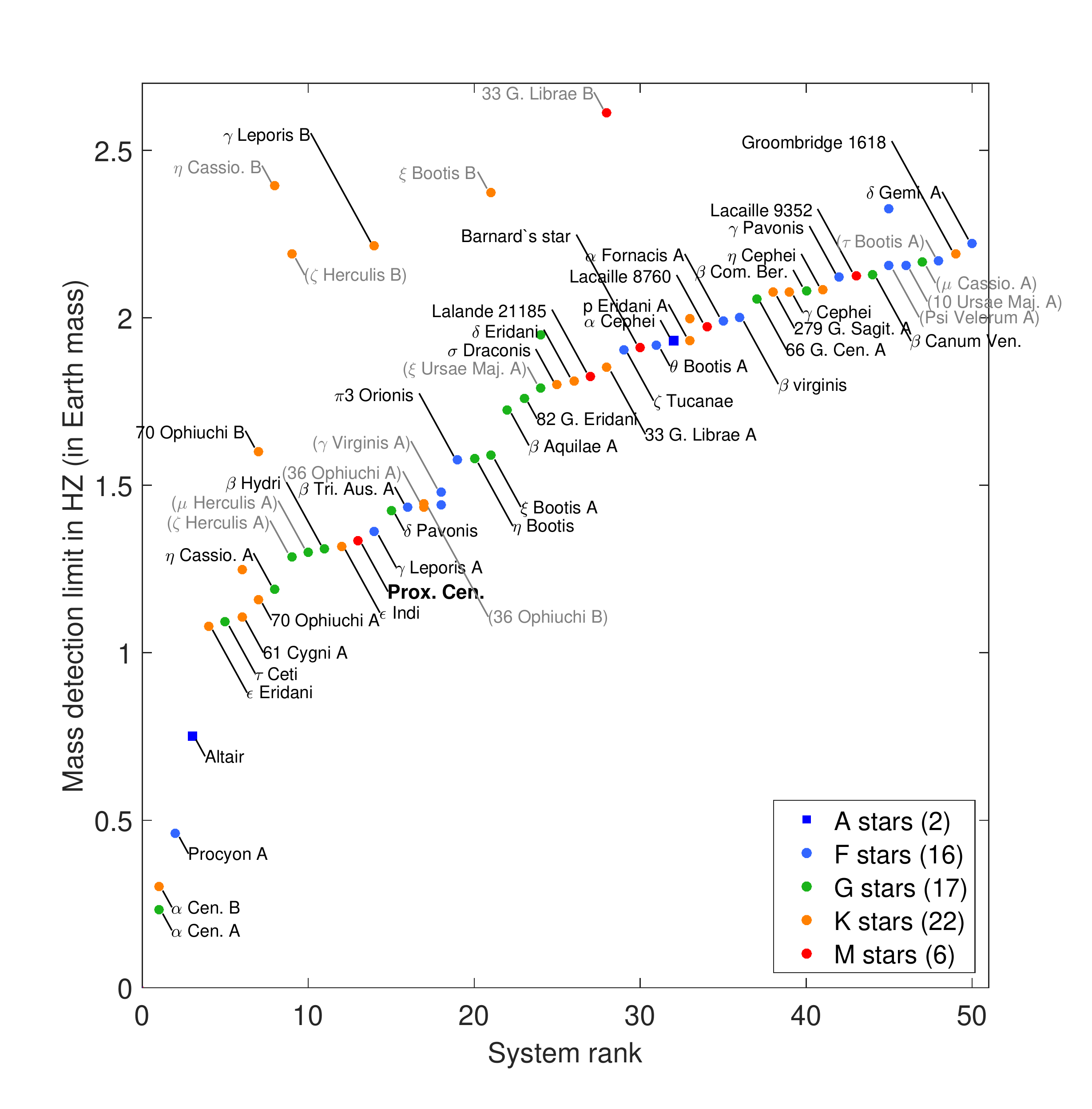}
\caption{Minimum masses of planets that can be detected at the center of the HZ of their star for the 63 best nearby A, F, G, K, M target systems. The target systems (either single or binary stars), are ranked from left to right with increasing minimum detectable mass in HZ around the primary system component, assuming equal observing time per system. Thus for binary stars, A and B components are aligned vertically, as they belong to the same system they share the same rank. When the A and B mass thresholds are close the name is usually not explicitly written down to avoid overcrowding. B components that have mass thresholds above 2.2 $M_\oplus$ are named in gray and binaries that are estimated too close for follow-up spectroscopy are named in gray and in parenthesis. These binaries are expected to be only part of the secondary science program (planet formation around binaries). The star sample that is best for astrometry is similar to that of the best stars for spectroscopy in the visible, or in thermal IR (see text for explanations). Earths and super-Earths with $M_{\rm p} \geq 1.5$ $M_\oplus$ can be detected and characterized (actual mass and full orbit) around 22 stars. All Super-Earths with $M_{\rm p} < 2.2$ $M_\oplus$ can be detected and characterized around 59 stars.
}
\label{fig:exoplan1}
\end{figure*}

\begin{table*}
  \caption{
  Target sample of binaries, including projected and linear separation, distance, spectral types, visual magnitudes, flux ratio, critical separation for direct imaging, corresponding flag for inclusion in the core program (see text for details) and the orbital semi-major axis if known.}
  \label{tab:sci.expl.binaries}    
  \centering
  \begin{center}
    \begin{footnotesize}
  \begin{tabular}{cccccccccccc} 
    \hline 
    Name & $\varrho$ [as] & $a$ [AU] & $D$ [pc] & SpT A & SpT B & VmagA & VmagB & Flux ratio & $\vartheta_\mathrm{AB}$. [as] & Core? & sma[as]\\
    \hline 
    \hline 
$\alpha$ Cen.       & 10.0&   	13.4&  	1.34&   	G2 V   &  	K0 V   &   	0.0&  	1.3&  	3.4&  	5.3& Y& $17.57$ \\ 
61 Cygni            & 31.4&   	108.8&  3.46&   	K5 Ve  &  	K7 Ve  &   	5.2&  	6.0&  	2.1&  	6.2& Y& $24.2$ \\ 
70 Ophiuchi         & 6.6&   	33.2&   5.03&   	K0 Ve  &  	K5 Ve  &   	4.2&  	6.0&  	5.2&  	4.6& Y& $4.56$ \\ 
$\eta$ Cassio.      & 12.9&   	76.6&   5.94&   	G3 V   &  	K7 V   &   	3.5&  	7.5&  	42.1&  	2.3& Y& $11.994$ \\ 
$\zeta$ Herculis    & 1.3&   	13.1&   10.10&   	G0 IV  &  	K0 V   &   	2.9&  	5.4&  	9.9&  	3.7& N& 1.33 \\ 
$\gamma$ Leporis    & 95.0&   	760.6&  8.01&   	F6 V   &  	K2 V   &   	3.6&  	6.1&  	10.5&  	3.7& Y& - \\ 
36 Ophiuchi         & 5.1&   	27.2&   5.33&   	K1 Ve  &  	K1 Ve  &   	5.1&  	5.1&  	1.0&  	7.9& N& 14.7 \\ 
$\gamma$ Virginis   & 2.3&   	23.3&   10.13&   	F0 V   &  	F0 V   &   	3.5&  	3.5&  	1.1&  	7.9& N& $3.66$ \\ 
$\xi$ Bootis        & 5.6&   	37.6&   6.71&   	G8 Ve  &  	K4 Ve  &   	4.7&  	7.0&  	8.1&  	4.0& Y& $4.904$ \\ 
$\xi$ Ursae Maj.    & 1.8&   	18.8&   10.42&   	G0 Ve  &  	G0 Ve  &   	4.3&  	4.8&  	1.5&  	6.9& N& $2.536$ \\ 
33 G. Librae        & 32.0&   	183.7&  5.74&   	K5 Ve  &  	M2 V   &   	5.8&  	8.0&  	7.9&  	4.0& Y& $32.34$\\ 
p Eridani           & 11.4&   	76.6&   6.72&   	K2 V   &  	K3 V   &   	5.8&  	5.9&  	1.1&  	7.8& Y& $7.817$\\ 
$\alpha$ Fornacis   & 5.4&   	74.3&   13.76&   	F7 IV  &  	G7 V   &   	4.0&  	6.7&  	12.6&  	3.4& Y& $4.36$\\ 
Psi Velorum         & 1.0&   	15.6&   15.58&   	F3 IV  &  	F0 IV  &   	4.1&  	4.7&  	1.6&  	6.8& N& $0.862$\\ 
10 Ursae Maj.       & 0.6&   	9.0&   	15.08&   	F3 V   &  	G5 V   &   	4.1&  	6.2&  	6.7&  	4.2& N& $0.6470$\\ 
$\delta$ Gemi.      & 5.7&   	91.3&   16.03&   	F1 IV-V&  	K3 V   &   	3.5&  	8.2&  	73.8&  	1.9& Y& $6.0$\\ 
    \hline 
  \end{tabular}
  \end{footnotesize}
  \end{center}
\end{table*}

Our core program is focused on the use of \theia's surgical single-measurement positional precision in pointed, differential astrometric mode ($<1\mu$as), in order to exploit the mission's unique capability to search for the nearest Earth-like planets to the Sun. The amplitude $\alpha$ of the astrometric motion of a star due to an orbiting planet is (in micro-arcseconds):


\begin{equation}
\alpha = 3\left(\frac{M_{\rm p}}{M_\oplus}\right)\left(\frac{a_{\rm p}}{1\,\mathrm{AU}}\right)\left(\frac{M_\star}{M_\odot}\right)^{-1}\left(\frac{D}{1\,\mathrm{pc}}\right)^{-1}\,\,\mu\mathrm{as}
\end{equation}
\noindent where $M_\star$ is the stellar mass, $M_{\rm p}$ is the mass of the planet, $a_{\rm p}$ is the semi-major axis of the orbit of the planet, and $D$ is the distance to the star. For a terrestrial planet in the HZ of a nearby sun-like star, a typical value is 0.3 $\mu$as (an Earth at 1.0 AU of a Sun, at 10 pc). 
This very small motion (the size of a coin thickness on the Moon as measured from the Earth) is accessible to \theia by measuring the differential motion of the star with respect to far-away reference sources. 

The sample selected for our core program is comprised of 63 of the nearest A, F, G, K, and M stars (Fig. \ref{fig:exoplan1}). Many of them are found in binary and multiple systems. Binary stars are compelling for \theia for a number of reasons. First, they are easier targets than single stars. For close Sun-like binaries, the magnitude of both components is lower (and sometimes much lower) than $V=9$ mag, which is the equivalent magnitude of a typical reference star field composed of 6 $V=11$ mag stars. 

Furthermore, as the photon noise from the references is the dominant factor of the error budget,  the accuracy for binaries increases faster with telescope staring time than around single stars. For binaries, the references only need to provide 
the plate scale and the reference direction of the local frame, the origin point coordinates are constrained by the secondary/primary component of the binary. 
Finally, when observing a binary, the astrometry on both components is obtained simultaneously: the staring time is only spent once as both components are 
within the same FoV. These two effects combined cause the observation of stars in binary systems to be much more efficient (as measured in $\mu$as$\times h^{-1/2}$) than that of single stars.

The binary star sample has projected separations $\theta_\mathrm{AB}$ in the range 0.6-100 arcsec.  We have evaluated the limiting values of $\vartheta_\mathrm{AB}$ 
for which the companion does not constitute a problem for the direct detection by e.g. a 10-m space coronagraph in the visible, a near-infrared space interferometer 
with four arms equipped with 1-m mirrors, and an extremely large telescope observing at 2.2 $\mu$m. We find that for all configurations 
$\vartheta_\mathrm{AB} \geq 8.0\times(F_B/F_A)^{1/3}$ arcsec, where $F_A$, $F_B$ are the visible fluxes of the primary (A) and secondary (B), respectively. In summary (see caption to Fig.~ \ref{fig:exoplan1}), $62\%$ of the binary sample can have both components observed as part of the core program. For the 6 systems not fulfilling the requirement, we include the targets in one of the components of our manifold secondary program (see below). 

We further stress that the complete census of small and nearby planets around solar-type stars is unique to high-precision astrometry. 
On the one hand, Sun-like stars have typical activity levels producing Doppler noise of $\sim1$ m/s (or larger), which is still 10 times the signal expected from an Earth-analog (\citealt{2011arXiv1107.5325L}). 
 We have run detailed simulations of detectability of the 9 cm/s RV semi-amplitude of the planetary signal induced by an Earth-twin around a solar analog, using 10 cm/s per-measurement precision appropriate for an instrument such as ESPRESSO, and in the presence of representative values ($1-2$ m/s) of uncorrelated and correlated RV jitter of stellar origin. We simulated 5- through 10-years observing campaigns with very intensive monitoring 
($\geq 100$ observations per season), similar to the time sampling adopted by \cite{Anglada-Escude+16}
to detect the HZ terrestrial planet candidate 
to Proxima Centauri. We then ran different periodogram analysis algorithms (GLS, BGLS, FREDEC), and estimated the recovery rate of the signals. With a bootstrap false alarm probability $< 1\%$, we found that there was $<40\%$ chance of a clear detection. 
On the other hand, \theia astrometry will be almost insensitive to the disturbances (spots, plages) due to stellar activity, having typical activity-induced astrometric signals with amplitude below 0.1 $\mu$as (\citealt{2011A&A...528L...9L}). 

For the full sample of the nearest stars considered in 
Fig.~ \ref{fig:exoplan1} we achieve sensitivity (at the $6-\sigma$ level) to planets with $M_p \leq 3$ $M_\oplus$ (See section 3.6). If we consider $\eta_\mathrm{Earth}\sim10\%$, for the sample of 63 
stars closest to our Solar System we thus expect to detect $\sim6$ HZ terrestrial planets. Of these, 5 would be amenable for further spectroscopic characterization of their atmospheres. 
\theia can perform the astrometry of the relevant stars and make a  thorough census (95\% completeness) of these planets by using less than 10\% of a four years mission. 
As indicated above, this program will also be valuable for understanding planetary diversity, the architecture of planetary systems (2-d information plus Kepler's laws, 
results in 3-d knowledge) including the mutual inclination of the orbits, a piece of information that is often missing in our exploration of planetary systems. 

\subsubsection{Secondary Program}

\begin{figure*}[htb]
\centering
\includegraphics[width=0.95\textwidth]{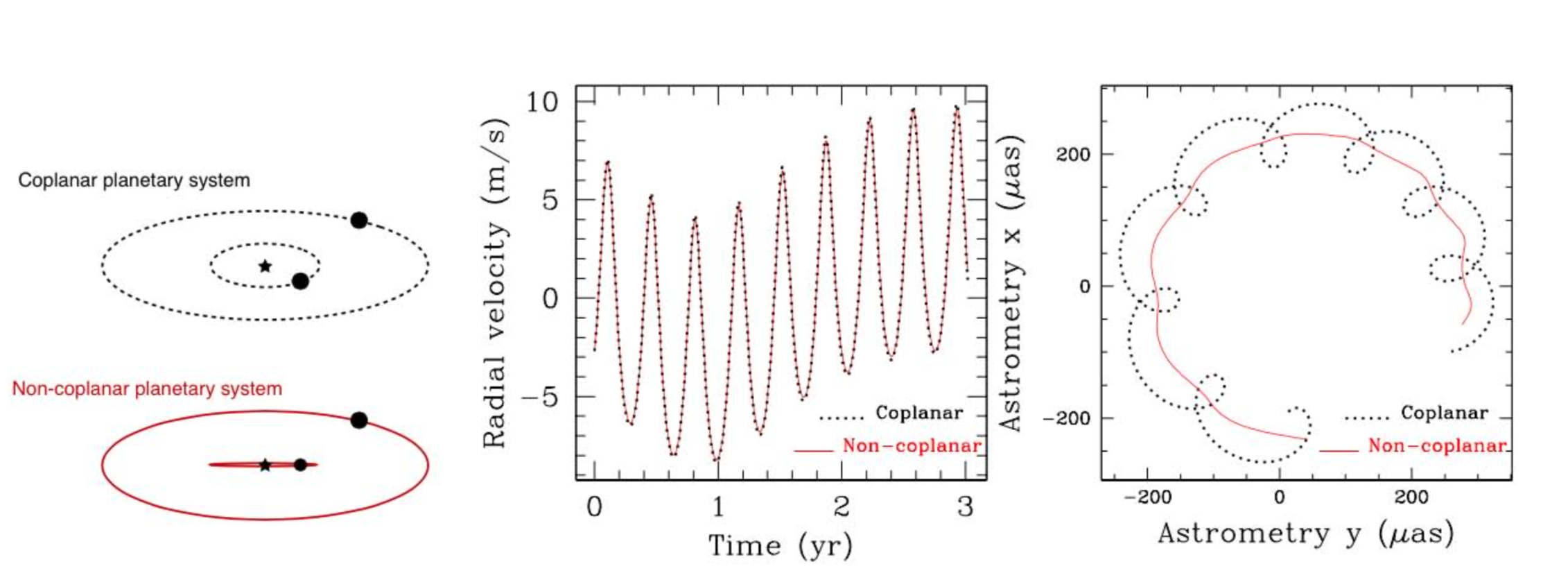}
\caption{An example where astrometry breaks the degeneracy. Two simulated planetary systems are around a solar-type star at 10 pc, with two Jupiter-like planets at 0.5 and 2.5 AU (\textit{left}). One is co-planar (\textit{dotted black line}), the other has a mutual inclination of 30$^\circ$ (\textit{full red line}). The two corresponding RV curves are shown (\textit{middle}), as well as the two astrometric ones (\textit{right}). Curves are identical in the former case, but clearly separated in the latter revealing the inclined orbits.}
\label{fig:exoplan2}
\end{figure*}

We envision a fourfold secondary program to exploit the \theia potential to 
elucidate other important questions in exoplanetary science. It will make use 
of an additional $\sim7\%$ of mission time.

\noindent {\bf a) Planetary systems in S-Type binary systems}
 \theia's performance for exoplanet detection around nearby binaries will be of crucial importance in revealing  planet formation in multiple stellar systems, the environment in which roughly half of main-sequence stars are born.

The unexpected discovery of numerous giant planets in binaries has indeed sparked a string of theoretical studies, aimed at understanding how planets can form and evolve in these environments (see \citealt{2014arXiv1406.1357T}). 
Of particular interest is the subsample of close binaries with separation 10-40AU, for which 5 exoplanets have been detected relatively close to the habitable zone around one stellar component (S-type orbits), but on orbits which are also close to the dynamical stability limit imposed by the companion's perturbations (\citealt{2004AIPC..713..269H}, \citealt{2011CeMDA.111...29T}, \citealt{2016AN....337..300S}). In such a highly perturbed environment, these exoplanets are very difficult to form following the standard planet-formation scenario, and their very existence presents a challenge to theoretical studies (\citealt{2014arXiv1406.1357T}). Several scenarios have been proposed in recent theoretical studies to solve this apparent paradox, such as the outward migration of growing embryos (\citealt{2009MNRAS.400.1936P}), the accretion by sweeping of small debris (\citealt{2010ApJ...724.1153X}), early orbital evolution of the binary (\citealt{2009MNRAS.393L..21T}), and even more radical solutions such as a binary-specific channel for planet formation (\citealt{2010ApJ...709L.114D}).
The contribution of \theia could be of great value for these ongoing studies, because it will survey at least 6 systems in this crucial 10-40AU separation range (11 in the 5-100AU range, see Table \ref{tab:sci.expl.binaries}), with a sensitivity down to terrestrial planets in the habitable zone, which is out of reach for present observation facilities. The expected additional constraints on the occurrence rate of planets in tight binaries could prove decisive in helping to discriminate between the different planet-formation-in-binaries scenarios that have been proposed.

\noindent {\bf b) Follow-up of known Doppler systems}
 
Another unique use of \theia is the study of non-transiting, low-mass multiple-planet systems that have already been detected with RVs. \theia astrometry will confirm or refute controversial detections, remove the $\sin i$ ambiguity and measure actual planetary masses. Furthermore, it will directly determine mutual inclination angles, which are critical to study a) the habitability of exoplanets in multiple systems, since they modify the orientation of the spin axes and hence the way the climates change across time (e.g.
\citealt{1993Natur.361..608L,2014MNRAS.444.1873A})) the dynamical evolution history of multiple systems, as e.g. coplanar orbits are indicative of smooth evolution, while large mutual inclinations and eccentricities point toward episodes of strong interactions, such as planet-planet scattering.  
Fig.~\ref{fig:exoplan2} illustrates a case where degeneracy in RV can be removed by astrometry.

\noindent {\bf c) Planetary systems on and off the main sequence}

\gaia has the potential to detect thousands of giant planetary companions around stars of all ages (including pre- and post-main-sequence), spectral type, chemical abundance, and multiplicity (\citealt{2008A&A...482..699C,2014MNRAS.437..497S,2014ApJ...797...14P,2015MNRAS.447..287S}). \theia will cherry-pick on \gaia discoveries and identify systems amenable to follow-up to search for additional low-mass components in such systems, particularly in  the regime of stellar parameters difficult for radial velocity work (e.g., early spectral types, young ages, very low metallicity, white dwarfs). Some of the systems selected 
might also contain transiting companions identified by \textit{TESS} and \textit{PLATO} (and possibly even \gaia itself), or planets directly imaged by \textit{SPHERE} or E-ELT.

\noindent {\bf d) Terrestrial planets around Brown Dwarfs}

To-date, among the few planetary mass objects that have been associated to brown dwarf (BD) hosts using direct imaging and microlensing techniques, only one is likely to be a low-mass planet (\citealt{2015ApJ...812...47U}, and references therein).  However, there are both observational 
\citep{2008ApJ...681L..29S,2012ApJ...761L..20R,2014ApJ...791...20R} as well as theoretical 
\citep{2007MNRAS.381.1597P,2013ApJ...774L...4M} reasons to believe that such systems could also be frequent around BDs. The recent identification of a trio of short-period Earth-sized planets transiting a nearby star with a mass only $\sim10\%$ more massive than the Hydrogen-burning limit \citep{2016Natur.533..221G} is a tantalizing element in this direction.

In its all-sky survey, \gaia will observe thousands of ultra-cool dwarfs in the backyard of the Sun with sufficient astrometric precision to reveal any orbiting companions with masses as low as that of Jupiter (\citealt{2014MmSAI..85..643S}). 

\theia will push detection limits of companions down to terrestrial mass. If the occurrence rate of $P\leq1.3$ d, Earth-sized planets around BDs is $\eta=27\%$ as suggested by He et al. (\citeyear{2016arXiv160905053H}) based on extrapolations from transit detections around late M dwarfs, the \theia measurements, probing for the first time a much larger range of separations with respect to transit surveys with sensitivity to low-mass planets, will unveil a potentially large number of such companions, and place the very first upper limits on their occurrence rates in case of null detection.

\subsection{Compact objects}

\subsubsection{Orbital measurements}
The brightest Galactic X-ray sources are accreting compact objects in binary systems.  Precise optical astrometry of these X-ray binaries provides a unique opportunity to obtain quantities which are very difficult to obtain otherwise.  In particular, it is possible to determine the distances to the systems via parallax measurements and the masses of the compact objects by detecting orbital motion to measure the binary inclination and the mass function.  With \theiax, distance measurements are feasible for $>$50 X-ray binaries (in 2000h), and orbital measurements will be obtained for dozens of systems.  This will revolutionize the studies of X-ray binaries in several ways, and here, we discuss goals for neutron stars (NSs), including constraining their equation of state (EoS), and for black holes (BHs).

\begin{figure}[h]
\centering
\includegraphics[width=0.8\textwidth]{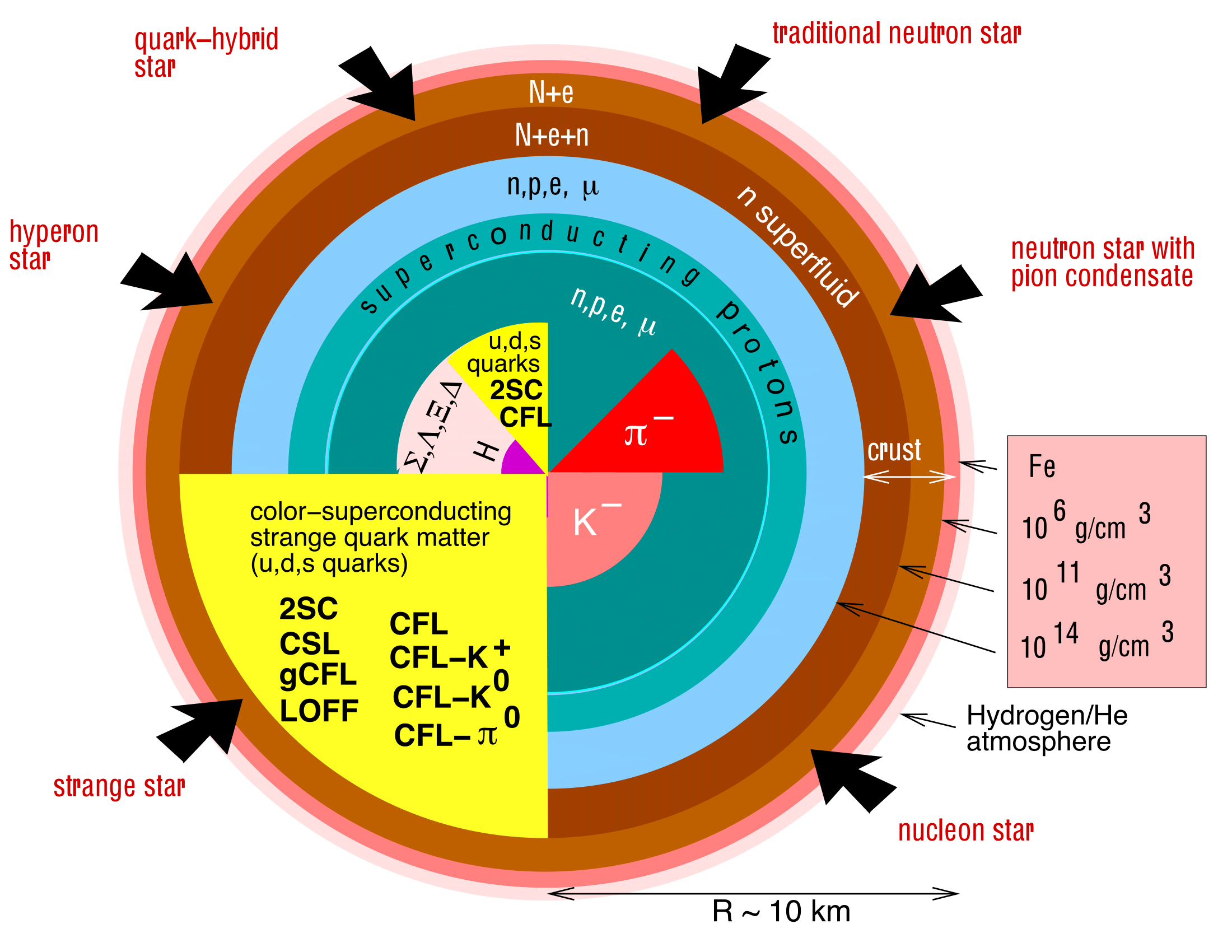}
\caption{Sketch of the different existing possibilities for the internal structure of a neutron star.  Figure courtesy of Fridolin Weber.}
\label{fig:nsstructure}
\end{figure}

Matter in the {\bf NS} interior is compressed to densities exceeding those in the center of atomic nuclei, opening the possibility to probe the nature of the strong interaction under conditions dramatically different from those in terrestrial experiments and to determine the NS composition.  NSs might be composed of nucleons only, of strange baryons (hyperons) or mesons in the core with nucleons outside (a hybrid star), or of pure strange quark matter (a quark star).  A sketch of the different possibilities is given in Fig.~\ref{fig:nsstructure}. Via the equation of state (EoS), matter properties determine the star's radius for a given mass. In particular, since general relativity limits the mass for a given EoS, the observation of a massive NS can exclude EoS models. Presently, the main constraint stems from the measurements of two very massive NSs in radio pulsar/white dwarf systems which have been reported with high precision \citep{2010Natur.467.1081D,2013Sci...340..448A,2016ApJ...832..167F}.

\begin{figure*}[htb]
\includegraphics[width=0.45\textwidth]{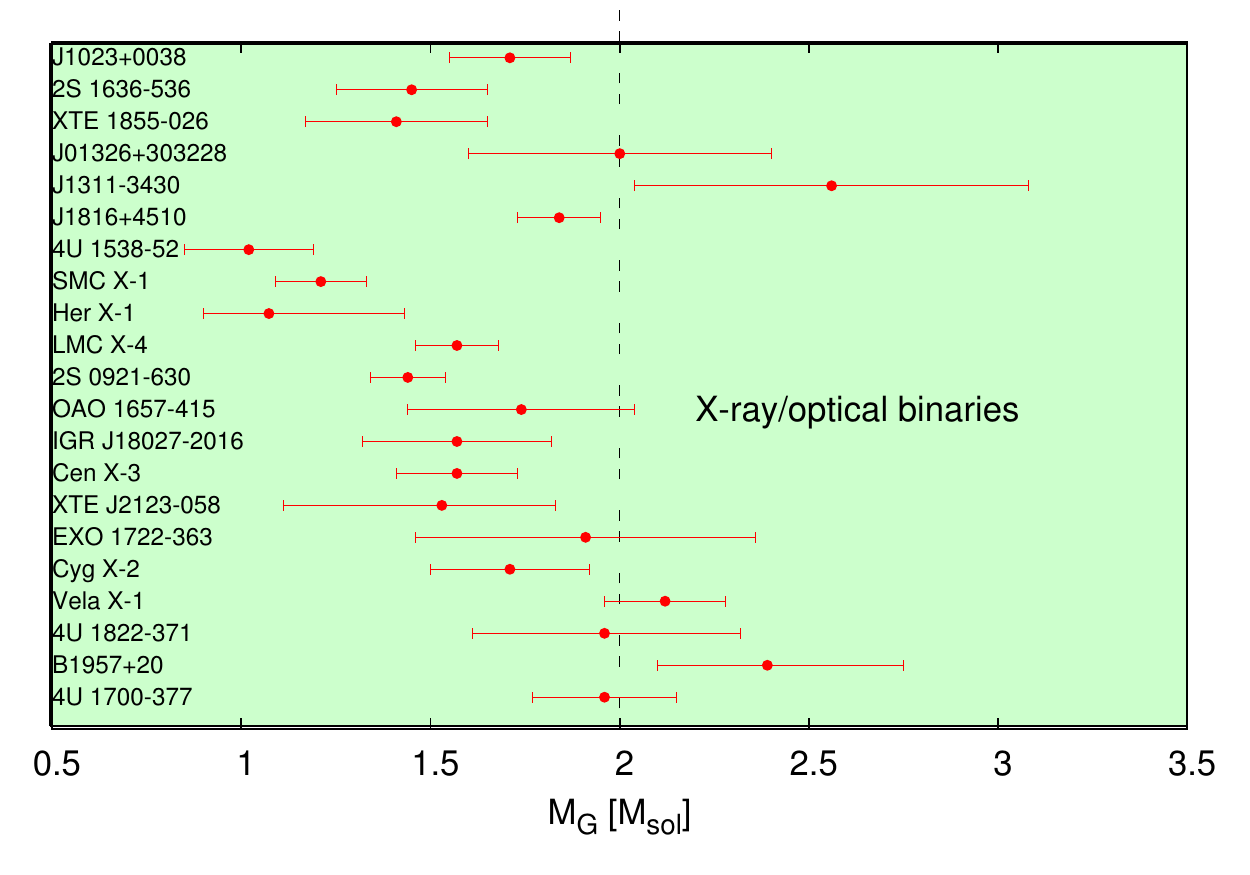}
\includegraphics[width=0.45\textwidth]{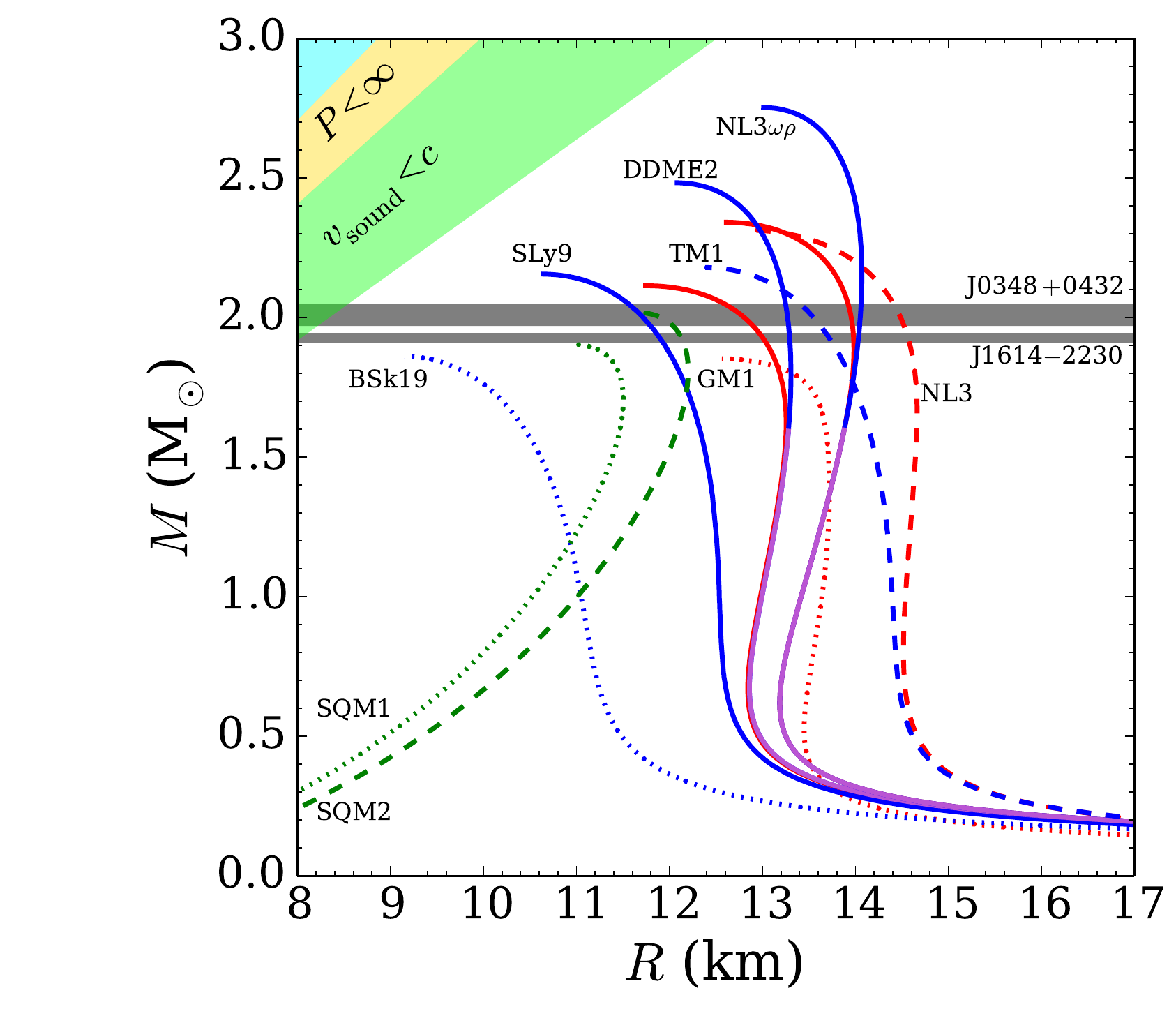}
\caption{Left: Neutron star mass measurements in X-ray binaries, update from
\citet{2005PhRvL..94k1101L}, \url{http://stellarcollapse.org}.  Right: $M$-$R$
relation for different EoS models \citep[adapted from][]{2016PhRvC..94c5804F}:
NS with a purely nucleonic core (in blue), with a core containing hyperons at
high density (in red), and pure strange quark stars (in green). The horizontal
grey bars indicate the masses of PSR J1614$-$2230 and PSR J0348+0432. The
models indicated by dotted or dashed lines are either not compatible with NS
masses or nuclear physics constraints. Note that a transition to matter
containing hyperons is not excluded by present constraints.}
\label{fig:nsmasses}
\end{figure*}

The key to constraining the NS EoS is to measure the masses and radii of NSs.
While masses have been measured for a number of X-ray binary and radio pulsar
binary systems \citep[e.g.,][]{2012ARNPS..62..485L,2016ARA&A..54..401O}, the
errors on the mass measurements for most X-ray binaries are large (see
Fig.~\ref{fig:nsmasses}, left).  The ultimate constraint on the EoS would be a
determination of radius and mass of the same object, and a small number of such
objects might be sufficient to pin down the entire EoS
\citep[e.g.,][]{2009PhRvD..80j3003O}, see Fig.~\ref{fig:nsmasses} (right),
where several $M$-$R$ relations for different EoSs are shown.  Current
techniques to determine radii rely on spectroscopic measurements of accreting
neutron stars, either in quiescence \citep{2014MNRAS.444..443H} or during
thermonuclear (type I) X-ray bursts \citep{2016ARA&A..54..401O}, and also timing
observations of surface inhomogeneities of rotating NSs
\citep{ 2016EPJA...52...63M,2016EPJA...52...59H}.

\theia will contribute by obtaining precise mass constraints with orbital measurements \citep{Tomsick&Muterspaugh10} and by improving distance measurements.  Distances must be known accurately to determine the NS radii.  For that purpose, \theia data can be combined with existing and future X-ray data, e.g., from \textit{Athena}, which is scheduled as an ESA L2 mission.  The \textit{Athena} Science Working Group on the endpoints of stellar evolution has observations of quiescent neutron star X-ray binaries to determine the NS EoS as its first science goal; however, their target list is restricted to systems that are in globular clusters.  \theia will enable distance measurements for many more NS X-ray binaries, allowing Athena to expand their target list.

Other techniques for constraining the NS EoS might also be possible in the future: detecting redshifted absorption lines; determining the moment of inertia of the double pulsar J0737$-$3039; and the detection of gravitational wave emission from the inspiral of a NS-NS merger.  However, the mass and distance measurements that \theia would obtain use techniques that are already well-established, providing the most certain opportunity for greatly increasing the numbers of NSs with mass or radius determinations.

In addition to the goal of constraining the NS EoS, NS masses are also relevant to NS formation and binary evolution.  Current evolutionary scenarios predict that the amount of matter accreted, even during long-lived X-ray binary phases, is small compared to the NS mass.  This means that the NS mass distribution is mainly determined by birth masses.  Determining the masses of NSs in X-ray binaries, therefore, also provides a test of current accretion models and evolutionary scenarios, including the creation of the NSs in supernovae.

{\bf BHs} are, according to the theory of general relativity, remarkably simple objects.  They are fully described by just two parameters, their mass and their spin. Precise masses are available for very few BHs and the recent detection of gravitational waves (Abbott et al. 2016) has demonstrated that BHs can have considerably higher masses than expected based on our understanding of stellar evolution and the fate of massive stars. Although BHs leave few clues about their origin, one more parameter that can be determined is the proper motion of BHs in X-ray binaries. Measurements of proper motions provides information about their birthplaces and formation, including whether they were produced in a supernova (or hypernova) or whether it is possible for massive stars to collapse directly to BHs.  A few BH X-ray binaries have proper motion measurements \citep[e.g.,][]{2001Natur.413..139M}, but this number will rise dramatically with the astrometry measurements that \theia will provide.

Currently, the cutting edge of research in BH X-ray binaries involves
constraining BH spins, including the rate of spin and the orientation of the
spin axis.  Techniques for determining the rate of spin include measuring of
the relativistic broadening of the fluorescent iron $K_\alpha$ line in the
X-ray emission and the study of the thermal continuum X-ray spectra
\citep{2006ARA&A..44...49R,2007ARA&A..45..441M}. Concerning the direction of their spin axes,
there is evidence that the standard assumption of alignment between the BH spin
and orbital angular momentum axes is incorrect in some, if not many, cases
\citep{ 2002MNRAS.336.1371M,2014ApJ...780...78T, 2016ApJ...826...87W}, likely requiring a
warped accretion disc. Theoretical studies show that such misalignments should
be common \citep{2016MNRAS.462..464K}. However, binary inclination measurements rely
on modeling the ellipsoidal modulations seen in the optical light curves
\citep{2011ApJ...742...84O}, which is subject to systematic uncertainties, and \theia will be
able to provide direct measurements of orbital inclination for many of the BH
X-ray binaries that show evidence for misalignments and warped discs (see
Sec.\ref{sec:compactobjects_requirements} for targets).

\subsubsection{Astrometric microlensing}
\label{sec:Compact objects in the GC}
About thirty years ago Bohdan Paczy{\'n}ski (Paczynski 1986) proposed a new method for finding compact dark objects, via photometric gravitational microlensing. This technique relies on continuous monitoring of millions of stars in order to spot its temporal brightening due to space-time curvature caused by a presence and motion of a dark massive object. Microlensing reveals itself also in astrometry, since the centre of light of both unresolved images (separated by $\sim$1 mas) changes its position while the relative brightness of the images changes in the course of the event. Astrometric time-series at sub-mas precision over course of couple of years would provide measurement of the size of the Einstein Ring, which combined with photometric light curve, would directly yield the lens distance and mass. Most microlensing events are detected by large-scale surveys, e.g., OGLE and, in future possibly also the LSST.  At typical brightness of V=19-20mag only \theia would be capable at providing good-enough astrometric follow-up of photometrically detected microlensing events. Among 2000 events found every year, at least a couple should have a black hole as the lens, for which the mass measurement via astrometric microlensing would be possible with \theia.  

\begin{figure}[htb]
\centering
\includegraphics[width=0.98 \textwidth]{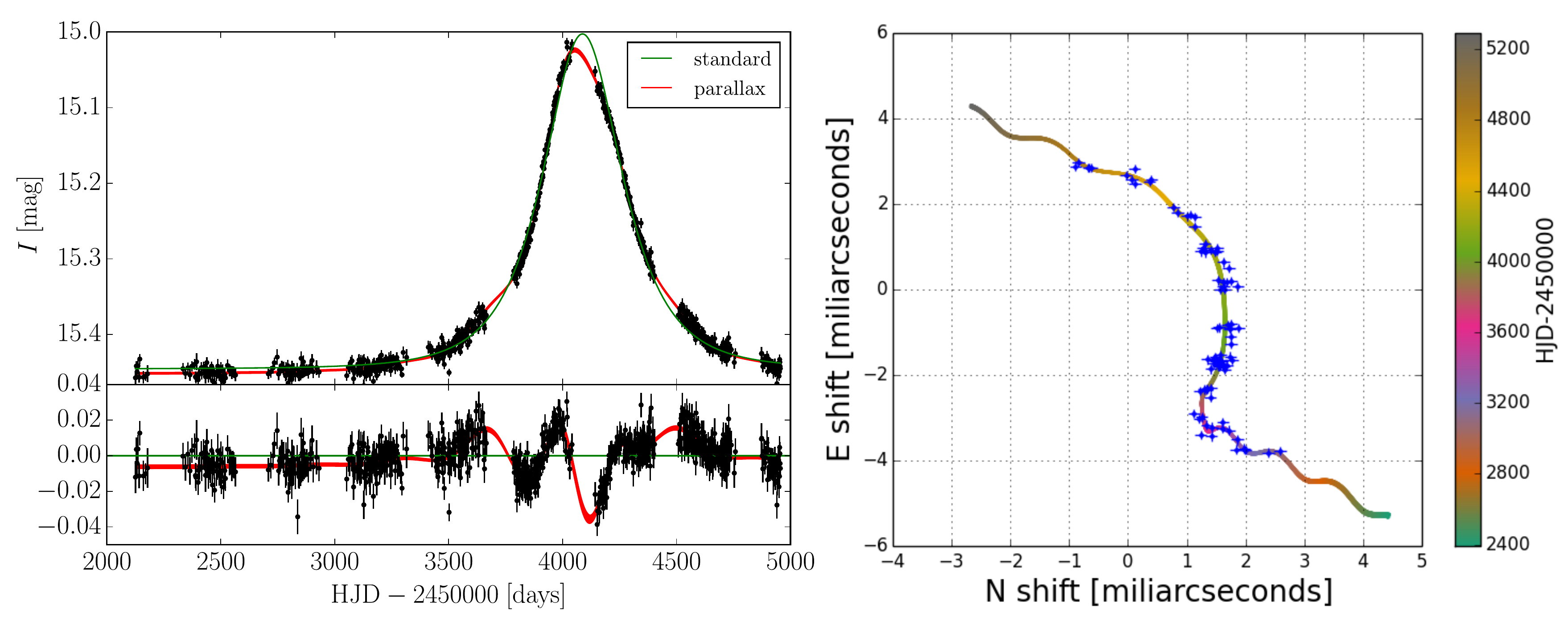}
\caption{Microlensing event, OGLE3-ULENS-PAR-02, the best candidate for a $\sim$10M$_\odot$ single black hole. Left: photometric data from OGLE-III survey from 2001-2008. Parallax model alone can only provide mass measurement accuracy of 50-100$\%$. Right: simulated astrometric microlensing path for a similar event if observed with \theia. Combining superb \theia astrometric accuracy with long-term photometric data would yield mass measurements of black holes and other dark compact object to 1$\%$ even at faint magnitudes.}
\end{figure}

Detection of isolated black holes and a complete census of masses of stellar remnants will for the first time allow for a robust verification of theoretical predictions of stellar evolution. Additionally, it would yield a mass distribution of lensing stars as well as hosts of planets detected via microlensing.

\subsection{ Cosmic distance ladder}

Measuring cosmological distances has revolutionized modern cosmology and will continue to be a major pathway to explore the physics of the early Universe. The age of the Universe ($H_0^{-1}$) is a key measurement in non-standard DM scenarios.  Its exact value is currently strongly debated, with a number of scientific papers pointing at discrepancies in between measurements methods at the 2-3$\sigma$ level. But the most serious tension appears between CMB estimates ($H_0 = 67.8 \pm 0.9 \, \rm{km/s/Mpc}$) (or for that matter BAO results from the SDSS-III DR12 data, combined with SNIa which indicate $H_0 = 67.3 \pm 1.0 \, \rm{km/s/Mpc}$, see \citet{Alam2016}) and measurements based on Cepheids and SNIa ($H_0 = 73.24 \pm 1.74 \rm{km/s/Mpc}$), with a discrepancy at the 3-4 $\sigma$ level. 

The tension between the methods can be due to unknown sources of systematics, to degeneracies between cosmological parameters, or to new physics (e.g. \citealt{Karwal&Kamionkowski16}, Boehm et al. 2014). It is therefore of crucial importance to consider methods capable of measuring $H_0$ with no or little sensitivity to other cosmological parameters. 

 Uncertainties can be drastically reduced by measuring time delays (TD) in gravitationally lensed quasars \citep{Refsdal64}, as this technique only relies on well-known physics (GR). With enough statistics, and a good modeling the mass distribution in the lensing galaxy, TD measurements can lead to percent-level accuracy on $H_0$, independently of any other cosmological probe (e.g. Bonvin et al. 2016a, Suyu et al. 2013, 2014).  

In practice, TDs can be measured by following the photometric variations in the images of lensed quasars. As the optical paths to the quasar images have different lengths and they intersect the lens plane at different impact parameters, the wavefronts along each of these paths reach the observer at different times. Hence the notion of TD. 

Significant improvements in lens modeling combined with long-term lens monitoring should allow measurements  of $H_0$ at the percent level. The H0LiCOW program ($H_0$ Lenses in COSMOGRAIL's Wellspring), which focuses on improving the detailed modeling of the lens galaxy and of the mass along the line of the sight to the background quasar, 
led to $H_0 = 71.9 \pm 2.7 \, \rm{km/s/Mpc}$ (that is 3.8$\%$ precision) in a flat LCMD Universe by using 
deep HST imaging, Keck spectroscopy and AO imaging and wide field Subaru imaging (Suyu et al. 2016, Rusu et al. 2016, Sluse et al. 2016, Wong et al. 2016, Bonvin et al. 2016a). This value is in excellent agreement with the most recent measurements using the distance ladder (though in tension with the CMB measurements from Planck) but still lacks of precision. 

By performing photometric measurements with the required sensitivity and no interruption, the combination of \theia and excellent modeling of the lens galaxy, will enable to measure $H_0$ at the percent level and remove any possible degeneracies between $H_0$ and other cosmological parameters. This will open up new avenues to test the DM nature.

An alternative technique consists in using  trigonometric parallaxes. This is  the only  (non-statistical and model-independent) direct measurement method and the foundation of the distance scale. \theia has the potential to extend the  "standard candles" - the more distant pulsating variables: Cepheids, RR Lyrae, Miras  and  also Stellar Twin stars -   well beyond the reach of \gaia. 

These distance measurements can be transferred to nearby galaxies allowing us to convert observable quantities, such as angular size and flux,  into physical qualities  such as energy and luminosity. Importantly, these  distances scale linearly with $\hn$, which gives the temporal and spatial scale of the universe. With this improved knowledge, we will then be able to to better understand the structure and evolution  of both our own and more distant galaxies, and the {\it age} of our universe.

\begin{figure*}[thb] 
\centering
\includegraphics[width=0.75\textwidth]{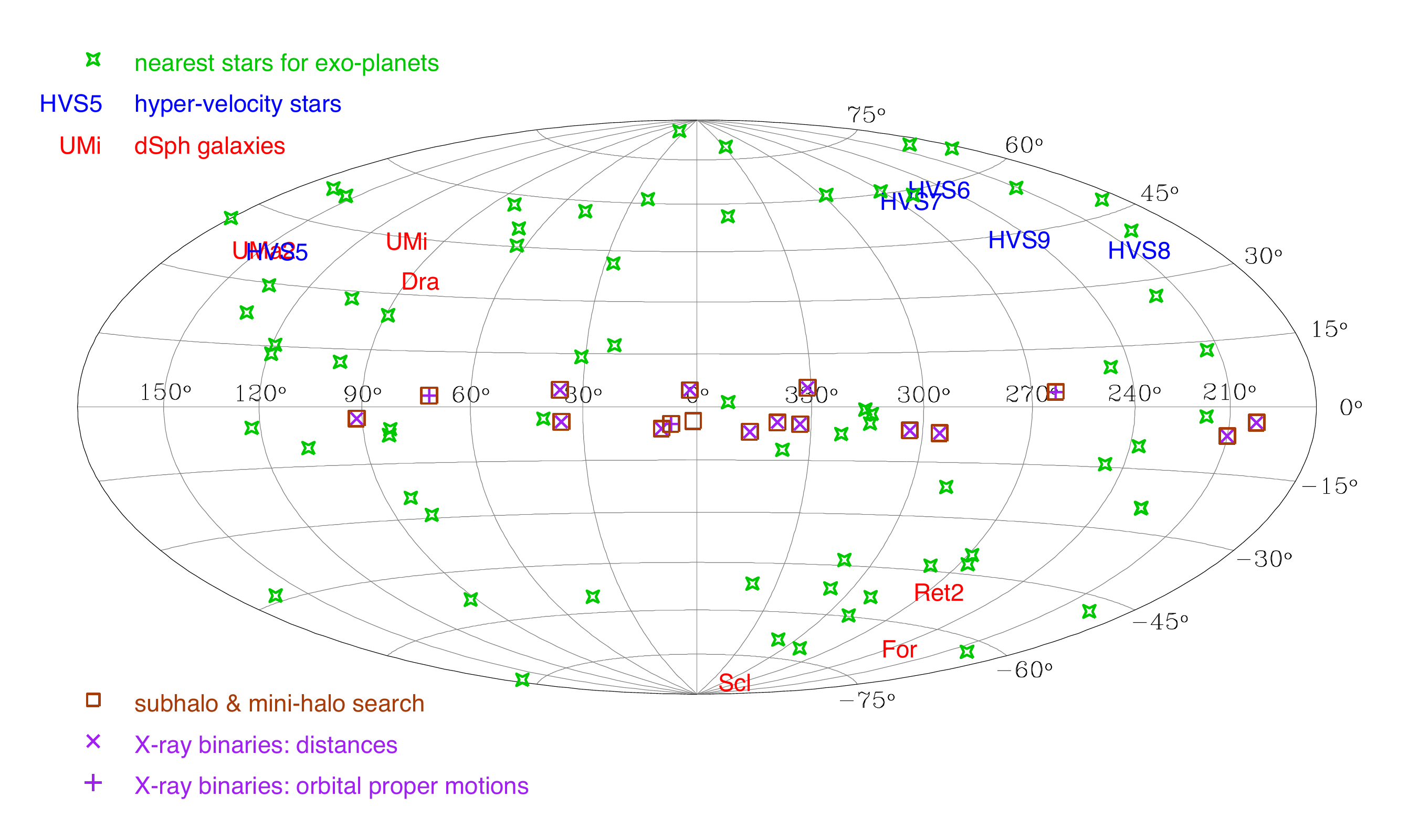}
\caption{Sky map of the different targets considered for observations with \theia} 
\label{fig:skyplot}
\end{figure*}

\subsection{Synergies}

\theia's observations will add significant value (and benefit from) a number of other ground and space missions operating in the 2030s, including ESA's \textit{Athena}, \textit{PLATO}, \euclid\ and \gaia, ESO's \textit{MICADO} and \textit{Gravity}, \textit{CTA}, \textit{SKA}, \textit{JWST} and \textit{LSST}. For example:
 
\textbf{\jwst:}
Estimates suggest that the \jwst will be able to detect Lyman Break galaxies with absolute magnitudes as faint as $M_{\rm{UV}} \sim -15$ at $z\sim7$, corresponding to halo masses of about $10^{9.5} \ M_{\odot}$. The combination of \theia's and the \jwst's observations will 
enable unambiguous tests of DM.

\textbf{\textit{PLATO}:} 
\plato  will look at planetary transits and star oscillations  in two fields (each covering 2250 deg$^2$), for 2-3 years each, in  
host stars brighter than 16 mag. 
\textit{PLATO}'s high cadence continuous monitoring of its target stars will provide 
information on the  internal structure of the stars, allowing   
determination of their stellar ages and masses. \theia will benefit from \textit{PLATO}'s characterization of many of \theia's core star samples.   For close `\textit{PLATO}' stars where transits were observed \theia can measure additional inclined planets.

\textbf{SKA:}
SKA aims to use radio signals to look for building blocks of life (e.g. amino acids) in Earth-sized planets.   \theia will  identify target planets  from their astrometric "wobble" that can be followed-up spectroscopically with the SKA. Furthermore, SKA aims to use its immensely fast sky coverage to detect transients, such as supernovae and gamma ray bursts. With its precise astrometry, \theia will help study the specific locations of such events in stellar clusters.

\textbf{ CTA: }
The Cherenkov Telescope Array (CTA) 
in the Northern and Southern Hemispheres will provide measurements of the gamma-ray flux with almost complete sky coverage and unprecedented energy and angular resolution, in the $\sim$ [0.02,100] TeV energy range. \theia's sub-microarcsecond performance will allow us to probe the so-called J-factor that defines the brightness of the gamma-ray flux in dSphs and thus determine the prime candidates for CTA's observations.
 
CTA also aims to observe star-forming systems over six orders of magnitude in formation rate, to measure the fraction of interacting cosmic-rays as a function of the star-formation rate. By combining \theia and CTA measurements, we will better understand the relative importance of cosmic rays and DM in places where star-formation is important. Furthermore, a small number of black-hole and neutron star binary systems in our Galaxy is known to emit gamma-rays. The mechanism by which the particle acceleration is achieved is not well-understood. \theia's sub-microarcsecond performance will allow us to probe the velocity structure of the nearby gamma-ray bright radio galaxies of NGC1275, IC310, M87 and Cen A, which combined with CTA's observations will enable important astrophysics breakthroughs.

\section{Scientific Requirements }
\label{sec:requirements}

To achieve our science goals, \theia will stare in the direction of 
\begin{itemize}
\setlength\itemsep{0em} 
\item Dwarf galaxies (Sphs), to probe their DM inner structure;  
\item Hyper-Velocity stars (HVSs), to probe the triaxiality of the halo, the existence of mini compact halo objects and the time delayed of quasars; 
\item the Galactic disc, to probe DM subhalos and mini compact halo objects; 
\item  star systems in the vicinity of the Sun, to find the nearest potentially habitable terrestrial planets;
\item known X-ray binaries hosting neutron stars or Black Holes.
\end{itemize}

\remove{\begin{figure*}[thb] 
\centering
\includegraphics[width=0.75\textwidth]{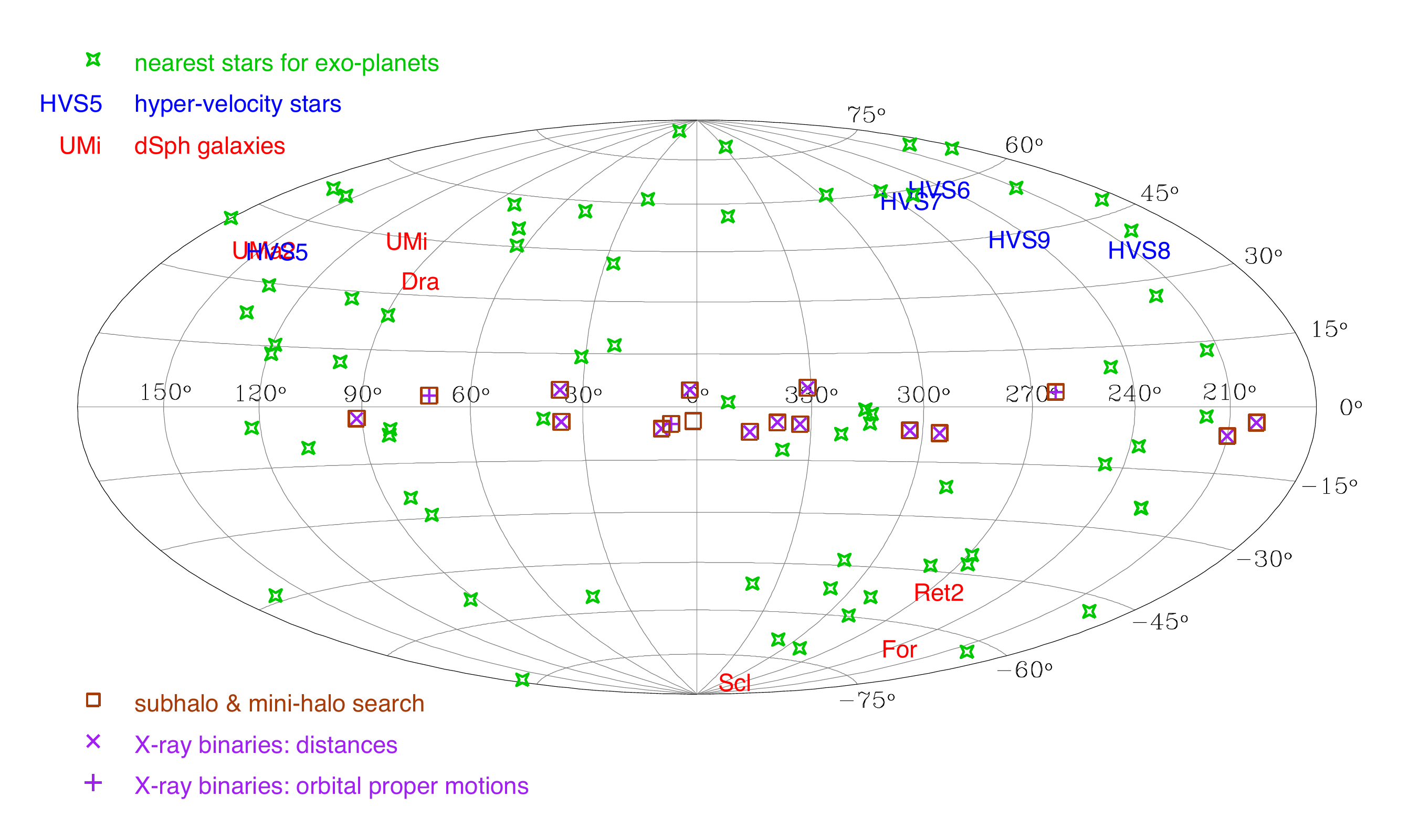}
\caption{Sky map of the different targets considered for observations with \theia} 
\label{fig:skyplot}
\end{figure*}}

\begin{figure}[thb]
\centering
{\includegraphics[width=0.8\textwidth,trim = 0.8cm 0cm 0cm 0cm, clip]{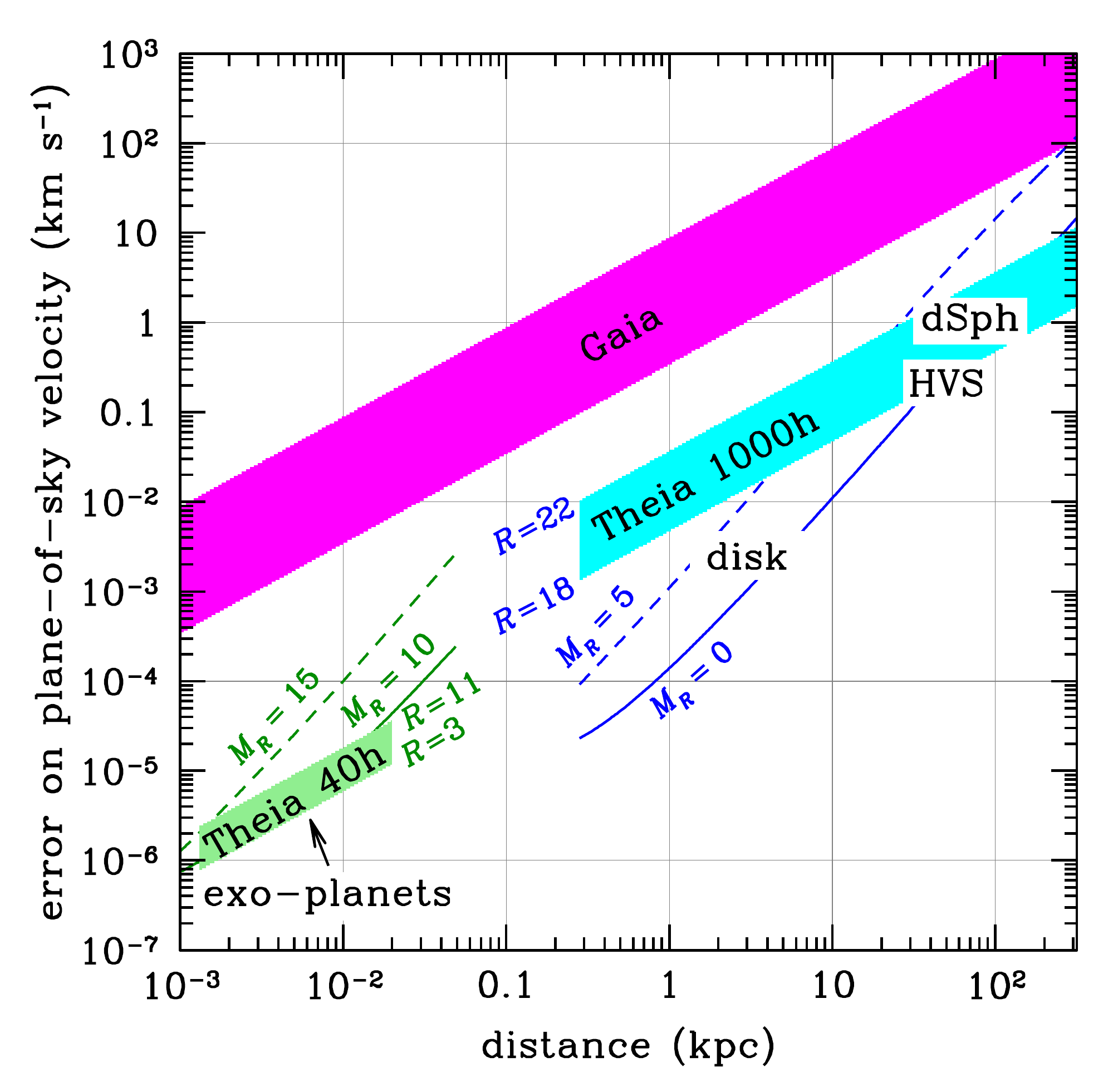}}
\caption{Expected plane-of-sky velocity errors from \theiap proper motions as a function of distance from Earth. These errors respectively correspond to 40 and 1000 cumulative hours of exposures for exo-planets (\textit{green}) and more distant objects (\textit{cyan} and \textit{blue}),  during a 4 year interval for observations, including the systematic limit from calibration on \gaia reference stars. The expected precision for specific objects are highlighted. The accuracy  for the 5-year \gaia mission is shown in \textit{magenta}.}\label{fig:vposerrvsDist}
\end{figure}

Observations will be done for a few hours, then we will slew to the next target object, and come back several times to the same fields during the mission time to sample the desired motion of the science targets. Target fields have been chosen ensure the maximal science outcome and minimise the amount of astrometric dead time. Photometric surveys, e.g. for measurements of $H_0$ by time delays will be performed after re-pointing the telescope and while waiting for stabilization.  Fig.~\ref{fig:skyplot} shows a sky map with the objects that we plan to observe.

As illustrated in  Fig.~\ref{fig:vposerrvsDist},  \theia will measure  the  plane-of-sky velocities of the faintest objects in the local Universe, with errors as small as a few mm/s in the case of the hosts of Earth-mass exo-planets in the habitable zone of nearby stars, a few m/s for stars in the Milky Way disc, i.e. for kinematical searches for dark matter sub-halos, micro-lensing searches for ultra-compact mini-haloes, and for the companions of neutron stars and black holes in X-ray binaries, 200 m/s for hyper-velocity stars whose line-of-sight velocities are typically $> 500\,\rm km/s$, and finally $1\,\rm km/s$ for $R=20$ stars in dwarf spheroidal galaxies.

\theiap expected astrometric precision  (which feasibility was demonstrated using an underway laboratory experiment in Grenoble) 
makes it a unique mission, as shown in Fig.~\ref{fig:astroComparions}, with LSST 10 yr and \gaia 5yr's accuracy not being able to catch up with \theiap precision. 
 \theia will therefore surpass the scientific goals set by any other mission planned for the next decade.

\subsection{Dwarf galaxies} 

We plan to observe at least 6 dSphs which display 1) a high mass-to-light ratio (so that most are dominated by DM, see Fig.~\ref{fig:NvsMoverL}); 2) long star formation histories (so that most of the objects have sufficiently old stellar populations that the duration of bursty star formation was too short to convert DM cusps into cores); 3) at least over 1000 stars with plane-of-sky velocity errors less than half the galaxy's internal velocity dispersion (see Fig.~\ref{fig:NvsMoverL}). These three criteria lead to 5 classical dwarf spheroidals (Draco, Fornax, Sculptor, Sextans, and Ursa Minor) as well as an Ultra-faint dwarf (Ursa Major~II), all of which would be observed for a total of 1000 h each. We also include 2 galaxies with low $M/L$ (Fornax and Sculptor) for testing the method, and we may extend our sample to poorer systems such as Reticulum~II. 
The $\sim$ 1000 brightest stars in Draco have magnitudes $R = 17.5$ to 20.5. \theiap field
of view allows the observation of an entire dwarf galaxy such as Draco in a single shot.  By performing $\approx 65\%$ of the measurements during the first and last years of the mission, a cumulative observation of 1000 h on a $R = 18 \ (R = 22)$ star result in a 1.0 (7.7) $\rm \mu as/yr$ astrometric uncertainty in its proper motion.  
This leads to plane-of-sky velocity errors $<3 \rm\, km\,s^{-1}$ for $R<22$ stars in  dwarfs such as Draco and Ursa Minor. This corresponds to less than half the  internal velocity dispersions (IVDs) of these galaxies, see Fig.~\ref{fig:vposerrvsDist}, allowing for an accurate recovery of the DM density profile.
While the ultra-faint dwarfs have lower IVDs, they are sufficiently close to also enable to measure the DM profile with \theia.


\begin{figure}[htb]
\centering
\includegraphics[width=0.8 \textwidth,clip]{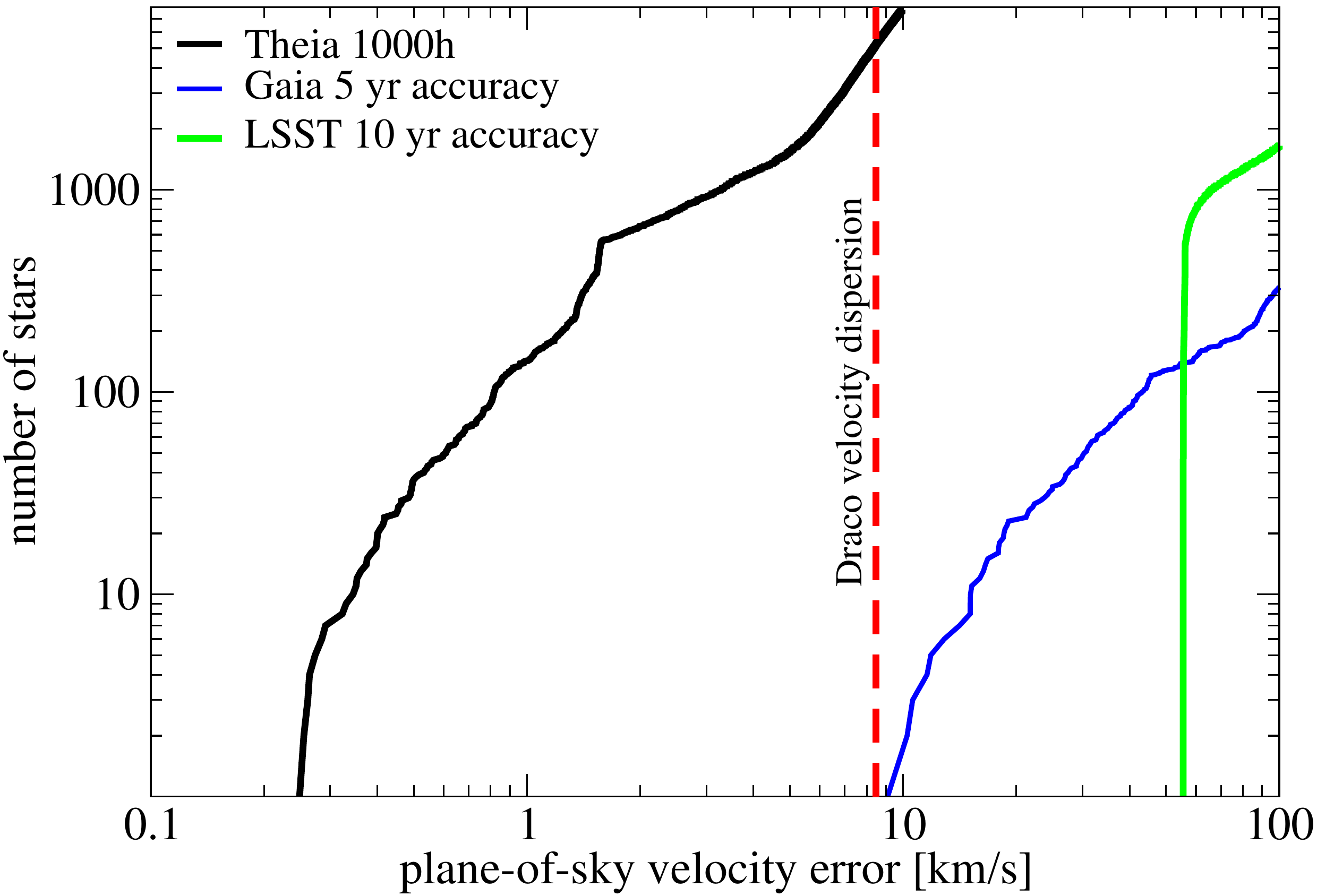}
\caption{Cumulative number of stars for given plane-of-sky velocity error for the Draco dSph.}
\label{fig:Nstarsvsvelerr}
\end{figure}

\subsection{Triaxiality of the Dark Matter halo} 
\label{sec:requirements_HVS}
The expected deflection of HVSs due to triaxiality is about 0.1-0.2 mas/yr. 
We expect to reach the proper motions precision of $5 - 15\,\mu$as/yr for a typical brightness of HVS stars ($R$=17-19 mag) for 125 hours of observation in 1 year.  With 500~h over 4 years per target, we will reach $1 - 4\, \mu$as/yr.  
Measuring several stars is critical, as each HVS provides an independent constraint on the Galactic potential. We thus propose to observe 5 HVS over 4 years, for a total of 2500 h. We note that \theia offers a unique window of opportunity for this science case, as such a measurement cannot be done from the ground. Laser-AO imagers have FoV of less than 1 arcmin, even smaller than HST, which at high Galactic latitudes contain few quasars. 

\subsection{Orbital distribution of Dark Matter particles}

This science case needs no additional observations.  After measuring the triaxiality (shape) of the DM halo and its radial density profile to a high precision with \theia, we estimate that  it will be possible to derive the orbits of $\sim$ 5000-10,000 field stars and thereby infer the orbital distribution of dark matter particles. This science case therefore does not have specific requirements other than those specified in Sec.~\ref{sec:requirements_HVS}.

\nocite{Stinson+12}
\nocite{valluri_etal_13}

\subsection{Dark Matter subhalos}
\label{Sec:clumps_requirements}

\remove{To  determine whether DM subhalos have interacted with the Galactic Disc, we need to detect anomalous bulk motions of the stars in the disc of the order of km/s (or even smaller).  To achieve this goal, we will focus on 9 observation fields (separated by 10$^\circ$ longitude), looking above the disc and the same number looking below the disc (at longitude $b = \pm 2^\circ$). In our minimal observation programme, we have allocated a total of 7000~h for these observations. We will focus on regions where the number of stars is very large and the variation of the \CB{dust?} extinction minimal to avoid sampling bias. We will use the parallax to determine which stars belong to observation fields and measure their proper motion. $\theiap$ accuracy is fundamental to this project and will allow us to measure both the bending mode in velocity space and the density anisotropy along the lines of sight. The combination of both measurements will enable us to determine the size  of the DM halos which have perturbed their orbits, and thus test the $\Lambda$CDM paradigm. }

To  determine whether DM subhalos have interacted with the Galactic Disc, we need to detect anomalous bulk motions of the stars in the disc of the order of km/s (or even smaller).  To achieve this goal, we will focus on 9 lines of sight, looking above and below the disc at longitude $b =
 \pm 2^\circ$ (18 fields) and separated by 10$^\circ$ in longitude. In our minimal observation programme, we have allocated a total of 7200~h (18$\times$400~h per field, with a scanning mode of 25$\times$16~h/field) for these observations. We will focus on regions where the number of stars is large and the variation of the interstellar extinction is minimal, to limit  any sampling bias. We will use the parallax to obtain the best distance estimator and measure their proper motion. $\theiap$ accuracy is fundamental to this project and will allow us to measure both the bending mode in velocity space and the density anisotropy along the lines of sight. The combination of both measurements will enable us to determine the mass of the DM halos which have perturbed the  Galactic disc, and thus test the $\Lambda$CDM paradigm.

\subsection{Ultra-compact Minihalos}

This science case requires no additional observations. It will be performed as we observe different directions towards the galactic disc and is thus based on the same requirements as defined in Sec.~\ref{Sec:clumps_requirements} (with an extra 800h available for directions that are further away from those chosen for the DM subhalos science case, thus leading to 8000h observational time available). We note that extensions of the observational program (either due to better systematics or use of open time) would allow us to set further constraints, by using extended HVS observations. The constraints thus obtained would probe even lower masses, and therefore smaller scales, than the Galactic disc projections shown in Figs.~\ref{fig:ucmh} and \ref{fig:ucmh2}.

\remove{\begin{figure*}[htb] 
\centering
\includegraphics[width=0.42\hsize,clip]{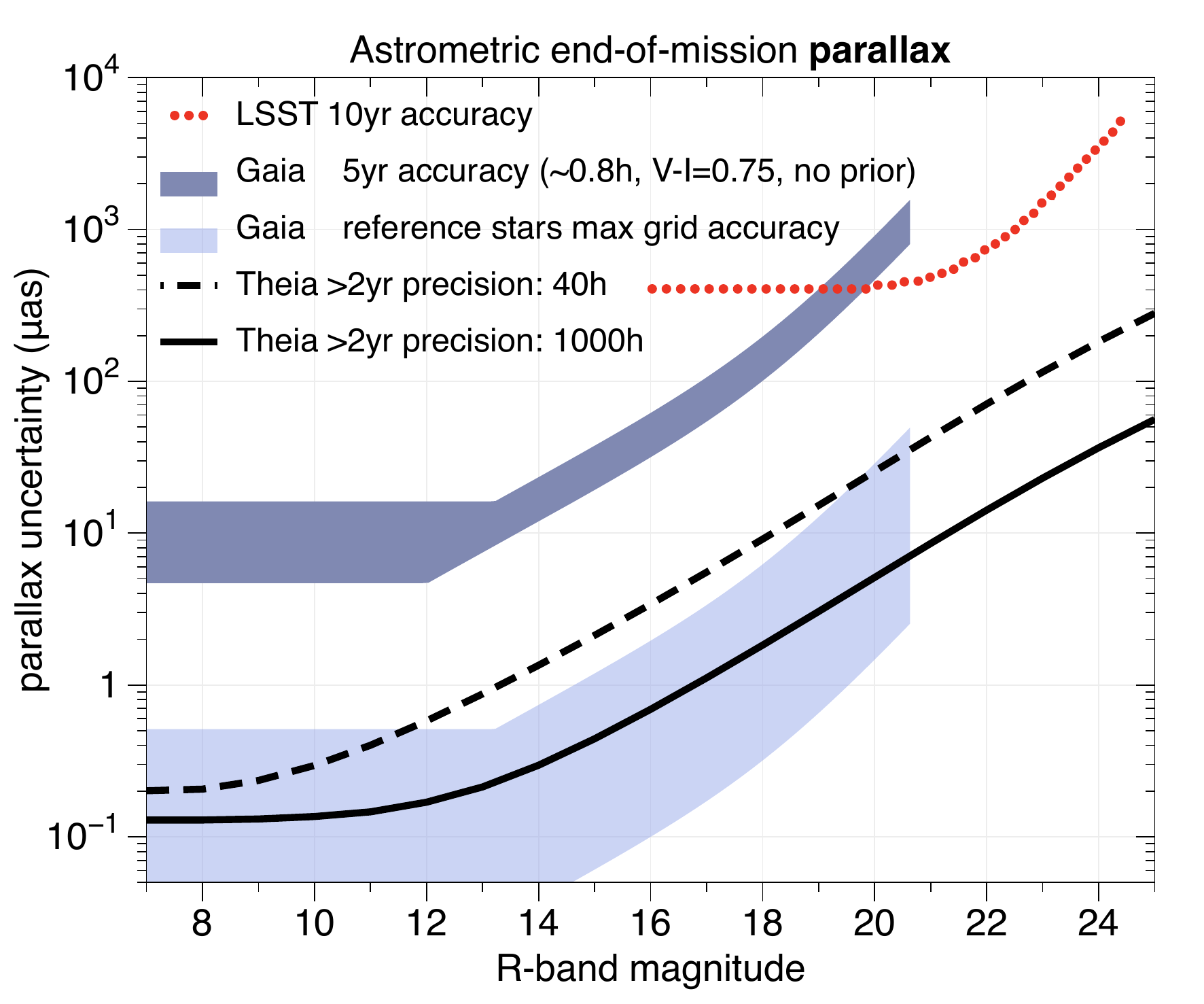}
\includegraphics[width=0.42\hsize,clip]{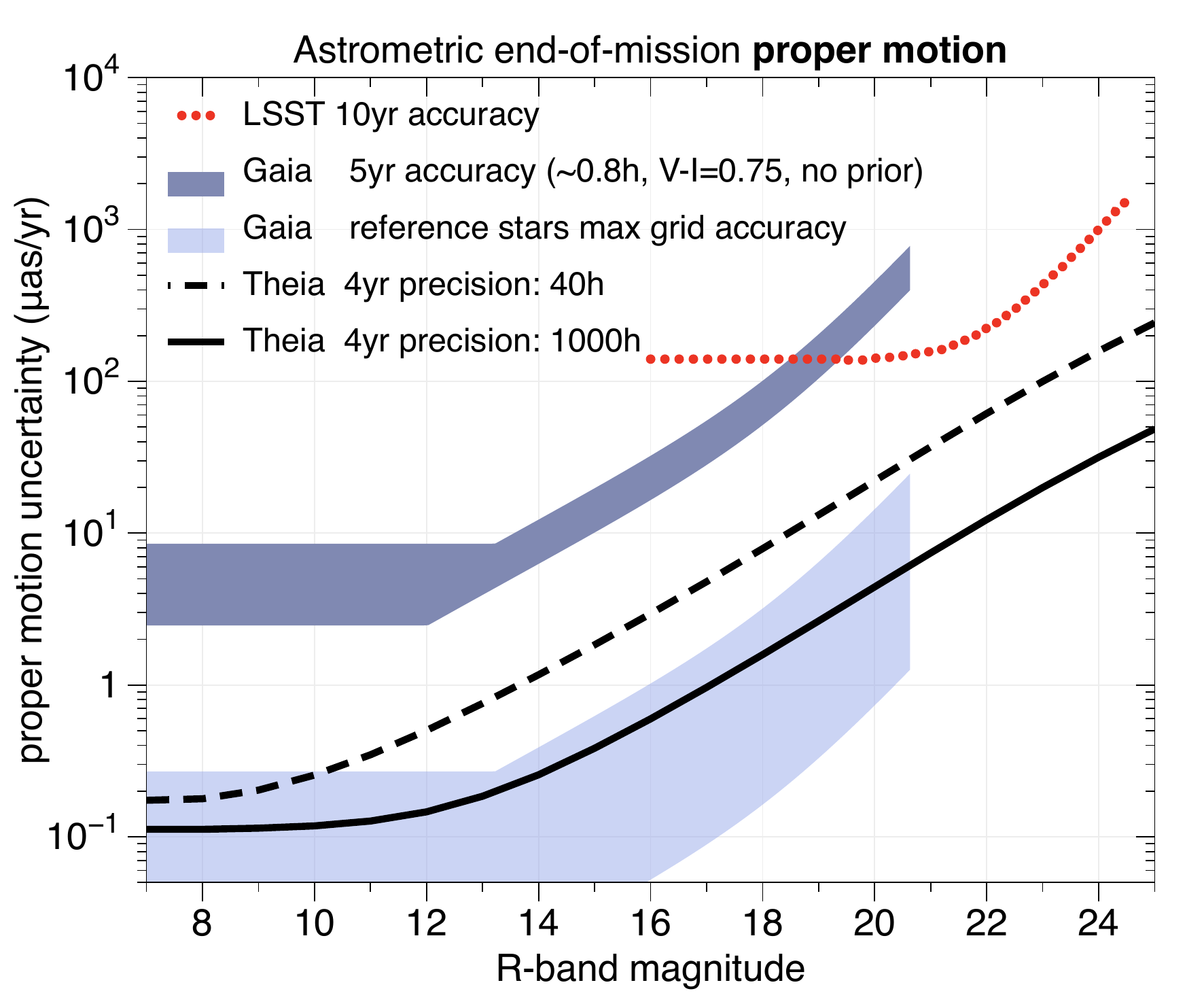}
\caption{Estimated RMS precision on \theia's relative parallax (\textit{left}, for ecliptic latitude $0^\circ$) and proper motion (\textit{right}) in the $R$-band. Also shown for comparison are the estimated accuracies for 10~years LSST \citep{2009arXiv0912.0201L} as well as the 5-year nominal \gaia mission \citep{deBruijne+15} (vertical spread caused by position on the sky, star colour, and bright-star observing conditions). Small-scale spatial correlations ($<$1$^\circ$) between \gaia reference sources will limit the maximum reachable absolute parallax and proper motion calibration for \theia, indicated by the light blue band for a range of assumed spatial correlations as function of reference star magnitude (see Sect.~\ref{sec:calib_priorsAbsAstrometry}). 
}
\label{fig:astroComparions}
\end{figure*}}

\subsection{Exoplanets} 

\subsubsection{Core program}

\begin{table}[thb]
  \caption{Abridged exoplanet target list (see text for details).}
  \label{tab:sci.exopl.list}    
  \centering
  \begin{center}
  \tabcolsep 2.3pt
  \begin{tabular}{cccccccc} 
    \hline 
    Star  & Gliese &  name &$D$  &$L_{\rm bol}$ &$M_{\rm st}$ & spectral &$M_{\rm p}$\\
 \#   & ID & & [pc] & [$\rm L_\odot$] & [$\rm M_\odot$] & type & $\rm M_\oplus$ \\
    \hline 
    \hline 
    1 & 559   &$\alpha$ Cen A    &1.3 &1.5 &1.1 &G2V        &0.2\\
    2 & 559   &$\alpha$ Cen B    &1.3 &0.5 &0.8 &K0V        &0.3\\
    3 & 280   &Procyon                &3.5 &6.8 &1.6 &F5 IV-V  &0.5\\
    4 & 768   &Altair                    &5.0 &9.4 &1.8 &A7 IV-V       &0.7\\
    5 & 144   &$\epsilon$ Eri           &3.3 &0.3 &0.8 &K2 V         &0.7\\
    (...)             &                            &(...)    &&&&&(...)\\    
    63 & 271 & $\delta$ Gem A  &16.0 &7.6 &1.7 &F1 IV-V &2.2\\
    \hline 
  \end{tabular}
  \end{center}
  
\end{table}

\theia will determine a complete census of potentially habitable terrestrial-type planets around a selected sample of the nearest stars achieving sensitivity to masses in the rocky Super-Earth regime ($1 < M_p < 5$ $M_\oplus$). These objects are identified as most amenable to spectroscopic follow-up with next-generation direct-imaging devices (both from the ground and in space) for identification of atmospheric bio-signatures indicative of a complex biology on the surface 
(\citealt{Kopparapu+14,2014IAUS..299..247R}). 
In collaboration with several groups across Europe and USA, we performed a Double Blind Test and reached the conclusion that a minimum condition to obtain detections of small planets (even in multiple planet systems), is a signal-to-noise ratio $S/N=6$. The signal $S$ is computed using the equation of the astrometric signal (see Section 2.2.2). The noise $N$ is given by the end of mission accuracy derived as $\sigma = \sigma_0\times[(t_\mathrm{vis}/1 h)^{-1/2}\times(N_\mathrm{vis})^{-1/2}+\sigma_\mathrm{sys}^2]^{1/2}$, where $\sigma_0 = 0.98$ $\mu$as is characteristic of the instrument (with a 0.80 m primary mirror. See Sect. 6.2 and Sect. 6.4.5), $N_\mathrm{vis}$ is the number of visits per star, $t_\mathrm{vis}$ is the duration of each visit, and $\sigma_\mathrm{sys}=0.125$ $\mu$as is a systematic noise floor term, which includes possible unmodeled jitter from spots with rms amplitude of 0.07 $\mu$as 
(\citealt{2011A&A...528L...9L}). 

Taking into account this requirement, the following program will be executed. Observations of the most suitable nearby 63 A-F-G-K-M stars in 50 individual target fields 
(including binary systems as described in Section 2.2.2) will be obtained with $N_\mathrm{vis} = 50$; $t_\mathrm{vis} = 0.8$~h, after deduction of the 30\% overhead for slews between targets ($\sim10$ deg). The total duration for such a program is $\sim$0.3~yr ($50\times50\times0.8 = 2000$~h), or 10\% of the 3.5 year observing mission time minus slew overhead. Table \ref{tab:sci.exopl.list} 
gives an extract of the target list where stars are ranked by increasing detectable planetary mass. 

\subsubsection{Secondary program}

\textbf{a) Planetary systems in S-type binary systems}  
We will use \theia to survey the 16 most suitable stellar systems, with separations in the overall range 5-100 AU, with sensitivity to terrestrial planets in the HZ of each component, which is out of reach for present observational facilities. Note that this secondary program comes at no cost in terms of observing time, but it constitutes a natural and valuable byproduct of the core program described above. The sample size is large enough to investigate the impact of close-in stellar companions in the formation and evolutionary history of such systems.

\textbf{b) Follow-up of known Doppler systems}  

A sample of $\sim20$ bright stars ($R\leq10$ mag) hosting Doppler-detected systems with low-mass planets will be observed at 50 epochs to determine actual masses and mutual inclination angles. We will invest 1/2 h of integration time per visit, excluding overheads ($20\times50\times 0.5 = 500$~h). The sample size is large enough to allow for in-depth studies of the dynamical evolution history as well as possible habitability of such systems. 

\textbf{c) Planetary systems on an off the main sequence}
We envisage some $\sim500$ hrs of observing time (excluding overheads) devoted to this program among 20-30 highly valuable systems with \gaia detections of intermediate-separation giant planets (transiting and not) around relatively bright ($R\leq13$ mag) young stars, very metal-poor stars, and early-type dwarfs. For the specific case of young systems hosting large circumstellar dust or 'older' debris discs, artifacts that might mimic planetary signals will be removed by means of follow-up measurements such as direct imaging or astrometry at radio frequencies with SKA 
(\citealt{2016A&A...592A..39K}). 

\textbf{d) Terrestrial planets around Brown Dwarfs} 
Approximately 500 hrs of mission time (excluding overheads) will be allocated to the program of low-mass planet astrometric detection with \theia around brown dwarfs. The target sample will be constituted of $\sim20$ benchmark early L-type dwarfs ($R\leq18$ mag) with and without detected companions by \gaia. In absence of detection, the sample size will allow to determine a lower frequency $f_p$ of terrestrial-mass companions with respect to the He et al. (\citeyear{2016arXiv160905053H}) estimates at the $\sim5 \, \sigma$ level.

\subsection{Compact objects \label{sec:compactobjects_requirements}} 

\begin{figure}[thb]
\centering 
\includegraphics[width=0.99\textwidth]{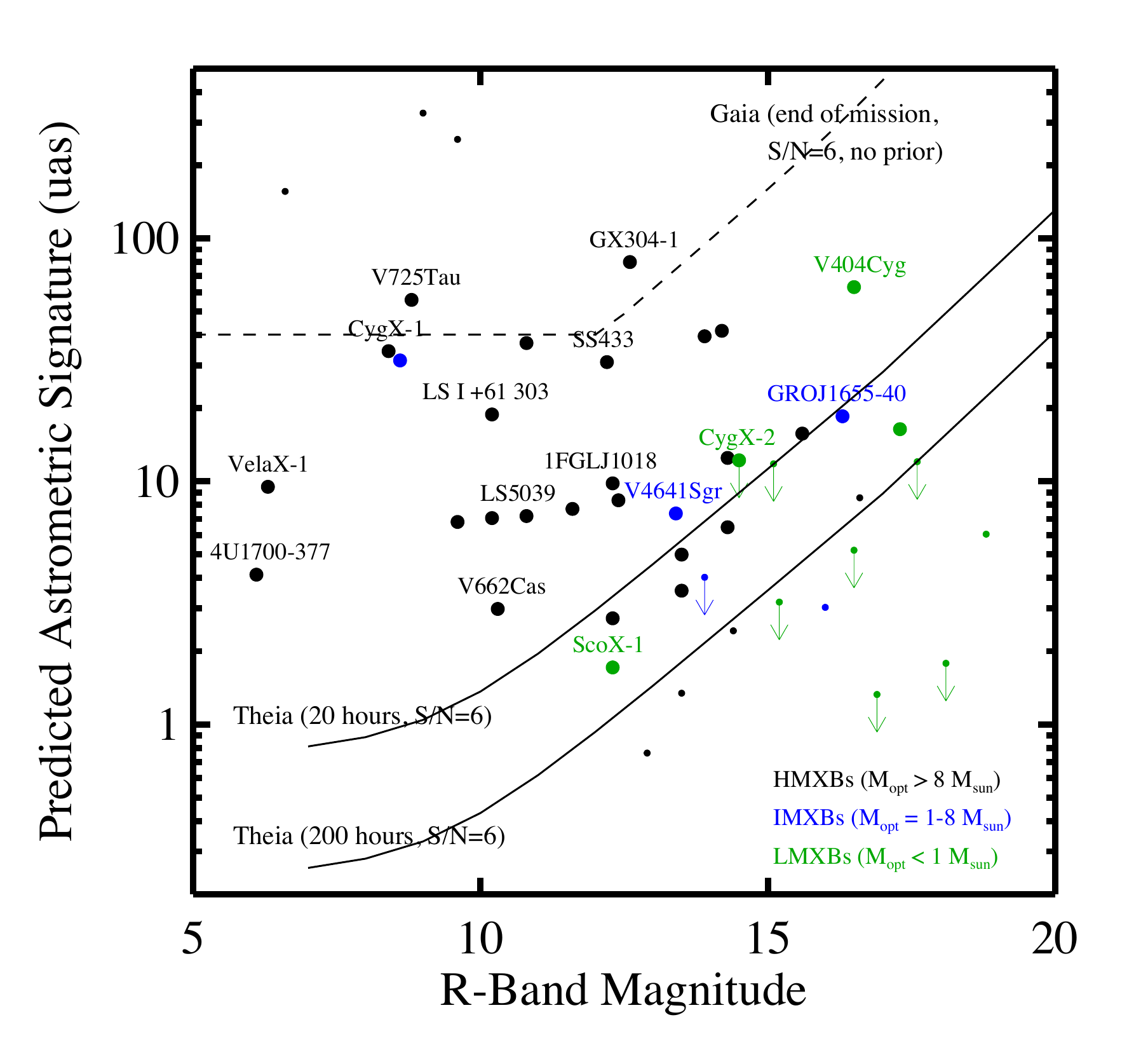}
\caption{Expected astrometric amplitude from orbital motion vs. $R$-band magnitude for NSs and BHS in high-mass (\textit{black circles}), intermediate-mass (\textit{blue circles}), and low-mass (\textit{green circles}) X-ray binaries.  The \textit{solid lines} show the threshold for detection at a signal-to-noise level of 6 for \theia in 20 and 200 hours of observation time.  The \textit{dashed line} shows the threshold for \gaia.  The \textit{larger points} indicate \theia targets.}
\label{fig:amplitude}
\end{figure}

Fig.~\ref{fig:amplitude} shows the expected astrometric amplitude and $R$-band magnitudes for a large number of NS and BH X-ray binaries. 
Precision on the microarcsecond level is required, which is not possible with other missions such as \gaia.  Compared to \gaia, \theia can provide a huge improvement with a dedicated strategy consisting of a series of pointings and observing times adapted to the target magnitudes and orbital phases.  Fig.~\ref{fig:amplitude} shows that while \gaia might measure orbital motion for a few of the largest or closest systems, \theia will provide measurements for a significant fraction of the X-ray binary population.  This includes nearly 30 High-Mass X-ray Binaries (e.g., Vela X-1), and most of these are accreting pulsars, allowing for direct NS mass measurements 
\citep{Tomsick&Muterspaugh10}.  
We also highlight the BH systems where binary orbital inclination measurements will be possible (Cyg X-1, V404 Cyg, GRO J1655--40, and V4641 Sgr).  In addition, \theia measurements will provide definitive answers to long-standing questions about the nature of the compact objects (BH or NS) in SS 433 as well as gamma-ray binaries such as LS 5039, LS I +61 303, and 1FGL J1018.6-5856 by improving constraints on the compact object masses.


For distances, one example of a NS X-ray binary where a \theia distance will lead directly to an improvement in the measurement of the NS radius is Cen X-4.  While \gaia is expected to obtain a distance measurement that is good to $\sim$16\%, a 15 hour observing program for Cen X-4 ($R$=17.6) with \theia will constrain the distance to 0.6\%.  While improving distances by a factor of nearly 30 is impressive, \theia will provide the first parallax measurement for the vast majority of the $>$50 sources on our target list. This science case will be done in conjunction with the search for disc perturbations  by DM subhalos and ultra compact halo mini halos and will not use additional time.


If time allows and depending on the characteristics of the final instrument,  astrometric microlensing observations in the direction of the galactic centre (targeting areas with low amount of interstellar dust, cf  Baade window) will be performed to search for dark compact objects. The frequency of these scans will depend on the final observational strategy, after including open-observatory time.

\subsection{Age of Universe through photometry}

This science case is based on photometry and does not have particular astrometric requirements. It makes use of the periods of thermal stabilization after large variations of the Sun Aspect Angle, where astrometric performance would be degraded, to perform photometric measurements and maximize scientific outcomes.  

\begin{figure}[thb]
\includegraphics[width=0.95 \textwidth,trim=0.8cm 0.5cm 0.8cm 0.2cm ,clip]{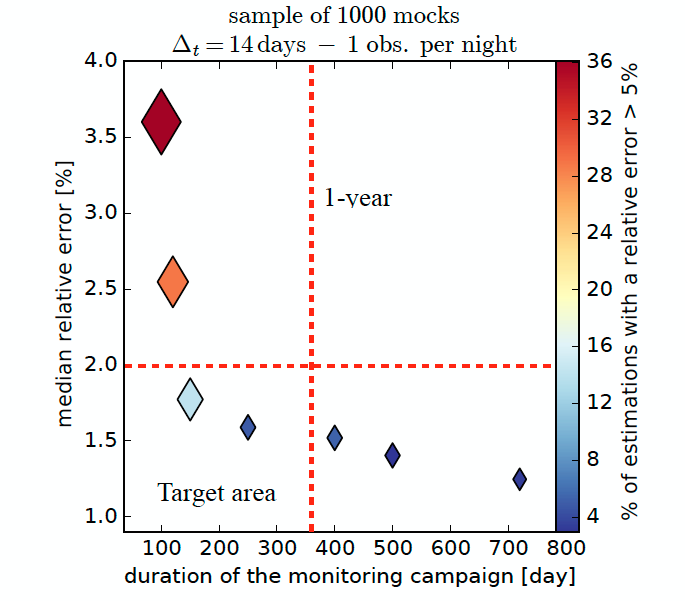}
\caption{Predicted accuracy of the TD measurement for a fiducial TD$=$14 days and a monitoring cadence of 1 point per day to 2 mmag rms. The expected relative error on the TD (\textit{left}) as well as the fraction of objects for which this  precision is actually reached (\textit{right}) are shown as a function of the length of the monitoring period. Essentially 90$\%$ of the observed objects will have TDs to 1.5$\%$ accuracy in 1 year of monitoring.}
\label{fig:mocks} 
\end{figure}

Measuring $H_0$ to 1$\%$ precision requires to measure 50 time delays of quasars to 2\% precision each (e.g. \citealt{Jee+16}). Assuming 1 year measurement of time delays per object, the strong requirement on the data is a mean SNR of 1000 per observing point and per quasar component to ensure that we are not limited by photon-noise. This allows us to measure any object down to R$\sim$19.5 with \theia. 
Using mock light curves with the same properties as the COSMOGRAIL ones (but much higher SNR) and PyCS curve-shifting technique \citep{Tewes2013a}, we  predict the expected precision on the TDs for different fiducial TDs in the range 10-60 days and for a range of lengths of monitoring campaign. The results are presented in Fig.~\ref{fig:mocks}, where the fiducial TD is 14 days, i.e. a pessimistic case, as shorter TDs are harder to measure. 

Fig.~\ref{fig:mocks} also gives the failure rate  (color code), i.e. the fraction of objects for which a minimum precision of 5$\%$ is not achieved for a given realization of the light curve. For a monitoring campaign of 250 days, this is only 10$\%$, meaning that at least 90$\%$ of the objects will be measured to the precision quoted in the figure.

\begin{figure*}[thb] 
\centering
\includegraphics[width=0.42\hsize,clip]{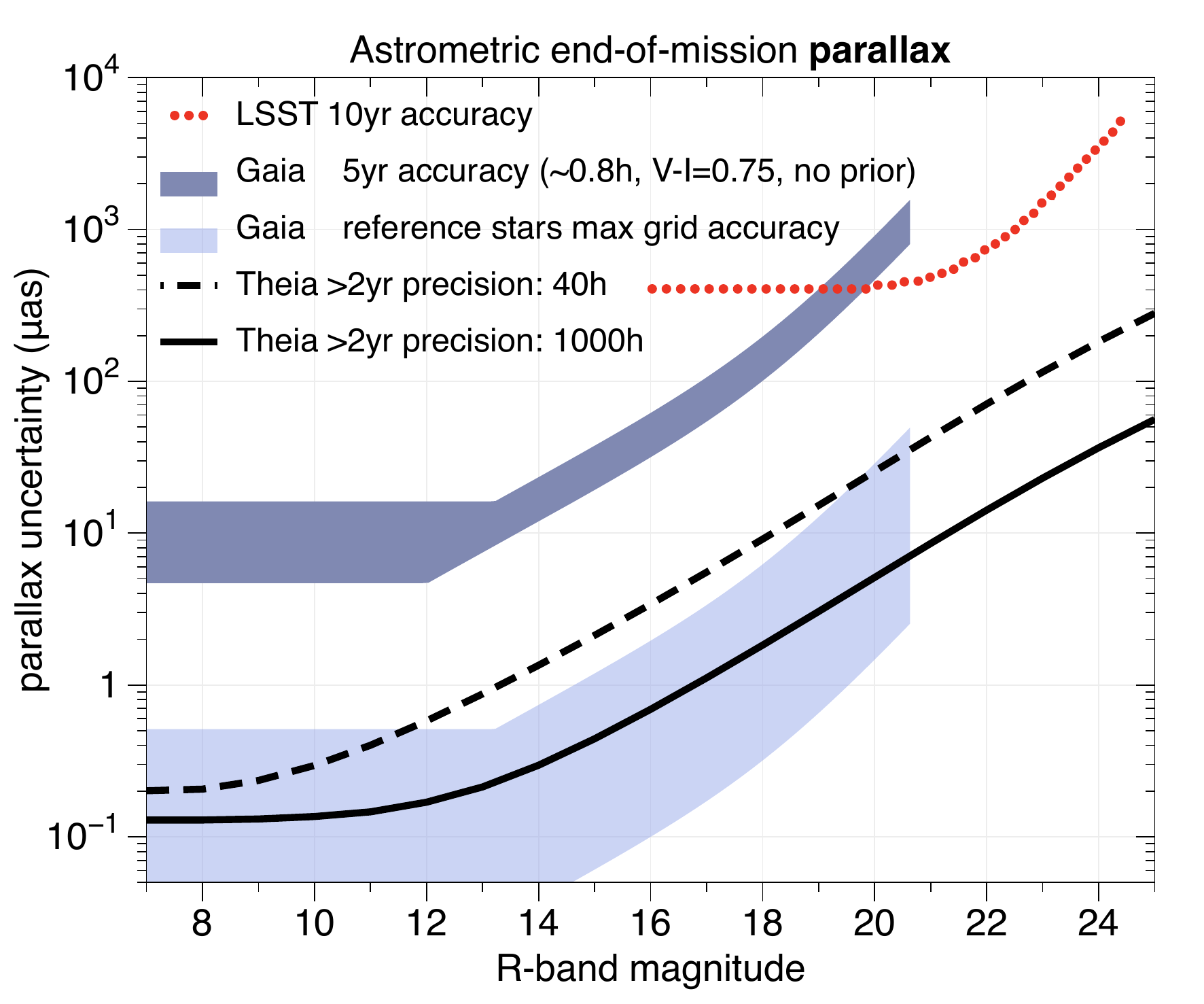}
\includegraphics[width=0.42\hsize,clip]{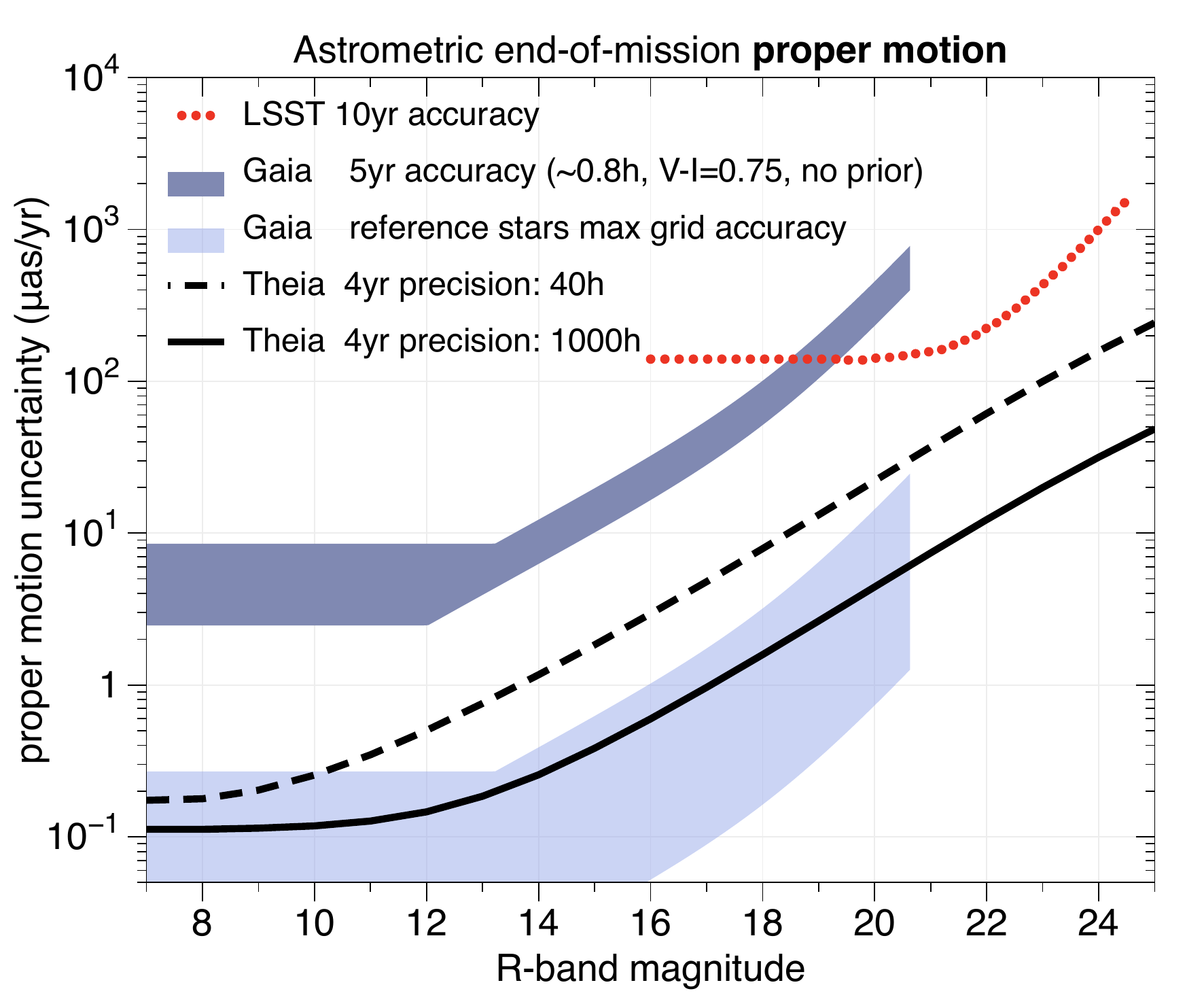}
\caption{Estimated RMS precision on \theia's relative parallax (\textit{left}, for ecliptic latitude $0^\circ$) and proper motion (\textit{right}) in the $R$-band. Also shown for comparison are the estimated accuracies for 10~years LSST \citep{2009arXiv0912.0201L} as well as the 5-year nominal \gaia mission \citep{deBruijne+15} (vertical spread caused by position on the sky, star colour, and bright-star observing conditions). Small-scale spatial correlations ($<$1$^\circ$) between \gaia reference sources will limit the maximum reachable absolute parallax and proper motion calibration for \theia, indicated by the light blue band for a range of assumed spatial correlations as function of reference star magnitude (see Sect.~\ref{sec:calib_priorsAbsAstrometry}). 
}
\label{fig:astroComparions}
\end{figure*}

\subsection{Top Level Requirements}

Table \ref{tab:tech.summary.science} summarizes the science cases with
most stringent performance requirements set in each case.
To cover the science cases of the \theia proposal requirements for two observing modes were derived, as detailed below.

\begin{table*}[t]
  \small
  \caption{Summary of science cases with most stringent performance requirements set in each case. Figures are based on a 4 year mission, thermal stabilisation ($+$slew time) is assumed to take 30$\%$ of the mission time.} 

  \label{tab:tech.summary.science}    
  \centering
  \begin{center}
  \begin{tabular}{lrrrrr} 
    \hline 
Program 
& Used
& Mission 
& Nb of objects
& Benchmark target 
& EoM precision \\
& time (h)
& fraction 
& per field 
& $R$ mag (and range) 
& (at ref.\ mag.) \\
    \hline 
    \hline 
    Dark Matter & 
    17\,000 &
    0.69 &
    10$^2$--10$^5$ &
    20 (14--22) &
    10 $\mu$as  \\ 
    $\&$ compact objects  
    \\
    \hline
    Nearby Earths           
    & 3\,500 &
    0.14 &
    $<$20 &
    5 (1--18)  &
    0.15 $\mu$as  \\ 
    $\&$ follow-up \\
    \hline
    Open observatory & 
    4\,000 &
    0.17 &
    10-10$^5$ &
    6 (1-22) &
    1.0 $\mu$as  \\
    \hline 

    Overall requirements& 
    24 \,500 &
    1.00 &
    10$^1$-10$^5$ &
    6 (1-22) &
    0.15-10 $\mu$as \\  
    \hline 
  \end{tabular}
  \end{center}
\end{table*}
%


\subsubsection*{Deep Field Mode} Required by: Dark matter studies, compact objects, and general astrophysics observations (open time).
\begin{itemize}
\setlength\itemsep{0em} 
\item{\textbf{R1A} Differential centroids must be precise to 10\,\uas per epoch.}
\item{\textbf{R1B} Field of views must encompass a diameter of $0.5^\circ$ for reference system materialization.}
\item{\textbf{R1C} The mode must provide sensitivity to faint objects ($R>20$).}
\item{\textbf{R1D} The mode must allow measurements of up to $10^5$ objects per FoV.}
\item{\textbf{R1E} The mode must provide good quantum efficiency in visible wavelengths (400-900 nm).}
\end{itemize}

\subsubsection*{Bright Star Mode} Required by: Exoplanets and general astrophysics observations (open time).

\begin{itemize}
\setlength\itemsep{0em} 
\item{\textbf{R2A} Differential centroids must be precise to 1\,\uas per epoch.}
\item{\textbf{R2B} Field of views must encompass a diameter of $0.5^\circ$ to provide enough reference objects.}
\item{\textbf{R2C} The mode must provide sensitivity to bright objects ($R<10$).}
\item{\textbf{R2D} The mode must allow measurements of up to $10^2$ objects per FoV.}
\item{\textbf{R2E} The mode must provide good quantum efficiency in visible wavelengths (400-900 nm).}
\end{itemize}

\subsubsection*{Pointing \& reproducibility}
\begin{itemize}
\setlength\itemsep{0em} 
\item{\textbf{R3A} The spacecraft must allow up to 20\,000 re-pointings over the mission life-time depending on the final target list.}
\item{\textbf{R3B}} Calibrations must allow the control of systematics to a level better than 150 nanoarcseconds per epoch for the Exoplanet observations.
\item{\textbf{R3C}} Calibrations must allow the control of systematics to a level better than 10 $\mu$as per epoch for the Dark Matter observations.
\end{itemize}

\subsubsection*{Engineering \& programmatics constraints}
\begin{itemize}
\setlength\itemsep{0em} 
\item{\textbf{R5A} The spacecraft and mission profile must be compatible with an Ariane 6.2 launch.}
\item{\textbf{R5B} Adopted technologies must be at $\mbox{TRL}\geq 5$ or reach that level before 2021.}
\item{\textbf{R5C} The mission must be launch-ready by 2029.}
\item{\textbf{R5D} The ESA CaC must be $\le$\euro550M.}
\end{itemize}

\subsection*{Derived Instrument Requirements}

Based on the above requirements and after an iterative process, the following instruments requirements were derived by the Consortium. The instrument must:

\begin{itemize}
\setlength\itemsep{0em} 
\item{\textbf{I1A}} have a FoV with a diameter $\ge 0.5$.
\item{\textbf{I1B}} have an aperture with diameter $\ge 0.8$m.
\item{\textbf{I1C}} be diffraction-limited.
\item{\textbf{I1D}} have a Nyquist sampled Point spread function (PSF).
\item{\textbf{I1E}} be capable of reading pre-determined pixel windows around objects.
\item{\textbf{I1F}} be capable of pixel readout at $>$kHz rates to prevent saturation of bright stars.
\item{\textbf{I1G}} be capable of integrating for up to $300$s to observe faint objects.
\item{\textbf{I1H} not suffer from optical variations that produce $\ge0.33$ $\mu$as differential centroid shifts during the frame acquisition of Exoplanet observations.}
\item{\textbf{I1I} not suffer from optical variations that produce $\ge2.2 \mu$as differential centroid shifts during the frame acquisition of Dark Matter observations.}
\item{\textbf{I1J} have a thermally stable orbit and a thermally stable spacecraft concept.}
\item{\textbf{I1K} be capable of performing observations for calibration purposes.}
\item{\textbf{I1L} allow monitoring of its long term variations.}
\item{\textbf{I1M} be capable of downlink, in a worst case-scenario, $\sim10^8$ windows per day.}
\end{itemize}

\subsection*{Derived Technical Specifications}

Based on the instrument requirements \textbf{I1A-M} the Consortium derived broad technical specifications that lead to the design of the proposed scientific instrument and mission. The main conclusions of this process are: 

\begin{itemize}
\setlength\itemsep{0em} 
\item That the instrument optics must be better than diffraction limited, have a primary mirror of 0.8\,m and provide a $0.5^\circ$ field-of-view.
\item The instrument camera must cover a $0.5^\circ$ field-of-view and have $\sim10$\,\microns pixels to adequately sample the PSF.
\item The instrument must comprise metrology units to monitor deformations of the optical surfaces positions and of the camera between observations to control deformations to scales of $10^{-5}$ pixel.
\item The spacecraft must provide relative pointing errors smaller than the FWHM of the instrument PSF profile for the longest individual frame acquisition time ($\sim300\,\rm s$).
\item The spacecraft must provide a downlink capable of attaining 74.7~Mbps in the worst case scenario and 52.5~Mbps in a mission-average scenario to transfer windows using 4h/day of antenna.
\item The payload, spacecraft and mission must optimize thermal stability aspects: L2-orbit, enhanced thermal shielding compared to \euclidx, avoid large variations of the sun aspect angle, adopt low-thermal expansion materials (Zerodur, SiC, Si3N4 and/or carbon fiber tubes), adopt active thermal control strategies.
\end{itemize}

The  scaled down and slightly adapted version of the \euclid design, with addition of metrological monitoring of the spacecraft optics and camera were adopted as one of the starting points of the design to fulfill these specifications and  to minimize mission configuration uncertainties

\section{Proposed scientific instrument}
\label{sec:instrument}

\subsection{General description of the payload and challenges}

The \theia Payload Module (PLM) is designed to be simple. It is composed of four subsystems: telescope, camera, focal plane array metrology and telescope metrology. These have been designed applying the heritage knowledge of the consortium members for space mission concepts like \gaia, \hst/FGS, \textit{SIM}, \NEATx/M3, \theiax/M4 and \euclidx. The PLM will be developed and delivered by the \theia consortium with an ESA contribution.

\begin{figure}[h]
\centering
\includegraphics[width =0.85\textwidth]{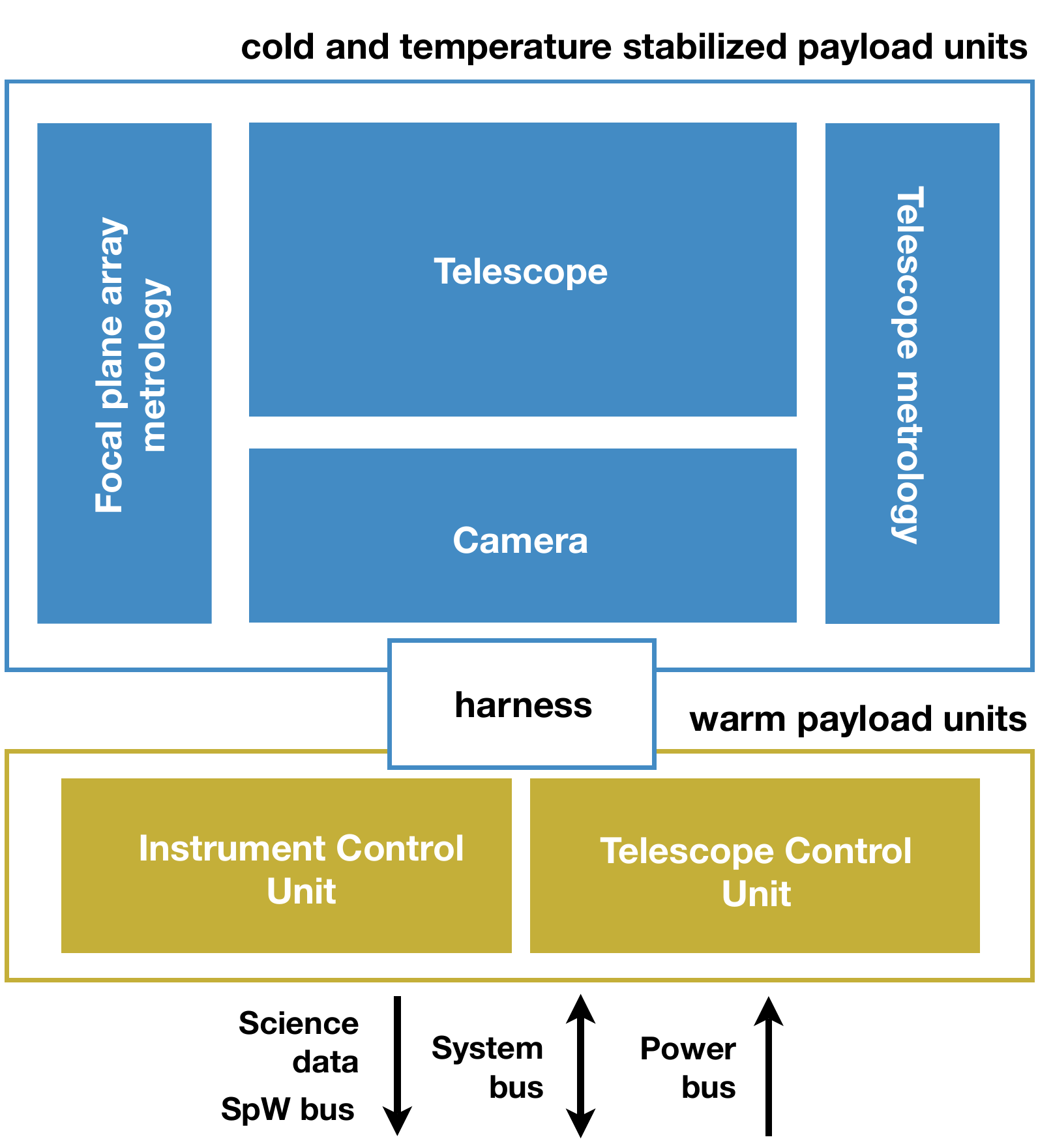}\\*[1em]
\vspace{-0.4cm}
\caption{Block diagram of the \theia PLM Hardware.}
\label{fig:theiam5-plm-blockdiagram}
\end{figure}

The PLM and the service module (SVM) will interface via thermally isolating support struts or \gaia-like bi-pods, and will be radiatively shielded from the SVM to ensure a stable operating thermal environment. A block diagram with an overview of the PLM Hardware can be seen in Fig.~\ref{fig:theiam5-plm-blockdiagram}.

\subsubsection{Instrumental challenges}
\label{sec:inst.challenges}

Achieving micro-arcsecond differential astrometric precision requires the control of all effects that can impact the determination of the relative positions of the point spread function. The typical apparent
size of an unresolved star corresponds to 0.2~arcseconds for a 0.8~m telescope operating in visible wavelengths. 

The challenge is therefore to control systematics effects to the level of 1 part per 200\,000. The precision of relative position determination in the Focal Plane Array (FPA) depends on i) the photon noise, which can be either dominated by the target or by the reference stars; ii)  the geometrical stability of the focal plane array, iii) the stability of the optical aberrations, iv) the variation of the detector quantum efficiency between pixels. These effects impair other missions that  could perform differential astrometry measurements, like \hst, {\textit Kepler}, \wfirst or \euclidx, posing fundamental limits to their astrometric accuracy. All these effects are taken into account in the \theia concept.

\subsubsection{Instrumental concept}
\label{sec:inst.concept}

\begin{figure*}[t]
  \centering
  \includegraphics[width=0.9\hsize]{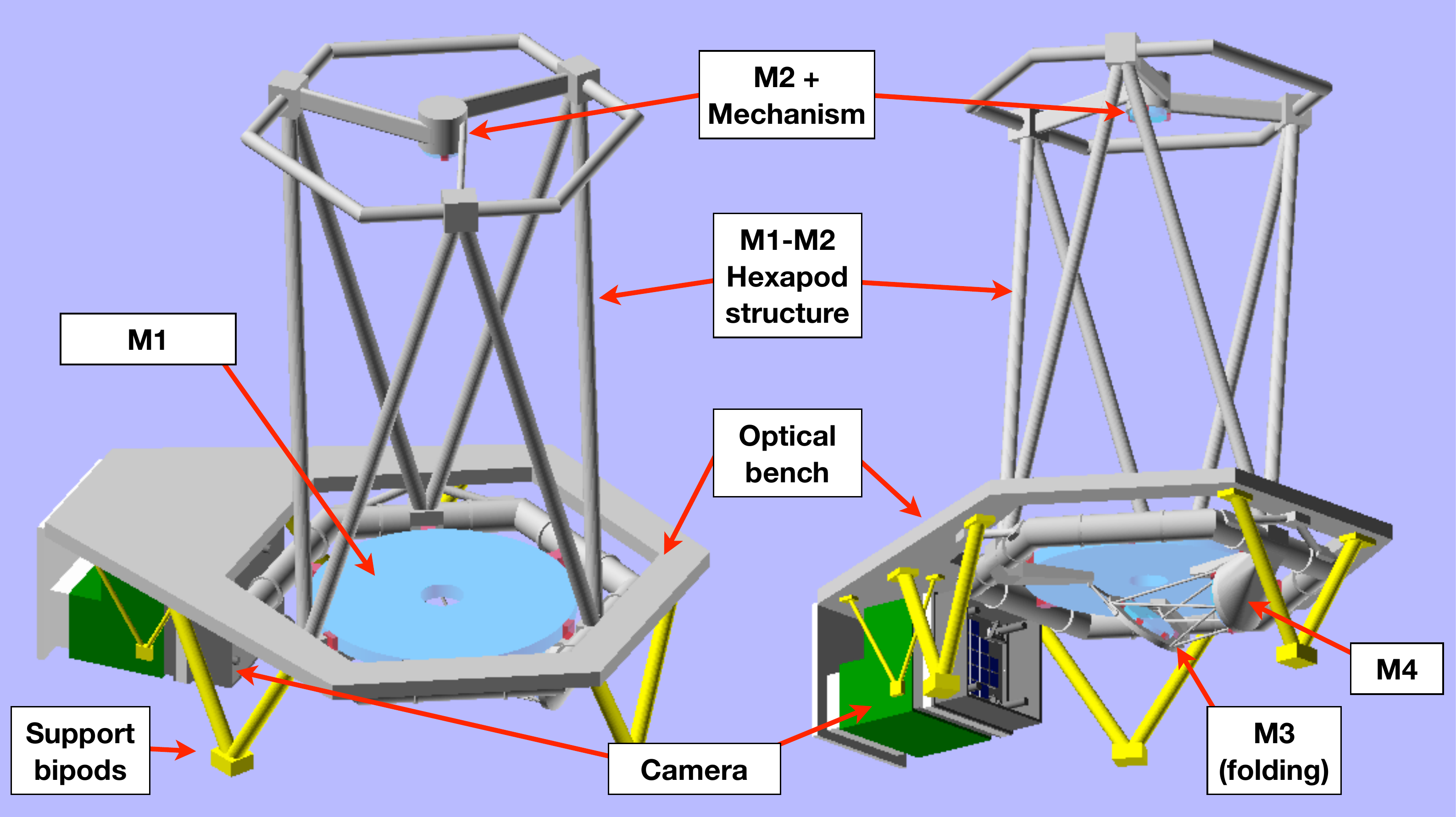}  
  \caption{Overall layout of the \theia Payload Module concept. Volume is estimated in $1.6\times1.9\times2.2$m$^3$.}
  \label{fig:theiam5-plmconcept}
\end{figure*}

To address the challenges outlined in section \ref{sec:inst.challenges} and fulfill the requirements from section \ref{sec:requirements}, two different possible concepts can be adopted. A \textit{NEAT}-like mission consisting on a formation flight configuration \citep{2012ExA....34..385M} or an \euclidx-like mission,\footnote{\euclid red book:
  \url{http://sci.esa.int/euclid/48983-euclid-definition-study-report-esa-sre-2011-12/}} but with a single focal plane and instrument metrology subsystems. Both concepts consist in adopting a long focal length, diffraction limited, telescope and additional metrological control of the focal plane array. The proposed \theia/M5 mission concept is the result of a trade-off analysis between both concepts. 

The \theia PLM concept consists on a single Three Mirror Anastigmatic (TMA) telescope with a single focal plane (see Fig.~\ref{fig:theiam5-plmconcept})  covering a $0.5^\circ$ field-of-view with a mosaic of detectors. To monitor the mosaic geometry and  its quantum efficiency, the PLM includes a focal plane metrology subsystem. While to monitor the telescope geometry, a dedicated telescope metrology subsystem is used.

\subsubsection{Telescope concept}
\begin{figure}[h]
\centering
\includegraphics[width = 0.8\hsize]{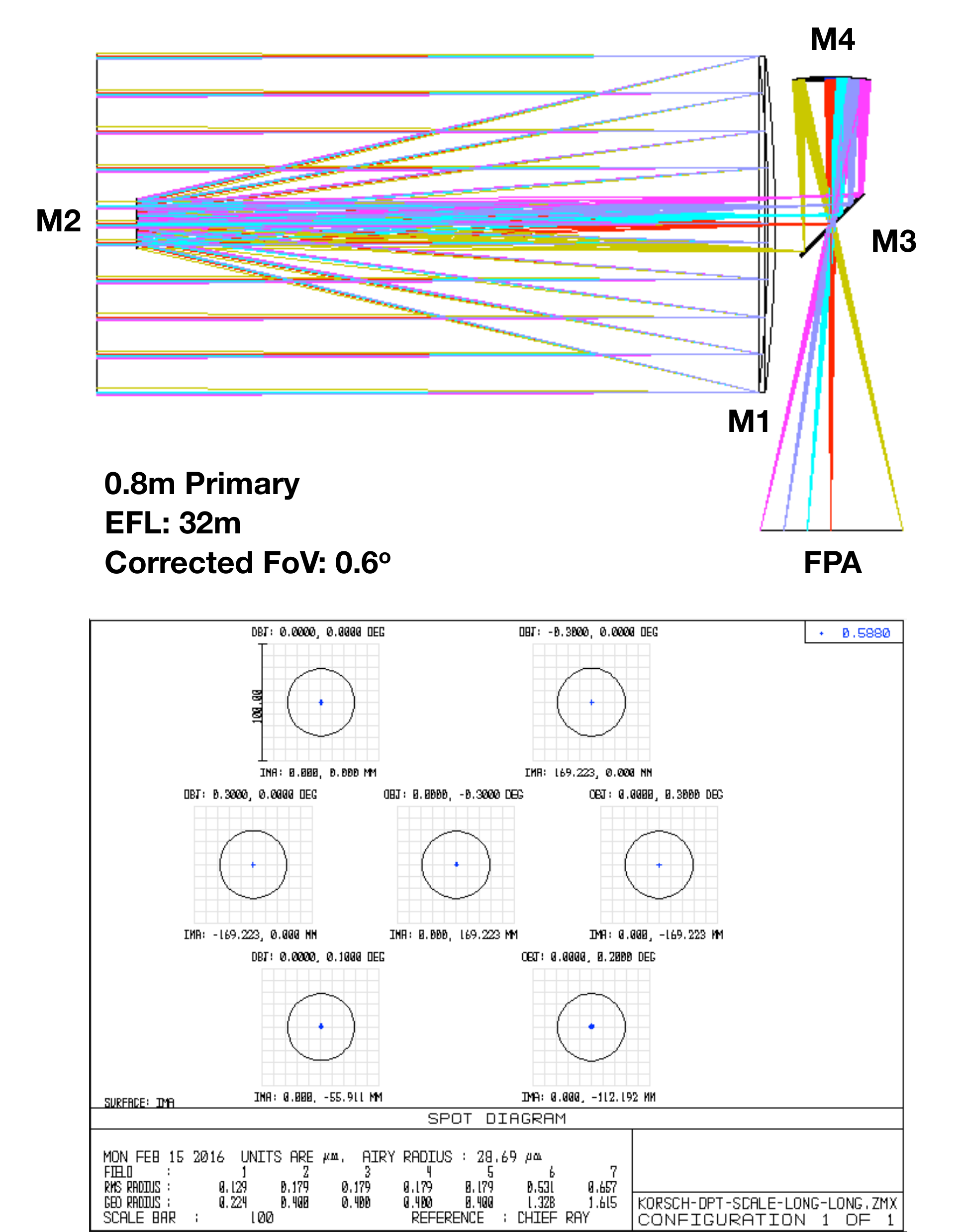}\\*[1em]
\vspace{-0.4cm}
\caption{On-axis Korsch TMA option. Raytracing and spot diagrams for the entire FoV. This design was adopted as the baseline for the \theia/M5 proposal.}
\label{fig:theiam5-onaxis-korsch}
\end{figure}

\begin{figure}[h]
\centering
\includegraphics[width = 0.9\hsize]{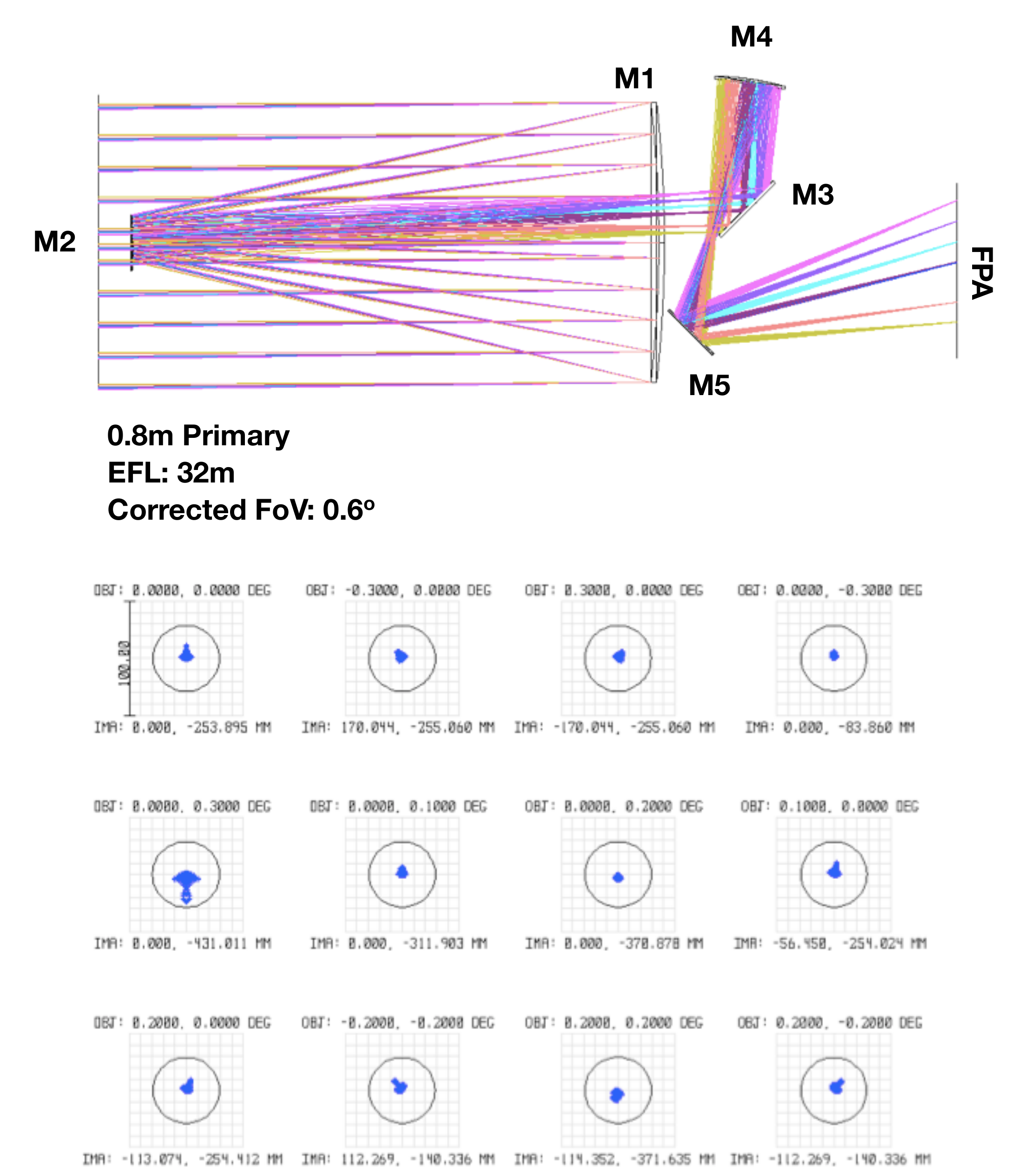}\\*[1em]
\vspace{-0.4cm}
\caption{Ray-tracing and spot diagrams for the entire FoV for the best off-axis Korsch TMA option, but which presents a worse control of the optical aberrations and requires an additional folding mirror.}
\label{fig:theiam5-offaxis-korsch}
\end{figure}

To reach sub-microarcsecond differential astrometry a diffraction limited telescope, with all aberrations controlled, is necessary. Controlling the optical aberrations up to third order in the large \theia Field-of-View required a large exploration of different optical design concepts. A trade-off analysis was performed between different optical designs, which resulted in two optical concepts that can fulfill  requirements. Both are based on a Korsch Three Mirror Anastigmatic telescope; one is an on-axis solution (Fig.~\ref{fig:theiam5-onaxis-korsch}) while the second is an off-axis telescope (Fig.~\ref{fig:theiam5-offaxis-korsch}). In both cases only three of the mirrors are powered mirrors. While the on-axis solution adopts a single folding mirror, the off-axis solution adopts two folding mirrors. For reasons discussed in Section \ref{design:telescope}, we choose the on-axis design as our baseline.

\subsubsection{Camera concept}
To achieve the precision by centroiding as many stars as possible, a mosaic of CCD or CMOS detectors will be assembled on the focal plane. The detectors must feature small pixels ($\sim 10 \mu$m) and well controlled systematic errors. Detailed in orbit calibration of their geometry and response will be monitored via a dedicated laser metrology system.

\begin{figure}[h]
\centering %
\includegraphics[width = \textwidth]{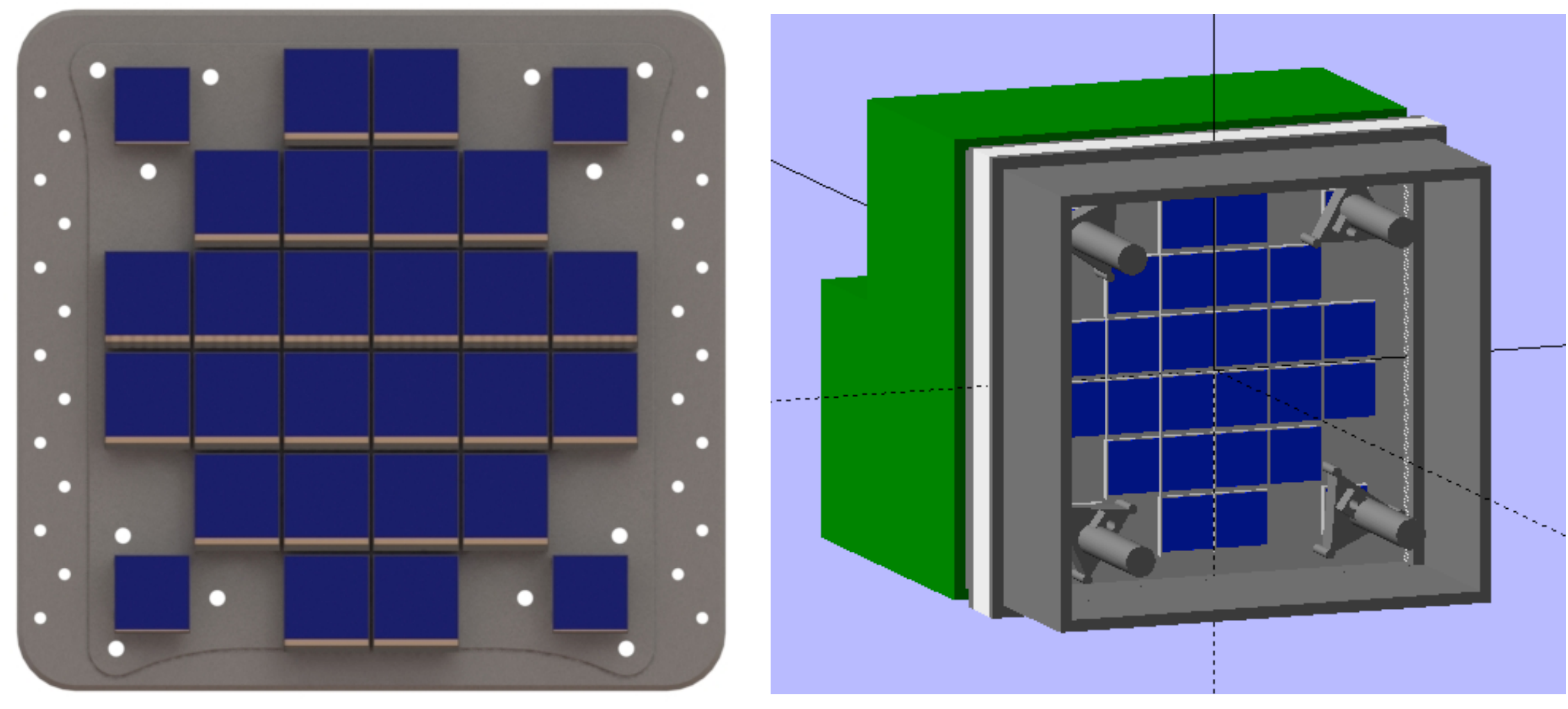}\\*[1em]
\vspace{-0.4cm}
\caption{Concept for the \theia/M5 Camera. Concept for the FPA detector plate at the left. Overall view of the camera concept on the right.}
\label{fig:theiam5-fpaconcept}
\end{figure}

\subsubsection{Focal Plane metrology concept}
\theia/M5 most stringent science requirement results in a centroid error calibration of $10^{-5}$ pixel. Even in the absence of optical system errors, systematics greater than $\mu$pixel are caused by the non-perfect detectors. These are caused by the non-uniform quantum efficiency and by pixel offsets. The pixel layout is not perfectly regular, and differences exists between the exact positions and a perfect regular grid structure. With CCD and CMOS detectors, non-uniform QE mitigation strategies are well known and are calibrated by flat fielding. But to reach a $\mu$pixel differential accuracy the pixel offsets and intra-pixel non-uniformity also have to be calibrated.

\begin{figure}[h]
  \centering
  \includegraphics[width=\hsize]{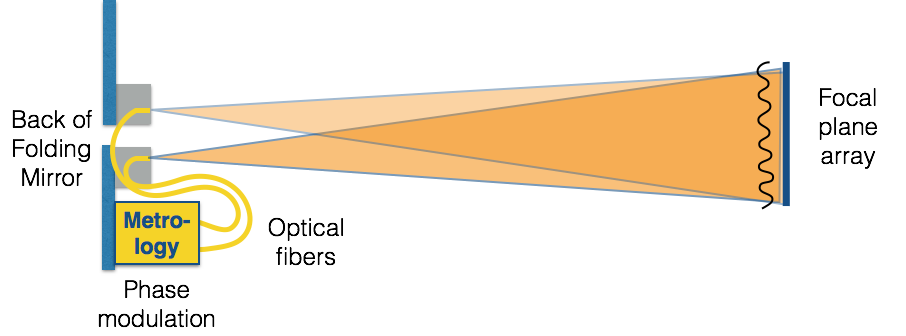}
 \caption{Focal-plane metrology system concept: pairs of optical fibers on the back of the
    folding mirror (M3) produce interference fringes on the focal plane
    detectors. One line is offset in frequency by
    a few Hz with respect to the other line, producing a
    continuous scan of the metrology fringes on the
    detectors at a rate of 10 fringes per second.}
  \label{fig:laser-metrology-system}
\end{figure}

To monitor such distortions of the focal plane array, and to allow the associated systematic errors to be corrected, \theia/M5 relies on metrology laser feed optical fibers placed at the back of the nearest mirror to the detectors. The fibers illuminate the focal plane and form Young's fringes that are observed simultaneously by all FPA detectors (Fig.~\ref{fig:laser-metrology-system}). These fringes allow to solve the XY position of each detector and pixel. To measure the QE (inter-pixel, and intra-pixel), the light beams have their phase modulated by optical modulators. The arrays are read at 50\,Hz providing many frames yielding high accuracy. By measuring the fringes at the sub-nanometer level using the information from all the pixels, it is possible to determine the QE map and solve the position of reference stars compared to the central target with a differential accuracy of $\sim$1\,\uas or better per hour.

\subsubsection{Telescope metrology concept}\label{sec:metrology}
In addition to measuring the FPA physical shape, the rest of the telescope needs monitoring to control time-variable aberrations at sub $\mu$as level. Even at very stable environments such as L2 the telescope geometry is expected to vary for different reasons: structural lattice reorganization (as the micro-clanks observed in ESA/\gaia), outgassing and most importantly, thermo-elastic effects due to the necessary variation of the Solar Aspect Angle during the mission for pointings to the different science targets. The telescope metrology subsystem is based on a concept of linear displacement interferometers installed between the telescope mirrors, with the role to monitor perturbations to the telescope geometry.

\subsection{Design of the payload subsystems}
\label{sec:inst.design-payload}

\begin{table}[t]
  \centering
  \caption{Payload key characteristics.}
  \label{fig:theia-payload-specs}
  \smallskip
  \includegraphics[width=\hsize]{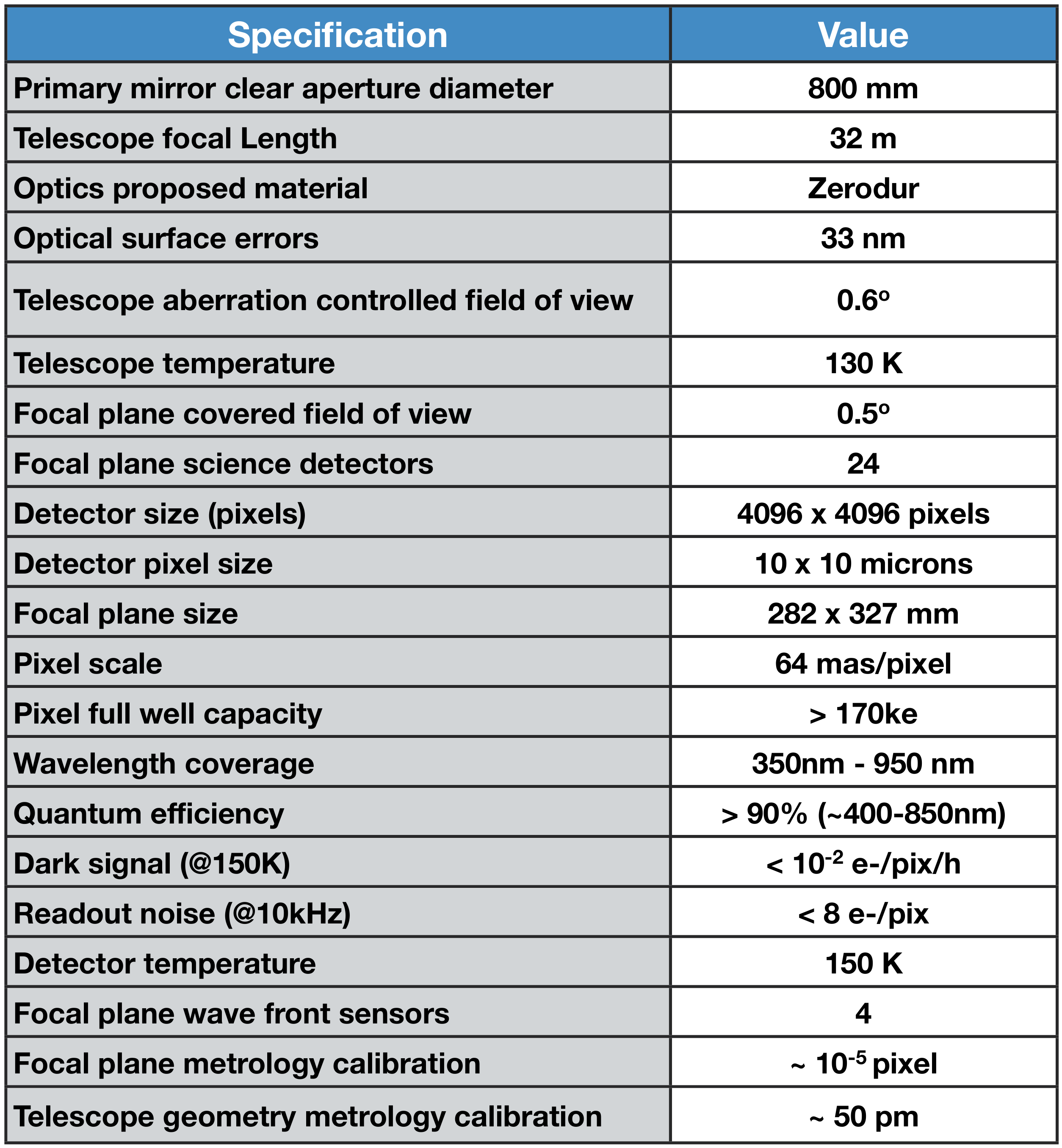}
\end{table}
In this subsection we provide specifications on the previously described concepts. The main specifications of the payload are summarized in Tab.~\ref{fig:theia-payload-specs}.

\subsubsection{Telescope assembly}
\label{design:telescope}
The demanding image requirements for the point spread function and the relative centroid motion due to temporal variations on the optics positions and shapes are the driving design parameters. An extensive trade-off analysis was performed with a series of optical designs, resulting in two concepts. As discussed earlier, the best solutions adopt a three powered mirrors, one configuration on-axis, and another off-axis.

The on-axis design (Fig.~\ref{fig:theiam5-onaxis-korsch}) is chosen as the baseline for the \theia/M5 proposal. The reasons are that the off-axis configuration does not achieve the same aberration control and requires additional folding mirrors with respect to the on-axis, thus adding complexity to the mechanics and the metrological controls. 

A detailed study of the astrometric sensitivity of the optical design to thermo-elastic effects is under way, and it is considering both designs. Results for the on-axis design show that it is possible to model and correct astrometric displacements on the entire FoV and keep the median residuals under $\sim125$ nas, defining the level of systematics control.

The baseline solution (Fig.~\ref{fig:theiam5-onaxis-korsch}) adopts mirrors operating on-axis and a fold mirror (M3). The fold mirror has a central hole to allow the on-axis light go through. The fold mirror is located at the exit pupil near the Cassegrain focus at the same time. The pupil is re-imaged on the same fold mirror by the M4 mirror, and it goes through it again to be imaged on the focal plane. Due to this hole, the centre of the FoV is lost. \theia scientific cases were verified and they do not required the central $0.06^\circ$ of the FoV. A further straylight impact assessment is required for future studies.

Static figure errors of the primary 0.8m mirror will produce centroid offsets that are mostly common-mode across the entire field of view. Differential centroid offsets are significantly smaller than the field dependent coma and are negligible.  Similarly, changes in the primary mirror surface error produce mostly common-mode centroid shifts and negligible differential centroid offsets. Static figure errors of the other mirrors will produce static biases on the centroids. However, temporal variation of the shape of the mirrors after the primary would produce non-common centroid shifts. As $\lambda/200$ temporal variations of the figure would produce sub-$\mu$as shifts, a monitoring of wavefront variations better than $\lambda/1000$ is required to ensure control of errors caused by secular mirror-deformations and optimal focusing of the telescope.

To ensure optimal focusing of the optics, the \theia concept adopts a 5 degrees-of-freedom, \gaia-like, mechanism at the secondary mirror \citep{2005ESASP.591...47U, Compostizo2011} to enable sub-micrometer repositioning after launch. And an active control of the temperature of the telescope is proposed (Sect.~\ref{sec:satel-design-descr}) by Thales Alenia Space. 

The mirrors can adopt low temperature optimized Zerodur \citep{doi:10.1117/12.2055086} or ULE, using light-weighting on M1 \citep{2009SPIE.7425E..0FK, 2010SPIE.7739E..2LK, 2014SPIE.9151E..0KK}. To minimize thermo-elastic impacts in the position of the mirrors, SiC, Si$_3$N$_4$ (CTE at $10^{-6}$K$^{-1}$) or CFRP based materials \citep[$10^{-8}$K$^{-1}$ was reached for Ti+CFRP for LISA, e.g.][]{doi:10.1117/12.925112} could be adopted for the telescope structure. Aluminum coatings can be adopted. The estimated total telescope mass considering Zerodur and SiC is 226.7\,kg (272.0\,kg with 20\% margins).

\subsubsection{Camera}
\label{sec:focal-plane-assembly}
The \theia science cases require a FoV of $\sim0.5^\circ$.  Our concept is based on filling the focal plane array with 24 detectors arranged in a circular geometry (Fig.~\ref{fig:theiam5-fpaconcept}). Each detector comprises at least  $4096$(H)$\times4096$(V) pixels of $\sim10$\,\microns. At the border of the FPA, four Shack-Hartmann wave front sensors sensitive to optical path differences of $\lambda/1000$ are placed -- performances are similar to \gaia WFS \citep{doi:10.1117/12.825240}. They can operate as a trigger by verifying that there is no variation of the shape of the optical surfaces before an observation starts, as an additional source of calibration information, and they enable a fully-deterministic and optimal focusing of the \theia telescope along the mission. To minimize straylight by bright objects, the \theia WFS telescopes could be treated with highly absorbing coatings such as NPL Super Black \citep{B204483H} or Surrey Nanosystems Vantablack \citep{Theocharous:14}.

Two detector technologies can be adopted: CCDs or CMOSes. CMOS detectors present a high QE over a larger visible spectral band, that can also reach infrared wavelengths depending on the sensitive layer, and programmable readout modes, faster readout, lower power, better radiation hardness, and the ability to put specialized processing within each pixel. On the other hand, there are many known CMOS detector systematics, even for advanced detectors  as the Teledyne H4RG10. The most challenging effects are: fluence-dependent PSF, correlated read noise, inhomogeneity in electric field lines and persistence effects \citep[e.g.][]{Simms2009}. For the \theia/M5 proposal we select CCDs due to their more mature status, including in astrometric missions such as ESA/\gaia. CCDs also have systematic effects but they are much better understood and have mitigation strategies in place (eg.
charge transfer inefficiency, \citep[e.g.][]{2012MNRAS.419.2995P, 2013MNRAS.430.3078S, 2014MNRAS.439..887M}.

In our baseline design we select CCDs, like the e2v CCD273-84 detectors originally developed for the \euclidx/VIS instrument \citep{2014SPIE.9154E..0RS}, with a minor modification to its pixel size, from 12\,\microns to 10\,\microns. Nevertheless, depending on the level of characterization of the CMOS detectors that are being developed for \euclidx, \jwst and \wfirst, and on  results from ESA TRP activities with European companies (as e2v, Leonardo and Selex) to further develop and characterize this type of technology, \theia's FPA could employ CMOSes. 

The typical characteristics of detectors required for the \theia/M5 Camera can be seen in Tab.\ref{fig:theia-payload-specs} indicating that complete feasibility of the focal plane data can be achieved with very high TRL device as the e2v CCD273-84. 

\begin{figure}[t]
  \centering
  \includegraphics[width=\hsize]{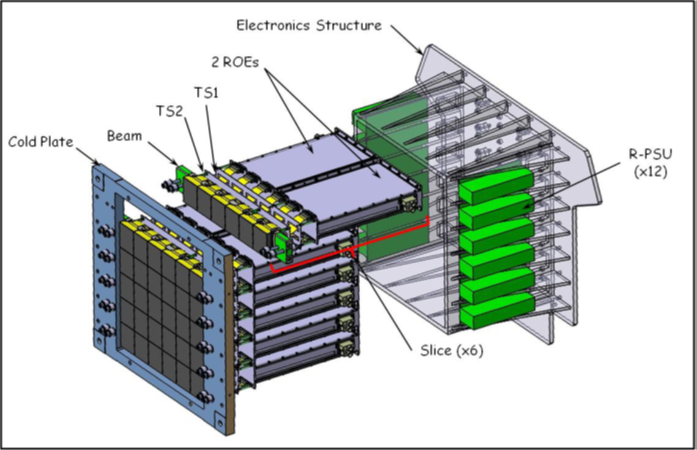}
  \caption{Optomechanical view of the Focal Plane Array of the \euclidx/VIS instrument. The structure of the \theia/M5 Camera backplane might be similar, albeit with a different geometrical arrangement.}
  \label{fig:inst.fpa}
\end{figure}

\begin{figure}[h]
  \centering
  \includegraphics[width=0.95\textwidth]{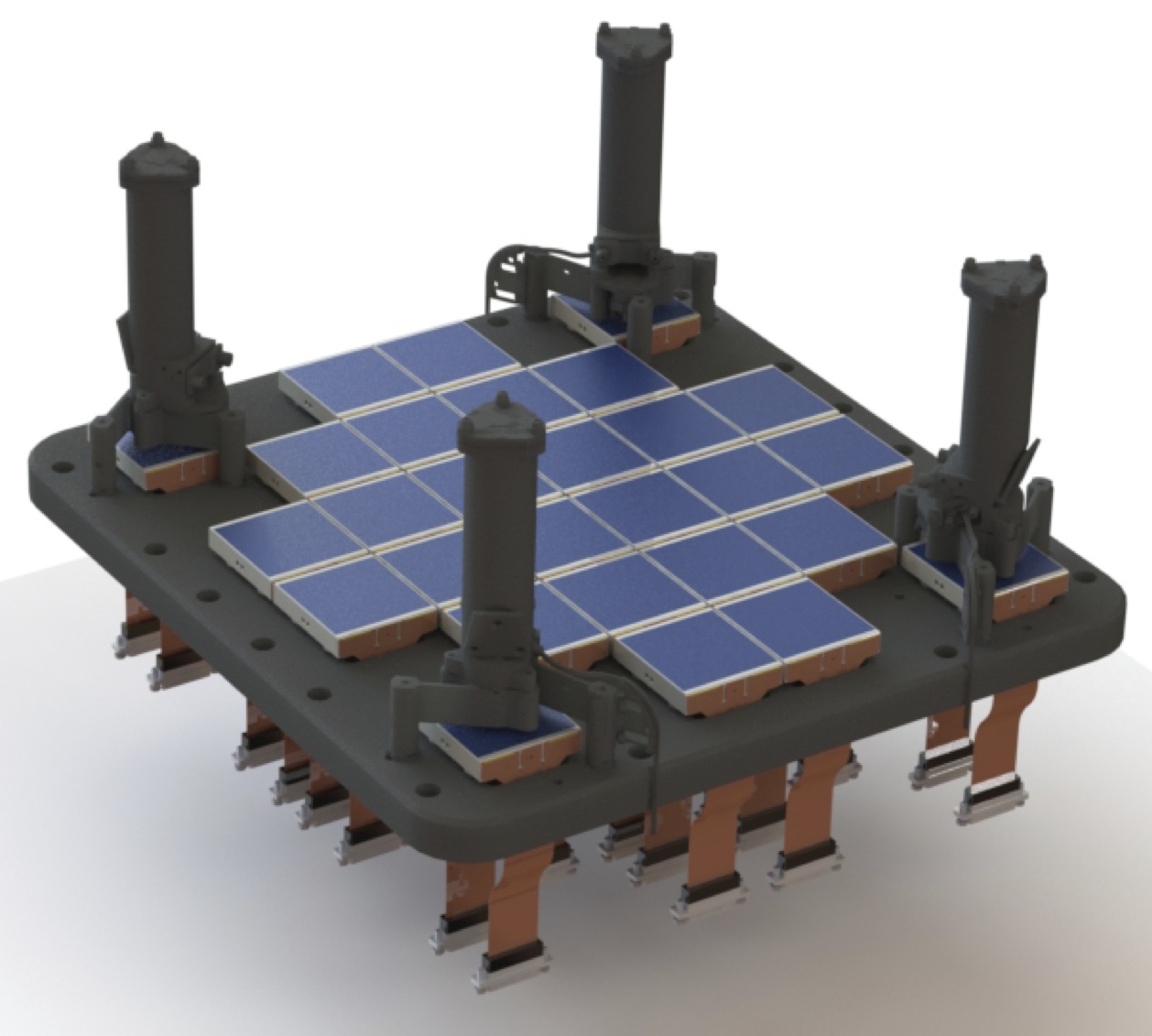}
  \caption{Indicative view of the \theia Focal Plane Array concept populated with e2v \euclid CCD273-84 detectors and TNO \gaia WFS. The actual orientation and FoV of the \theia WFS will be determined at Phase-A.}
  \label{fig:inst.fpa.WFS}
\end{figure}

The main \theia science cases have a wide dynamic range. As CCDs would not be able to readout the exoplanet target stars at a high enough rate, there are two possible solutions for a CCD-based FPA. Either two of the detectors in the focal plane array can be replaced by CMOS detectors, and thus individual windows could be read at high enough rates, or two of the CCD detectors of the focal plane could be covered with a filter to prevent saturation of the brightest target stars. We consider two detectors always to provide dual redundancy. During the exoplanet observation, the target star would always need to be located at one of such detectors. Also, to read out the CCD detectors while observing faint targets for DM related science cases, a shutter mechanism using a slow leaf like \euclid's design can be adopted \citep{2010SPIE.7739E..3KG}. We note that these solutions are only necessary if CCD detectors are chosen over CMOS detectors during the Phase-A studies.

\paragraph{Focal Plane Data Handling Architecture.}
\label{sec:focal-plane-data}

The number of large scale detectors to be supported and
processed requires a modular approach. A distributed data
handling has the advantage to locally and in parallel provide
the data acquisition, providing the proper buffering and computation
capability for each detector module of the FPA array. Only after the
local parallel processing of the data, the data will be forwarded to
the central DPU for final formatting, compression and ground transmission
minimizing resource-associated risks. The modular
approach when properly decoupled in terms of H/W resources can provide
additional redundancy and extended life time, at most paid in term
of focal plane reduction in the case of critical failure of one or more detectors (or an element of their direct processing chain).

Each detector will have its own Local Digital Processing (L-DPU) which
will power and drive the detector, sampling and processing its data
outputs. The combination of a detector and a L-DPU will constitute a
FPAM (Focal Plane Assembly Module) which will
\begin{itemize}
\setlength\itemsep{0em} 
\item control and manage the detector low level functionality;
\item support the acquisition at the maximum throughput rate, from up to 64 channels;
\item support buffering of at least 4 complete frames;
\item receive from the Instrument Control Unit (ICU) the
  roto-translational parameters to adjust the current data set of
  data;
\item accumulate (add) sequentially  roto-translated corrected ROI (Region Of Interest);
\item possibility to communicate with the closest 8
  FPAMs\footnote{located at North, North-East, East, South-East,
    South, South-West, West, North-West} to support the transmission
  of the pixels residing in overlapping ROIs.
\end{itemize}

Conversely the main ICU will coordinate the power distribution to the
FPAMs modules, the driving of the clock and synchronisms to minimize
EMC effects and keep time synchronization between all the
FPAMs. In general the ICU will:
\begin{itemize}
\setlength\itemsep{0em} 
\item Control and manage all the FPAMs  modules functions;
\item Transmit to the FPM the roto-translational parameters  to adjust the current data set of data;
\item Dispatch to each FPMA  the list of overlapping ROI data areas to be received or transmitted;
\item Compress / scale, format \& transmit sensors data according to
  the available S/C TLM allocation;  
\item Buffer the complete FPAM dataset records received from the sensors modules.
\end{itemize}

\paragraph{The L-DPU Data Handling Architecture}
\label{sec:l-dpu-data}

The derivation of the H/W parameters, representative of a L-DPU
suitable to be host in a FPAM module strongly depends on the expected
computational load. Since the early phase concept stage a high
computational capability solution has been addressed. The standard design
techniques for high performances DPU for typical space applications
relies on two strategies: base the development around a well proven
and qualified processor family or develop computational blocks in a
hardware description language scaling the performances on the
available Field Programmable Gate Array (FPGA) target technology. Such
approach has the advantage to scale the computational performances to
the latest radiation-tolerant technology available from the FPGA space
market.

\begin{figure}[t]
  \centering
  \includegraphics[width=0.8\textwidth,trim=0cm 0cm 0cm 0.1cm,clip]{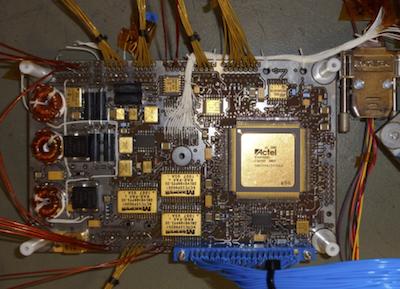}
  \caption{FM BepiColombo DPU SCU Board/Eurocard of $100 \times
    160$\,mm$^2$. SCU is a high-rel DPU FPGA processor card such as
    the ones matured on the BepiColombo Mission / SERENA DPU in which
    the DPU had to control and support a similar S/C data handling bus
    and four sensors units.}
  \label{fig:inst.fpa.DPU}
\end{figure}

Each \theia/M5 L-DPU would then have analog and digital front ends. The
whole L-DPU, as demonstrated by the SCU plus ELENA Main DPU based FPGAs 
in the SERENA package onboard BepiColombo ESA mission (Fig.~\ref{fig:inst.fpa.DPU}), and, by the new GR712RC Based DPU
developed for the SWA Package in the frame of Phase A/B developments
for the ESA Solar Orbiter Mission, may be hosted in a simple and
compact $16 \times 10$\,cm$^2$ Eurocard PCB cards.

A first estimate of the power required to run each L-DPU is about 5230\,mW (6277\,mW with 20\% contingency). So far the complete 28 array of FPAM (24 science detectors and 4 wave front sensors and the related L-DPUs) requires about 147\,W (176\,W) in total.

\paragraph{The ICU  Data Handling Processors}
\label{sec:icu-data-handling}

The space FPGA solution investigated for the FPAM has as constraint to have 25\,MHz as a typical upper limit of the system clock of the CPUs implemented in this way. Conversely high-rel COTS processors can run at much higher speeds but may require additional glue logic and H/W control logic which may penalize the whole power or dimension budgets of the digital electronics. Following the recent eruption in the space component market of the high-rel/rad-hard GR712RC AEROFLEX processor recently qualified up to the space levels standards, also having the very appealing advantage to be not controlled by ITAR regulation, we considered replacing the FPGA Leon3 FT synthetized Processors, present on the LDPU, by this processor. The rationale relies on lower power budget, smaller mechanical package, availability of 6 SpaceWire channels per chip and compatibility with the optimized data handling architecture of the L-DPU. In particular the AEROFLEX GR712RC processor is an implementation of the dual-core LEON3FT SPARC V8 processor using RadSafeTM technology. The addressed \theia CPU solution will natively support the memory bus Error Detection and Memory Correction (EDAC) on all the memory segments.

\paragraph{ICU Hardware Architecture}
\label{sec:icu-hardw-arch}

\begin{figure}[t]
  \centering
 \includegraphics[width=\hsize]{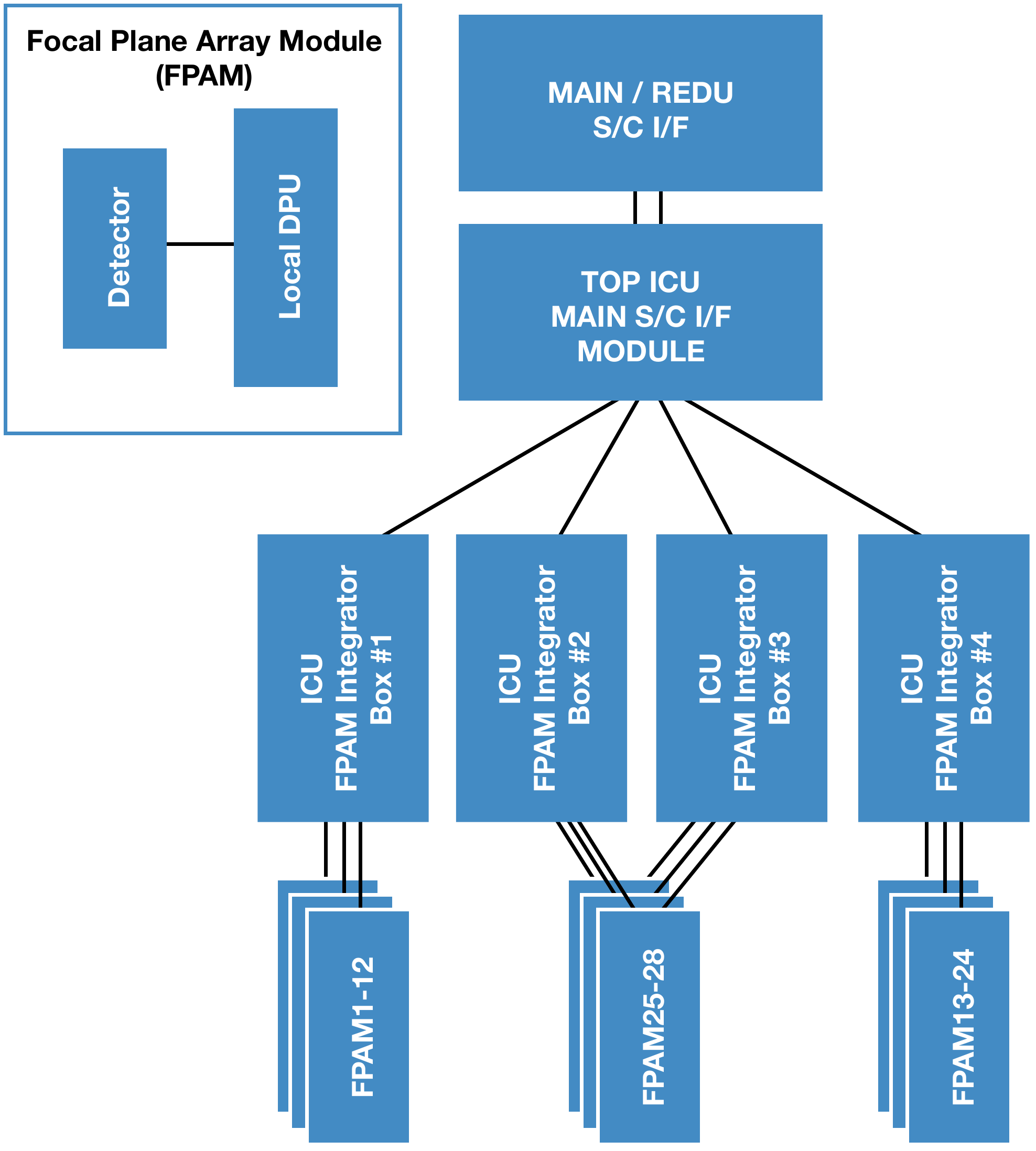}
  \caption{ICU Hardware Architecture}
    \label{fig:inst.fpa.icu-hw}
\end{figure}

An analysis of the general H/W and functional requirements for ICU data processing has led to verification that the same modular approach defined for the FPA processing could be pyramidal extended to the ICU assuming as basic tile the optimized architectures which were successfully verified and tested first in the frame of the BepiColombo Mission within the SERENA/SCU DPU and then in the frame of the ESA Mission Solar Orbiter for the Solar Wind Analyzer package DPU. Assuming these consolidated products cards, the DPU design capable to be ready to support the required 28 FPAM in such SpaceWire based topology (see Fig.~\ref{fig:inst.fpa.icu-hw}) where up to 12 FPAM at time can be developed on 5 standard 100x160mm2 euro cards for each ICU FPAM Integrator Box:

\begin{itemize}
\setlength\itemsep{0em} 
\item[(1)] Primary I/F card
\item[(2)] Primary Main DPU card interfaced to Ancillary 1,2,3 and Upper Main node
\item[(3)] Ancillary 1 interf. to Main \& to FPAM 1 to 4
\item[(4)] Ancillary 2 interf. to Main \& to FPAM 5 to 8
\item[(5)] Ancillary 3 interf. to Main \& to FPAM 9 to 12
\end{itemize}

\theia main Primary (card 2) and Ancillary (card 3 to 5) are based on the proven and validated designs described above. The Main DPU tasks running on card 2 guarantees data communication between the higher ICU Top box, to the all FAPM modules. Similarly the TOP ICU Main Module would be based similar five standard $100\times160$mm$^2$ euro cards. WFS detectors would have redundant  infrastructures.

The top-level ICU contains 8 Leon 3 FT processor + 8 FPU, 6 GBytes SDRAM EDAC protected memory for temporary storage of up to 50 frames (including FPA calibration data). These processors would be added to the $8\times 4$ Leon 3 FT processors and $8\times 4$ FPU available in the ICU FPAM integrator boxes, bringing the overall budget to 40 high-speed processors. According to the tested existing boards, the power is estimated to be $< 10$ W for each ICU box.

\paragraph{Camera budgets \label{sec:mechanics-budgets}}

Top ICU DPU Module structure according to the electronics layout
presented in the previous sections could  be accommodated in a typical box
structure similar.
This architecture would have the advantage to utilize the I/F board as 
a motherboard to route the signals to the Main and Ancillary DPU stack. 
The I/F board would host all the I/F connectors with a through hole 
mounting which would allow to eliminate completely the harness inside 
the box. Being modest the power dissipation of the boards, the thermal 
path would be completely assured by the 5 fixation turrets also 
providing a good stiffness of the box.

The total power budget for the camera, including the ICU, is estimated at 291\,W. The total mass budget for the total camera unit including the SVM deported units, radiator and a possible shutter unit is 95.2\,kg (114.24\,kg with 20\% margin). 

The camera is the PLM system that will drive the data transfer budget. Considering a worst case scenario, $\sim134.5$ Gbytes/day would be produced in a day filled with one-minute integrations of DM-like observations, each with $10^5$ windows of $18\times18$ 16-bits pixels will be created, three FPA metrology sessions/integration and WFS measurements, and considering a compression factor of 2.5 (assuming CCSDS 121.0 or FAPEC). The average value during the mission would be $\sim94.5$ Gbytes/day.

\subsubsection{Focal Plane Array metrology}
\label{sec:laser-metrology}

\begin{figure*}[h]
  \centering
  \includegraphics[width=0.8\textwidth]{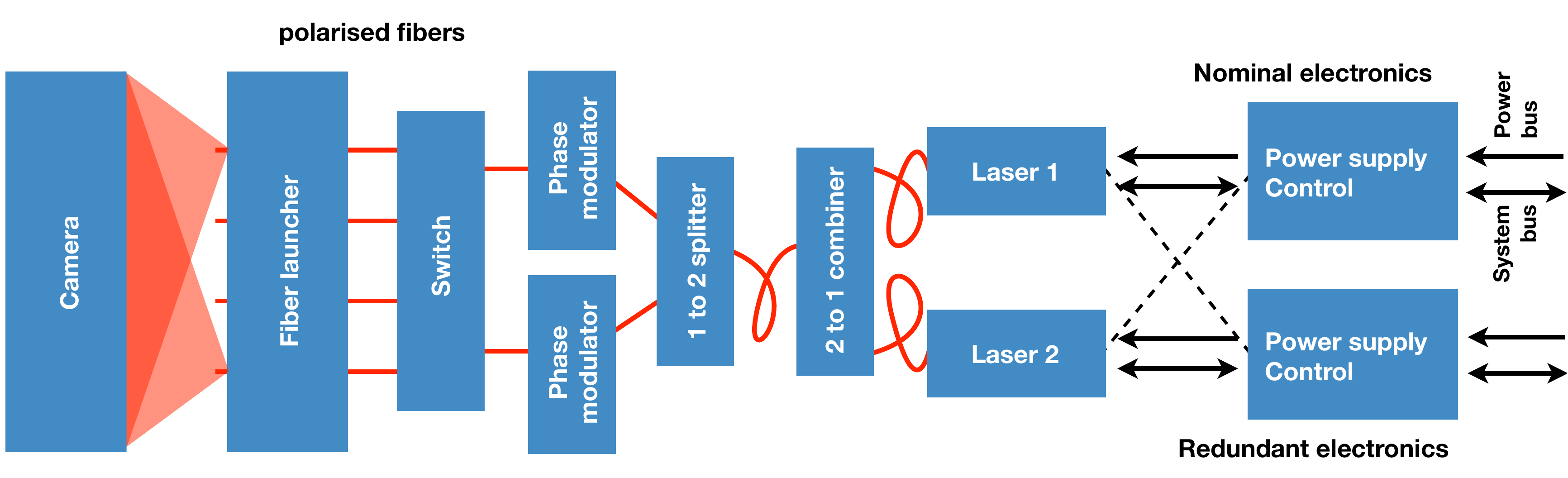}
  \caption{Conceptual block diagram of the focal plane array metrology subsystem.}
  \label{tab:inst.fpa.fpamet.diagram}
\end{figure*}

The focal-plane metrology is used to monitor the position of the detectors and pixels relative to each other to conduct the astrometric measurement. It is
also used to calibrate the inter- and intra-pixel response of the detector during periodic focal plane-calibration.

Three testbeds have been set up to demonstrate that this metrology concept can reach $10^{-5}$ pixel levels. Two were built in USA, at the NASA/JPL: the MCT (Micro-pixel Centroid Testbed) and the VESTA (Validation Experiment for Solar-system STaring Astrometry) \citep{Nemati+11}. The best results obtained at the JPL testbeds were $10^{-4}$ pixels after using flat field and pixel offsets corrections, measured respectively with a supercontinuum source (incoherent broadband) and a single mode stabilized laser. By averaging relative star positions over groups of detector positions (using 100 independent positions, i.e.\ a space of $\sim 40 \times 40$ pixels for Nyquist-sampled centroids), the final error went down to $5 \times 10^{-5}$ pixels.

\begin{figure*}[t]
\centering
\includegraphics[width = 0.4\textwidth]{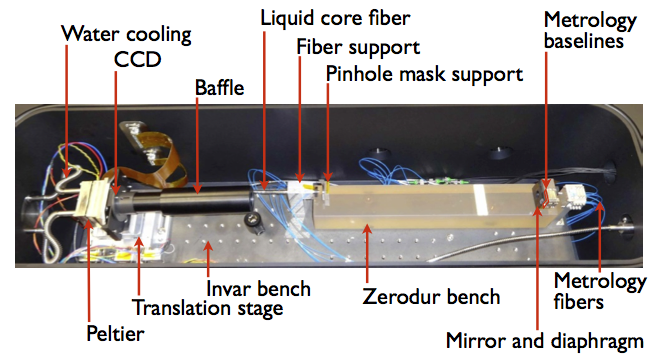}
\includegraphics[width = 0.55\textwidth]{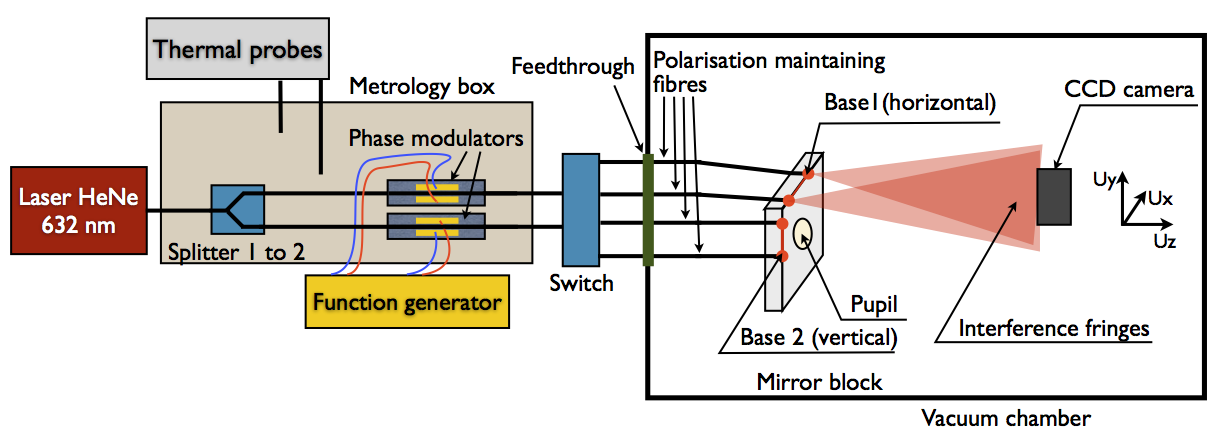}
\caption{\textbf{Schematic of the NEAT-demo setup of picture of the internal part of the vacuum chamber.}}
\label{fig:inst.calibration}
\end{figure*}

A metrology testbed was built in Europe, in France, at IPAG/Grenoble: the NEAT-demo\citep{2014SPIE.9143E..4SC}. A schematic of this testbed setup is shown in Fig.~\ref{fig:inst.calibration}. It has been determined that the level of stray light inside the vacuum chamber of the testbed was too high to obtain a useful measure of pixel offsets due to the difficulty to attenuate properly coherent stray light. Nevertheless, even without a stringent control of stray light the NEAT-demo testbed reached a calibration of $6\times10^{-5}$ the pixel size \citep{2016arXiv160800360C, 2016arXiv160902477C}.

The FPA metrology block diagram can be seen in Fig.~\ref{tab:inst.fpa.fpamet.diagram}. The system consists of the metrology source, the metrology fiber launchers and the focal plane detectors. The
focal plane detectors are the detectors of the \theia/M5 camera itself, described in Sect.~\ref{sec:focal-plane-assembly}; the detectors alternatively measure the stellar signal (54\,s observations) and the metrology (1\,s per
axis every minute). The metrology fiber launchers consist of a set of optical fibers (at least four) attached to a fiber launcher at the back of the M3 mirror. 

The fiber tips are at a distance of $\sim40$ cm from the FPA. Two single mode fibers with a numerical aperture of 0.4 and an optical power of 1 mW at the tips yield about $10^8 \ e^-$/s per pixel on the detector. Because of the proximity between the fiber tips and the focal plane, the large numerical aperture of 0.4 is required. Even with this aperture, the flux will be lower at the corners of the FPA, thus if full calibration accuracy is required for the extreme positions of the FPA, more power per fiber will be required. A power of 1 mW per fiber tip allows a characterization of the center of the focal plane at the required level in a relatively short period. To reach the full calibration accuracy $10^{10}$ photons are needed, this total is obtained every 100 minutes of operations (accounting for the 1 second of metrology per axis available every minute). Using two 15 mW lasers will ensure redundancy and a flux of at least 1 mW at the fiber tips, accounting for the losses in the metrology system. With this setup, an estimate of the power required to run the Focal Plane Array metrology subsystem is 29\,W (35\,W with 20\% margin). The total mass of the system is estimated to be 7.9\,kg (9.4\,kg with 20\% of margin).

\subsubsection{Telescope metrology \label{sec:telescopemetrologysection}}
\begin{figure*}[t]
\centering
\includegraphics[width = 0.8\hsize]{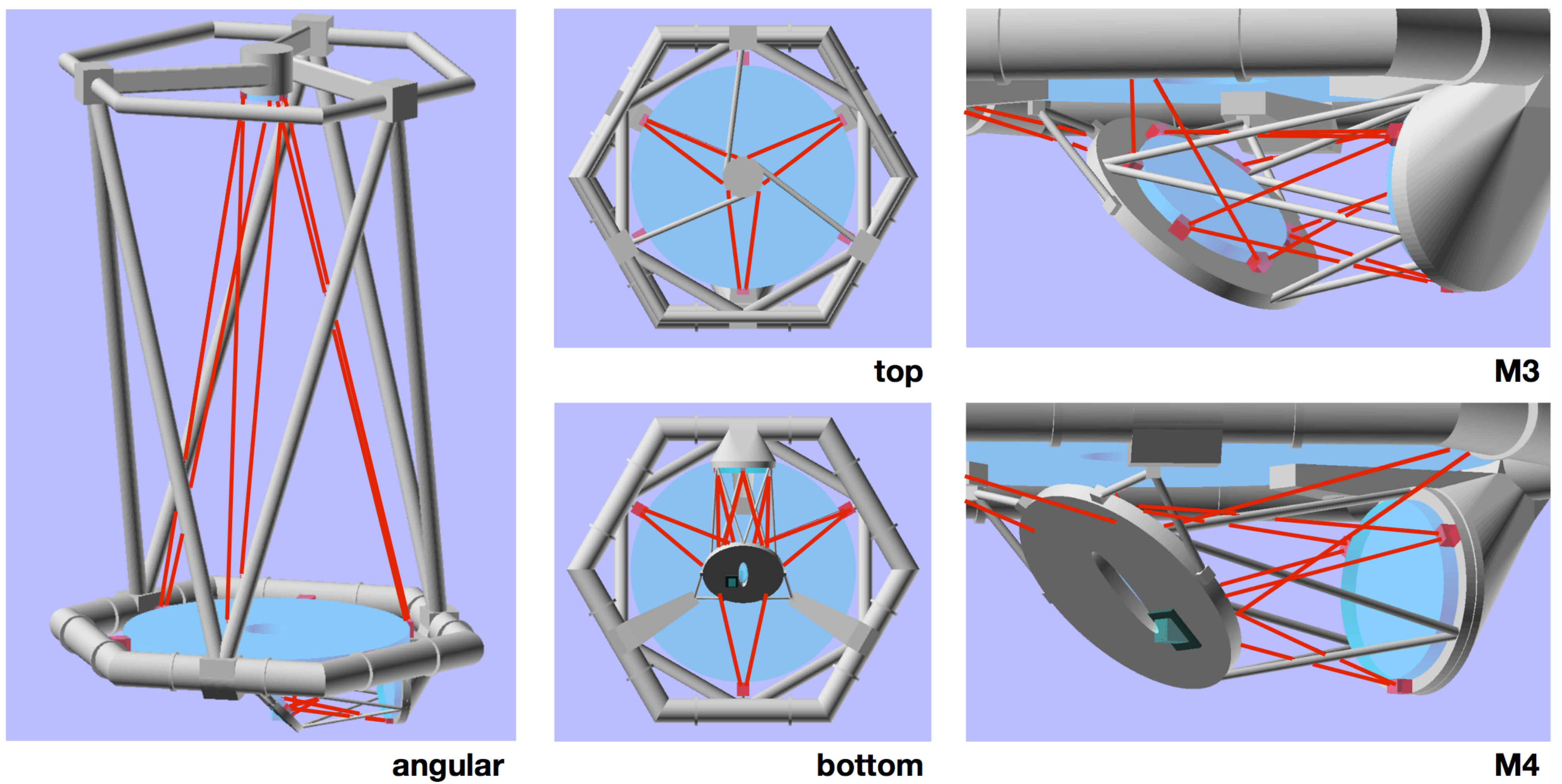}\\*[1em]
\vspace{-0.4cm}
\caption{Concept for a possible \theia/M5 Telescope metrology subsystem showing in red all the independent baselines forming laser hexapods. Retroreflectors are at M2 and M3. Microinterferometers at M1 and M4.}
\label{fig:theiam5-telmetconcept}
\end{figure*}

To monitor the distortions of the telescope geometry and to allow the associated systematic errors to be corrected up to sub-microarcsecond levels, \theia/M5 relies on a linear metrology concept. All pairs of mirrors of the telescope are virtually connected using a set of six independent interferometric baselines per mirror pair, each baseline measuring a distance variation. These baselines are organized into virtual hexapod structures (see Fig.~\ref{fig:theiam5-telmetconcept}). Accordingly, based on the independent distance determinations obtained by each baseline, the relative positions and angles between each pair of mirrors can be fully and rigidly determined, as any modification of the mirror-pair geometry impacts into all six interferometric baselines. This provides a small redundancy, considering that the mirrors are axis-symmetric in the ideal case.

The relative position variations of the mirrors must be known with a precision at the hundreds of picometers level to reach sub-microarcsecond level differential astrometric corrections in the whole \theia focal plane due to telescope geometry variations. Based on detailed analysis of Zemax simulations of perfect and thermally distorted SiC structure telescopes, to fulfill the most stringent \theia/M5 science case, and attain a soil of controlled systematics up to 125 nano-arcseconds over the FPA, the individual interferometer baselines between the pairs of mirrors must be capable of differential measurements of 50 picometers over the observation time (that can reach several minutes).

Existing space based interferometers from TNO, as the ESA/\gaia Basic Angle Monitor are already capable of reaching more precise measurements than those required by \theia/M5 -- BAM can perform $\sim1.5$ pm optical path difference measurements \citep{doi:10.1117/12.2026928}. A Thales telemeter developed for CNES can reach $\sim100$ pm, and the Thales interferometer produced for the MTG (Meteosat Third Generation) satellite can reach 1 nm per measurement \citep{Scheidel2011} -- higher precisions can be reached by averaging over many measurements. These already existing instruments shows that \theia/M5 requirement for the telescope metrology baselines can be fulfilled.

The proposed \theia telescope metrology subsystem has three main components: the laser source, the micro-interferometers and the associated electronics. 
The laser source can be derived from the MTG instrument 15~mW laser, with a power increase and improved long-term stability performances. Alternatively a TRL9 TESAT YAG laser like the LISA Pathfinder source can be adopted. This laser provides 45~mW and can be locked to a cavity to provide excellent mid-term stability. For multi-year stabilization it can be locked to an Iodine gas cell, providing absolute frequency precision to better than 5~kHz. \theia PLM will have redundancy in the laser sources, and they are estimated to weigh $\sim2$~kg each, or a total of 4~kg.

\begin{figure}[t]
  \centering
  \includegraphics[width=\hsize]{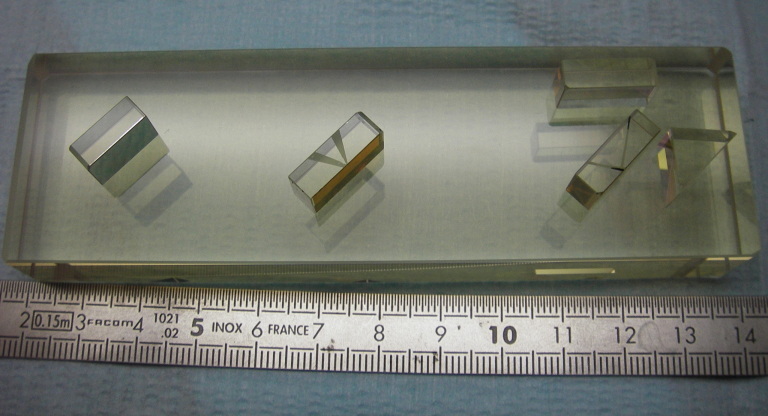} 
  \caption{A microinterferometer bench similar to the \theia/M5 concept (TAS).} 
  \label{tab:inst.fpa.telmet.microintexample}
\end{figure}

The laser sources will inject the beam into an optical switcher connected to mono-mode fibers that drive the beam to each micro-interferometer baseline.
Each micro-interferometer baseline consist in the micro-interferometer optical bench and an associated retro-reflector. The retro-reflector can be a classical corner cube produced from Zerodur \citep[e.g.][]{doi:10.1117/12.2055086}. The optical benches consist on a Zerodur bench and associated optics (e.g. Fig.~\ref{tab:inst.fpa.telmet.microintexample}). They fill a volume of $3.5\times3.5\times(8-10)$ cm each, with a first estimated mass of 200~g. Molecular adhesion can be used to fix the optics and it could also be used to fix the bench and the retroreflectors directly to the borders of the \theia/M5 telescope mirrors. The dimensions and materials of the design can be further optimized during Phase-A studies.

There are two options for the associated detection and readout electronics: it could be shared with the FPA electronics or be developed specifically for the telescope metrology subsystem. Although implementation could be more complex, the first option provides simpler qualification and procurement processes and results in higher homogeneity and stability. Thus it is considered as the baseline. Additional FPGA based electronics (e.g. Actel RTG4) for the on-board processing of the metrology signal would fit into a mechanical box with $30\times22\times3.5$\,cm, and has a mass estimate of $\sim1$ kg. The electronics can be shared between the baselines.

\begin{figure*}[h]
  \centering
  \includegraphics[width=0.8\hsize]{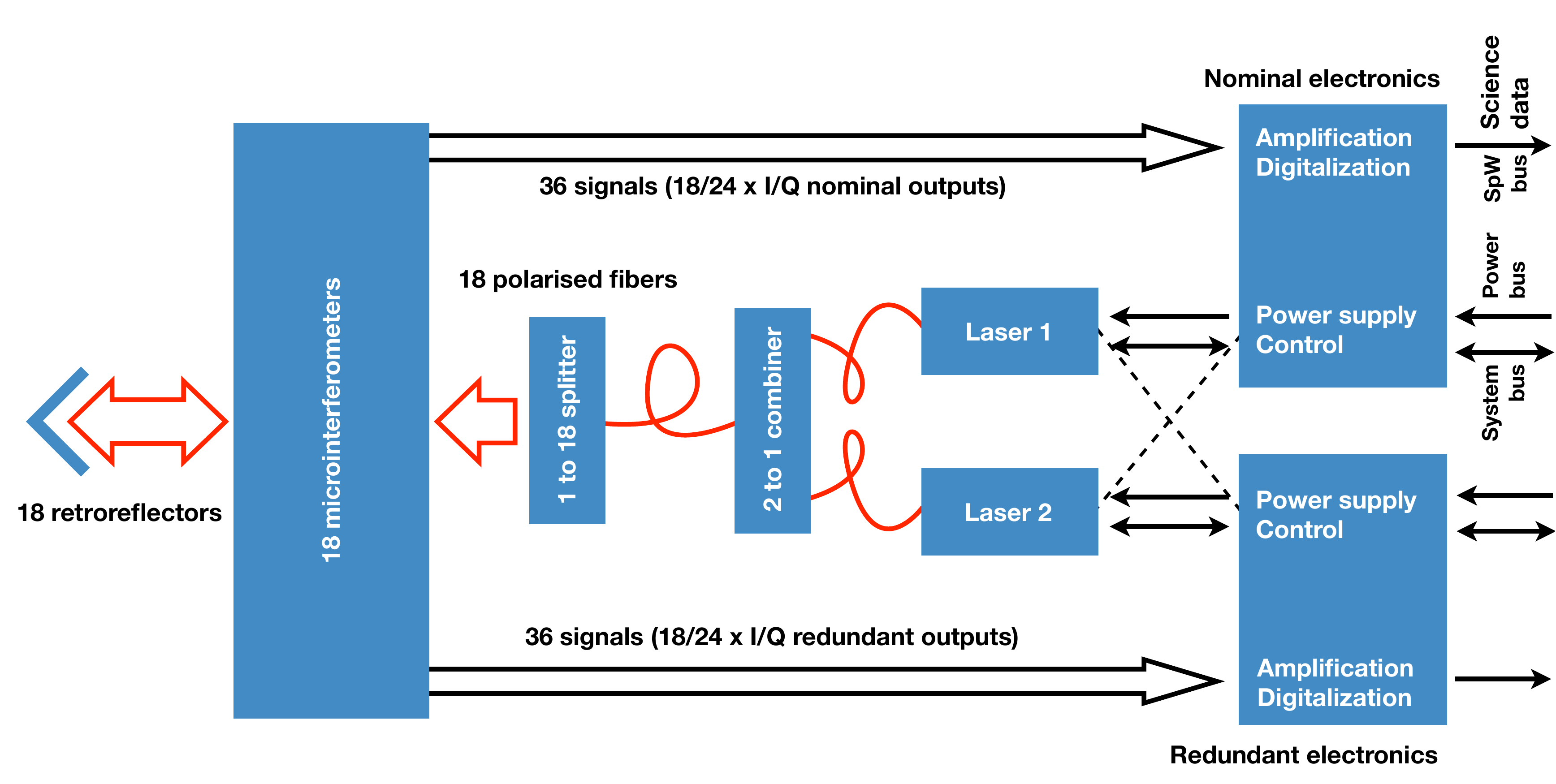} 
  \caption{Conceptual block diagram of the telescope metrology subsystem.} 
  \label{tab:inst.fpa.telmet.diagram}
\end{figure*}

The telescope metrology subsystem can be composed by 18 or 24 individual baselines. If only the telescope geometry is monitored, which is the minimum requirement for \theia/M5, three laser hexapods, each with six baselines, are required (see Fig.~\ref{fig:theiam5-telmetconcept}). Optionally the telescope to FPA geometry can also be monitored with an additional hexapod. A conceptual block diagram of this subsystem can be seen in Fig.~\ref{tab:inst.fpa.telmet.diagram}, and a  possible implementation concept in Fig.~\ref{fig:theiam5-telmetconcept}. Further optimization of this concept will take place during Phase 0/A studies. A small reduction of the number of baselines can be foreseen, for instance considering assurances of the TAS active thermal control and on detailed thermo-mechanical analysis of the spacecraft and payload structures.

An estimate of the power required to run all the telescope metrology subsystem lasers and electronics is $58.7$\,W ($\sim70.5$\,W including 20\% margins). The total mass reaches  $12.4$\,kg ($\sim14.9$\,kg with 20\% margin). 

\subsection{Performance assessment \& error budget}
\label{sec:perf-assessm-error}
\begin{figure*}[t]
  \centering
  \includegraphics[width=\hsize]{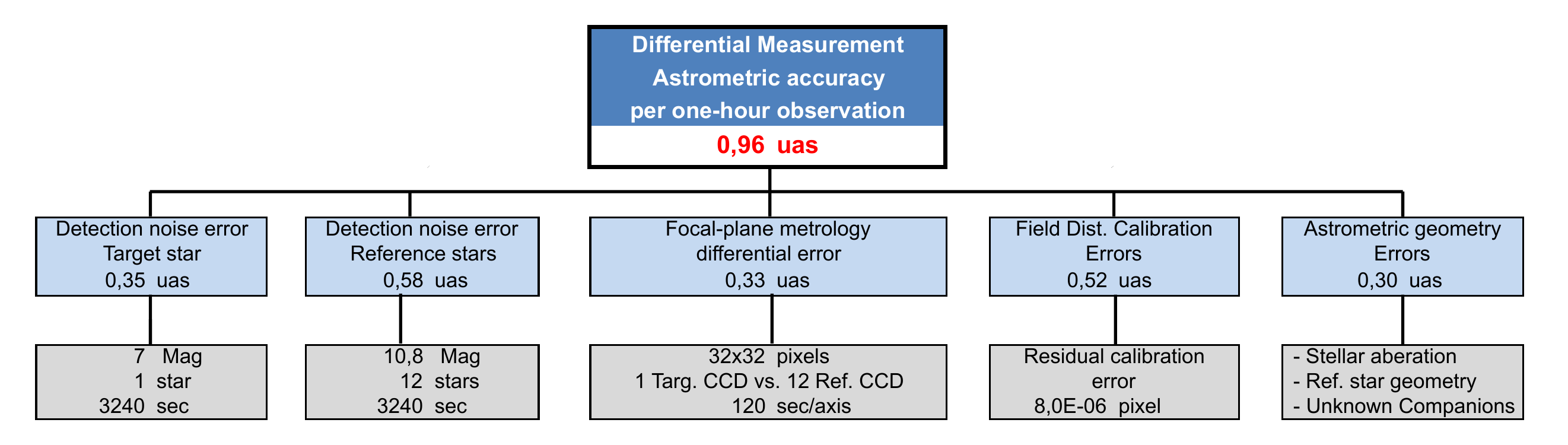}
 \caption{Top-level error budget for \theia.} 
  \label{fig:error-budget}
\end{figure*}

A detailed error budget was developed for this mission concept. The major errors terms are captured in the top level version shown in Fig.~\ref{fig:error-budget}.

The biggest term is the brightness dependent error for the set of
reference stars. There are thousands stars in the FOV. Assuming we
limit to no more than one star per detector, there are 24 potential
reference stars. However the error budget was created considering 12 stars
as a compromise: more stars improve astrometry and reduce systematic
errors but fainter stars have much lower SNR. 
After 1\,s of integration, $8\times10^5$ photoelectrons are detected
for each of the 10.8-mag reference stars. Since all the stars are
measured simultaneously, the stars do not need to be kept centered on
the detector at the sub-mas level, but only to a fraction of the PSF
width to avoid spreading of the photon outside of the PSF and
therefore cause the PSF effective width to be larger. A tenth of pixel
(1\,\microns) stability over the one-second frame integration is sufficient.
After 3240\,s of the observation integration, the statistical averaged position of the
barycenter of the set of reference stars (e.g.\ 12 stars with
$R\sim10.8$\,mag) will be measured with a residual 0.091\,nm
(0.58\,\uas) uncertainty.  Similarly, the position of the target star
($R\sim7$\,mag) will be measured with a residual 0.055\,nm
(0.35\,\uas) uncertainty. Although in the end of the whole observation the spacecraft might have, in the worst case, 
moved by arc-seconds depending on the AOCS, the differential position 
between the target star and the barycenter of the set of reference stars 
will be determined to 0.58\,\uas.

Similarly, the focal plane metrology system will have determined the
differential motion of the target detector relative to the barycenter of
the set of reference detectors with an error smaller than 0.33\,\uas after
$60\times 1$\,s metrology measurements in three directions. And the WFS and the telescope metrology 
will enable a field distortion calibration of at least 0.52\,\uas during 
the integration time. Finally, additional geometrical errors due to uncorrected
astrophysical effects as unknown terms of differential aberration, reference
star geometry (e.g. the existence of circumstellar discs), 
unknown multiple companions, etc., are expected to contribute with 0.3\,\uas.

\subsection{TRL assessment}
\label{plm.trl.assessment}
As shown in previous sections \theia relies on systems that are built on top of technologies with large heritage. This is made explicit in Tab.\,\ref{tab:technology-readiness}. The key \theia PLM profits from a series of developments performed for past missions, but phase-A activities will be necessary to raise the TRL of the FPA metrology system and specially of the electronics, and the identification or space qualification of optical fibers with NA$\sim4$ will be necessary. Phase-A activities will also be required to raise the TRL of the picometer microinterferometers (mainly for size reduction) and to breadboard the hexapod as an integrated system. Phase-A activities will be necessary for straylight assessment.

\begin{table}[t]
  \centering
  \caption{Payload TRL assessment.}
  \smallskip
  \includegraphics[width=\hsize]{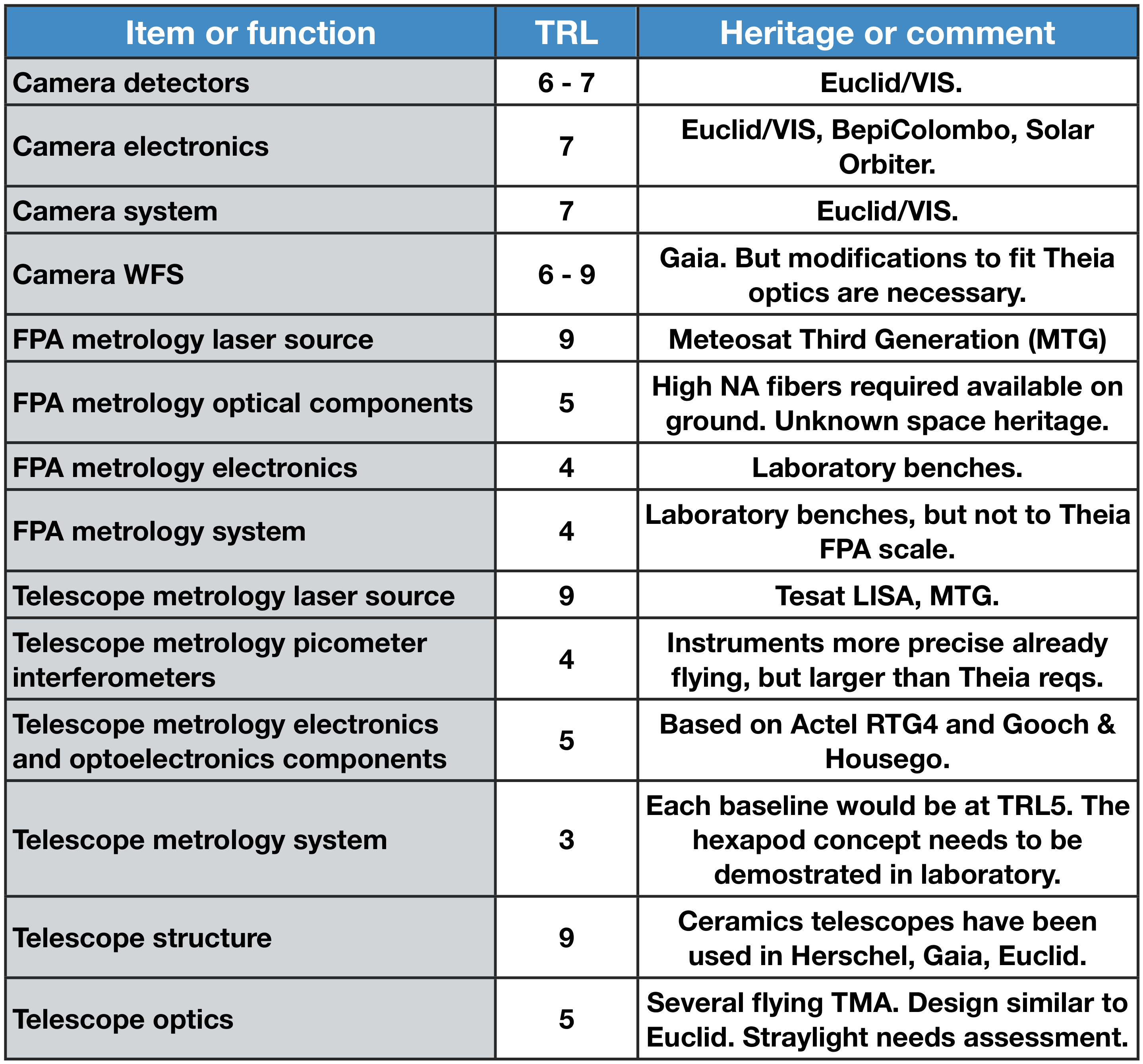}
  \label{tab:technology-readiness}
\end{table}


\subsection{Astrometric and photometric calibration}
\label{sec:calibration}
\subsubsection{Introduction}
At the heart of the \theia mission lies a careful and accurate calibration of the observations, and the use of  optimized methods to extract as much science out of the data as possible. In order to achieve this goal, our team consists of experts from various astrometric and photometric (proposed) missions like \gaia, \hst, \textit{SIM}, and \NEATx. The work is formally split up in several development units which are described below following processing order when applicable.

\subsubsection{Processing of raw satellite data }
This unit will take care of receiving, decompressing and combining the raw measurement and ancillary data packets from the focal plane and instrumentation in general into self-contained data records, so that each of it can be easily processed afterwards. A record of all on-board events (e.g. anomalous events) and telemetry (e.g. temperatures) is constructed.
Raw on-board attitude (pointing) from the satellite will also be processed here, although without any refinement yet, in order to tag the output records with not only their measurement time but also with their associated sky coordinates.  The satellite position and velocity, as determined from ground-tracking data will be compiled for downstream processing. 

\subsubsection{Processing of the raw frames}
The raw outputs from the previous stage include tags with the spacecraft pointing, position, velocity, temperature and time of observation data.  Raw frames will be processed, which include correction for known detector systematics: nonlinearity, charge transfer ineficiency, in case of CCDs or nonlinearity, inter-pixel capacitance (IPC) which can include small anisotropies, afterimage (persistence) and reciprocity failure (flux-dependent nonlinearity) in case of CMOSes. Correlated read-noise would then be corrected for in the next step. The refined centroid and flux of the stars will be determined by iteratively fitting a model of the point spread function to the local images of each star. 
Initial broad-band photometry of the stars will also be determined during this process, although it will be refined (see Sect.~\ref{sec:photCalibration}) after metrological pixel response calibration.

\subsubsection{Design of the Astrometric Solution}
After all calibrations have converged, relative astrometric parameters can be determined with the precision presented in the next section (\ref{sec:astrometricPrecision}).

Since \theia is aimed at the sub-microarcsecond accuracy for its differential observations over a field of about 0.5 degrees, a serious fully-relativistic model is indispensable. The fact that the accuracy is higher than for \gaia while the nature of the observations is differential largely compensate each other so that the relativistic model for \theia can have the same physical content as the model used for \gaia\/ \citep{2003AJ....125.1580K,2004PhRvD..69l4001K}. The latter has an accuracy of 0.1 microarcseconds for global astrometry. Nevertheless, the details of the \theia model and its optimal formulation still has to be investigated. The model of \theia should be based on the system of the hierarchy of the relativistic reference systems -- BCRS, GCRS, etc. -- recommended by the IAU and used for high-accuracy relativistic modeling over last 20 years \citep{2003AJ....126.2687S}. The main components of the models are as follows.

First, \theia needs a relativistic treatment of its barycentric orbit as well as its orientation. To calculate the differential aberration in the \theiap FoV, its velocity should be known with an accuracy of about 20 mm/s --- a rather relaxed requirement. To account for the parallax (or planetary aberration) when observing Near-Earth Objects, it is desirable to know its position with an accuracy of a few hundred meters. This should be better quantified at a later stage of the mission. 

Second, the propagation of light from the source to the satellite has to be accurately modeled. This includes effects due to several types of gravitation field - monopole, quadrupole and multipolar, gravitomagnetic due to translational and rotational motion of the gravitating bodies. The differential nature of observations relaxes the accuracy requirements in some cases, but in other cases (e.g. observations close to Jupiter or other giant planets) the effects should be computed with an accuracy that meets the observational accuracy. The physical content of these effects is well-known \citep[e.g.][]{2010CQGra..27g5015K,2012CQGra..29x5010T,2014PhRvD..89f4045H,2011rcms.book.....K}. Detailed optimal formulation still should be found.

Finally, \theia requires a high-accuracy definition of the motion of the observed sources with respect to the barycentre of the solar system. Here one can expect that most of the sources can be described either by the dynamical equations
of motion (e.g. for solar system objects or for components of non-single stellar systems in their motion relative to the common barycentre) or by the model assuming that the source moves with a constant velocity (the model used both for \hips and \gaia for single stars). In some special cases the latter model may require an update taking into account non-linearity of the relative motion as well as light-travel effects. Again, the physical content of such an update is well known and understood \cite{2011rcms.book.....K}.

The Optical Field Angle Calibration (OFAD) can only be calibrated in space, with observations of  a well known star field, with multiple observations at different pointings over a short period of time.   The observations are put into a simultaneous solution of the chosen mathematical equation for the mirror form and the terms representing the metrology components of the mirror. 

Thermo-mechanical deformations, due to the variation of the Solar Aspect Angle (SAA) in different repointings of the instrument and the non-zero Coefficient of Thermal Expansion (CTE) of the material chosen for the telescope and Focal Plane Array structures, will produce relative astrometric shifts along the lifetime of the mission. To monitor these deformations the Science Payload includes three different subsystems: wavefront sensors, a focal plane metrology subsystem and a linear metrology subsystem.  Additionally, continued monitoring of the calibration field throughout the mission allows for redundant monitoring of temporal changes in the mirror and instruments that would require further calibration to maintain precision and stability for astrometry.

The Shack-Hartmann wavefront sensors are responsible for a continuous monitoring of the focus of the telescope and for a continuous monitoring of the quality of the wavefront arriving at $\theia$ detectors due to deformations of the telescope mirrors from thermal variations.    

A $\gaia$-like WFS solution is able to sense $\lambda/1000$ variations of the wavefront, and $\theia$ FPA concept includes four of such WFS at each corner of the FoV to allow sub-$\mu$as monitoring.

The Focal Plane Array Metrology subsystem  includes 
an interferometer. It is responsible for  frequent monitoring and calibration of the deformations of the positions of each pixel of the detectors of the FPA at a level of $<10^{-5}$ \citep{2016arXiv160902477C}. This subsystem 
measures the geometrical parameters of the FPA with respect to M3, of each detector with respect to the center of the FPA and of each pixel with respect to the center of the detector.

The Telescope Metrology subsystem 
contains 18 linear interferometers that continuously measure the relative positions and angles of the mirrors of the telescope. Between each pair of mirrors, a  "Virtual Laser Hexapod" is created using retro-reflectors, monitoring  any thermo-mechanical effects. 

Detailed Zemax simulations of the $\theia$ telescope deformed under a set of worse case scenarios (dT = 100 mK, for different SAAs) and analysis of the astrometric displacements caused by such thermo-mechanical deformations show that the OFAD can be modelled in terms of an 8th order Chebyshev polynomial.

The existence of this low order expansion shows that it is possible to calibrate at sub-micro arcsecond level the astrometric displacements in the FPA caused by variations of the telescope structure by using the Telescope Metrology subsystem measurements at the $50 pm$ level.

After the OFAD calibration is completed, additional calibrations are made for color response using stars of different colors in the same field across the field of view of the instrument. 

\subsubsection{Precision of the Astrometric Solution}\label{sec:astrometricPrecision}
\begin{figure}[t]
\includegraphics[width=0.99\textwidth]{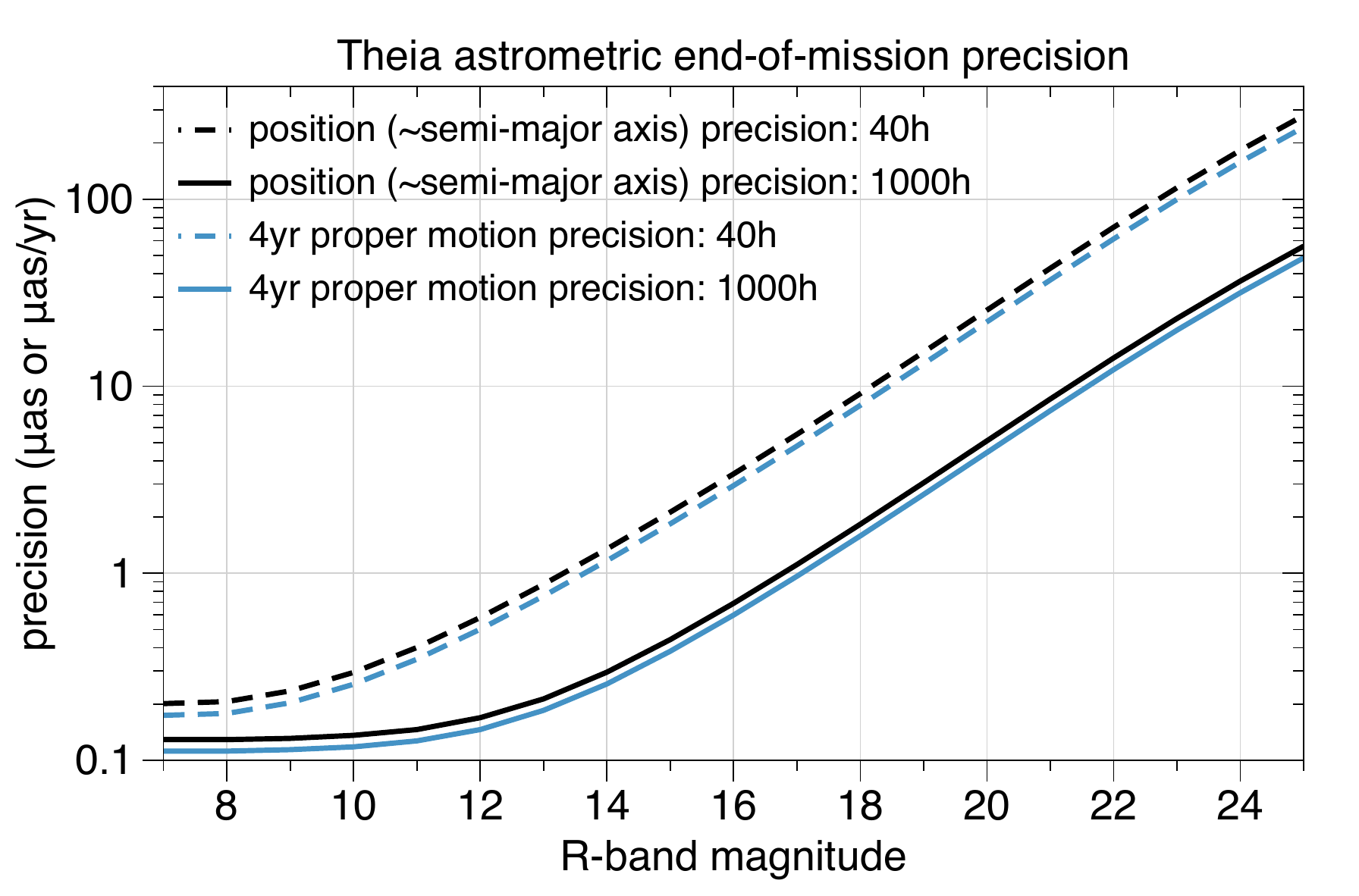}
\caption{End of 4-year mission astrometric differential measurement uncertainty for the detection of exoplanets due to stellar wobble (orbital semi-major axis), as well as proper motion measurements (for dark matter science). See  Tab.~\ref{tab:astroPrecision} for more details. 
}
\label{fig:astroPrecision}
\end{figure}

For each hour of observation the final relative positions precision (RMS) is estimated to be 1, 14, 94, 1100~\uas at R-band magnitudes 7, 15, 19, 24 respectively. The resulting astrometric parameters for a 40h and 1000h science case during a 4~year mission is shown in Table~\ref{tab:astroPrecision} and Figs.~\ref{fig:astroPrecision} and ~\ref{fig:astroComparions}. A position calibration noise floor of 0.125~\uas has been used (see Sect.~\ref{sec:telescopemetrologysection}).

\begin{table}[t]
\caption{End of 4~yr mission astrometric differential measurement uncertainty as a function of target star magnitude for a 40~h and 1000~h science case, see also Fig.~\ref{fig:astroPrecision}. The parallax precision ($\sigma_\pi$) is given for ecliptic latitude 0, being equal to the position precision. The position calibration floor is estimated at $0.125$~\uas. The parallax precision improves up to a factor of $\sqrt{2}$ at the ecliptic poles.
The proper motion precision ($\sigma_\mu$) is similar to position precision for a 4 year mission, but would improve linearly with (extended) mission time. }
  \label{tab:astroPrecision} \centering
  \begin{tabular}{l@{\hspace{0.8em}}l@{\hspace{0.8em}}l@{\hspace{0.8em}}l@{\hspace{0.8em}}l@{\hspace{0.8em}}l@{\hspace{0.8em}}l@{\hspace{0.8em}}l} \hline R (mag) & 10 & 15 & 18 & 20 & 22 & 24 & 25\\
  \hline
  \multicolumn{8}{l}{40h science case}\\
  $\sigma_\mu$ (\uas/yr) & 0.26 & 1.8 & 7.9 & 22 & 61 & 158 & 243 \\ 
  $\sigma_\pi$ (\uas)    & 0.30 & 2.1 & 9.1 & 26 & 71 & 183 & 281 \\
  \hline
  \multicolumn{8}{l}{1000h science case}\\ 
  $\sigma_\mu$ (\uas/yr) & 0.12 & 0.38 & 1.6 & 4.4 & 12 & 32 & 49 \\ 
  $\sigma_\pi$ (\uas)    & 0.14 & 0.44 & 1.8 & 5.1 & 14 & 37 & 56 \\
  \hline
  \end{tabular}
\end{table}

\subsubsection{Absolute astrometry \& reference frame}\label{sec:calib_priorsAbsAstrometry}

$\theia$ is a small-field relative astrometry mission, meaning that the derived astrometric parameters for the target stars in a field will have position, parallax and proper motion relative to a local reference frame tied to a global one. At the time of the $\theia$ mission, the most accurate and complete optical reference frame will be that of the $\gaia$ catalog\footnote{Note that for the hyper velocity stars science case the current SDSS quasars will already suffice to perform measurements (see Sect.~\ref{triaxial}).}. Typically hundreds to thousands of $\gaia$ sources will be visible in a single $\theia$ frame. By using $\gaia$ global astrometry parameters as priors, the astrometric solution of all the stars observed by $\theia$ will be automatically tied to the $\gaia$ frame, without the need of forcing physical priors on sources such as quasars or remote giant stars. Let us note that since most science cases require parallaxes and proper motions, it is possible to construct a \textit{kinematic} reference frame for $\theia$ at almost $\gaia$ accuracy.
While $\gaia$ positions will degrade linearly over time, the 
accuracy of proper motions and parallaxes will remain almost the same as for the $\gaia$ epoch\footnote{A \textit{positional} reference frame derived with \gaia stars will be degraded by a factor of $5-6$ at the \theia epoch, while proper motions and parallaxes themselves degrade to a much lesser degree, See Sect.~1.5.4 and 1.5.5 of the Hipparcos and Tycho catalogues \citep{1997A&A...323L..49P}.}.

Zonal systematic errors in the $\gaia$ absolute reference frame will set the uncertainty floor on each astrometric field. In what follows, we discuss the possible impact of these correlations into \theia's absolute proper motion and parallax measurements.

For $\gaia$, the correlations of astrometric parameters resulting from the determination of attitude parameters have been studied before launch  in detail \citep[see][]{2012A&A...543A..14H, 2012A&A...543A..15H} and are expected to induce correlations at angular separations $<0.7$\,deg, which could be a serious concern for $\theia$ as this corresponds to its proposed FOV. However, these correlations were estimated to be (much) below $r=0.5$\% \citep{2009IAU...261.1703H}. Bright stars ($V<13$) and
low star-density regions will have the highest correlations. Table \ref{tab:correlations} gives a general overview of the limit to which one can average $\gaia$ parallax and proper motions for various possible correlation coefficients (precise values of correlations will not be known until a few years into the $\gaia$ mission), and hence the ultimate accuracy by which parallax and proper motion measurement of $\theia$ can be expressed in the $\gaia$ reference frame. The limiting accuracy for the correlation coefficients between $5\times 10^{-3}$ and $5\times 10^{-5}$ is plotted as function of magnitude in Fig.~\ref{fig:astroComparions}. In principle all $\gaia$ reference stars observed in a $\theia$ fields can be used to fix the reference frame, which will be of varying magnitude and number, therefore the reachable absolute astrometry parameter accuracy of $\theia$ will differ from field to field.

\begin{table}
\begin{center} \small
\begin{tabular}{cccc}
 $V$  &  $\gaia$ $\sigma_{\pi}$  &\multicolumn{2}{c}{$\sigma^{\rm mean}_{\pi} \lim N\to \infty$} \\
\cline{3-4}
[mag]             &  for one star               & $r=5\times 10^{-3}$ & $r=5\times 10^{-5}$   \\
\hline
15    & 24\, $\uas$          & 1.7\,$\uas$       &  0.17\,$\uas$       \\
20    & 540\, $\uas$            & 38\,$\uas$       &  3.8\,$\uas$       \\
\hline
\\
$V$ &  $\gaia$ $\sigma_{\mu}$  &\multicolumn{2}{c}{$\sigma^{\rm mean}_{\mu} \lim N\to\infty$} \\
\cline{3-4}
[mag]          &  for one star               & $r=5 \times 10^{-3}$ & $r=5\times 10^{-5}$   \\
\hline
15    & 13\,$\uas$/yr      &0.92\,$\uas$/yr &  0.092\,$\uas$/yr \\
20    & 284\,$\uas$/yr    &20\,$\uas$/yr &  2.0\,$\uas$/yr \\
\hline
\multicolumn{4}{l}{\tiny $\gaia$ accuracies from \url{http://www.cosmos.esa.int/web/gaia/science-performance}}
\end{tabular}
\caption{
Limiting precision when averaging parallax and proper motion for two possible values 
of the correlations. The limit is effectively reached when averaging over $O(r^{-1}) $ stars. The corresponding positions accuracies at the $\theia$ epoch are about 5~times worse than tabulated for $\sigma_\pi$.}
\label{tab:correlations}
\end{center}
\end{table}

In pre-launch studies the basic-angle metrology specification of 0.5~\uas over 5 min, the global parallax
zero-point has been estimated to be around 0.1~\uas. With regard to the proper
motions, $\gaia$ itself will be tied to the radio ICRF with an expected precision
in proper motions of 0.2-0.3~\uas/yr \citep{2012A&A...547A..59M,
2012MmSAI..83..918M}. 
The \textit{Tycho-\gaia} solution of $\gaiap$ DR1 release contains systematic errors that are not representative for the final ($\gaia$ only!) solution due to incomplete calibration models, short $\gaia$ data span, and mixing with \textit{Tycho} data. Therefore it can currently not be reliably assessed if the calibration errors of the final $\gaia$ data release (that will be available at the $\theia$ launch) will cause any departure from the estimated pre-launch systematic errors.

In summary, anchoring $\theia$ observations to the $\gaia$ reference frame is a critical aspect to enable global astrometric capabilities of a differential astrometry mission. 

Since specific targets and general astrophysics
programs (e.g. very distant halo stars, objects in nearby galaxies)  might need to push global astrometric precision to the limit, future mission development plans will have dedicated work-packages and a team specifically devoted to this
task. Similarly, optimization of the pointing to capture good reference stars
and/or distant sources will be done for all science cases and observations at a
later stage.

\subsubsection{Derivation of photometric calibrations}\label{sec:photCalibration}
Astrometry and photometry are closely linked. In astrometry, the most accurate way to determine the centroid of a source involves the fitting of a PSF to the image data. Photometry can also be derived using such PSF fitting. The same error sources in astrometry also affect photometric accuracy.  Detailed calibration of the detector enable both more accurate astrometry and more accurate photometry.  DICE experiment \citep{2016arXiv160800360C} concludes that in order to reach an error below $5\times 10^{-6}$ pixel on the centroid, the pixel response non uniformity (or PRNU) must be know to better than $1.2\times 10^{-5}$. Such a precision can only be obtained if a precise photometric calibration is used.

\begin{figure}[thb] 

\includegraphics[width=0.99\textwidth]{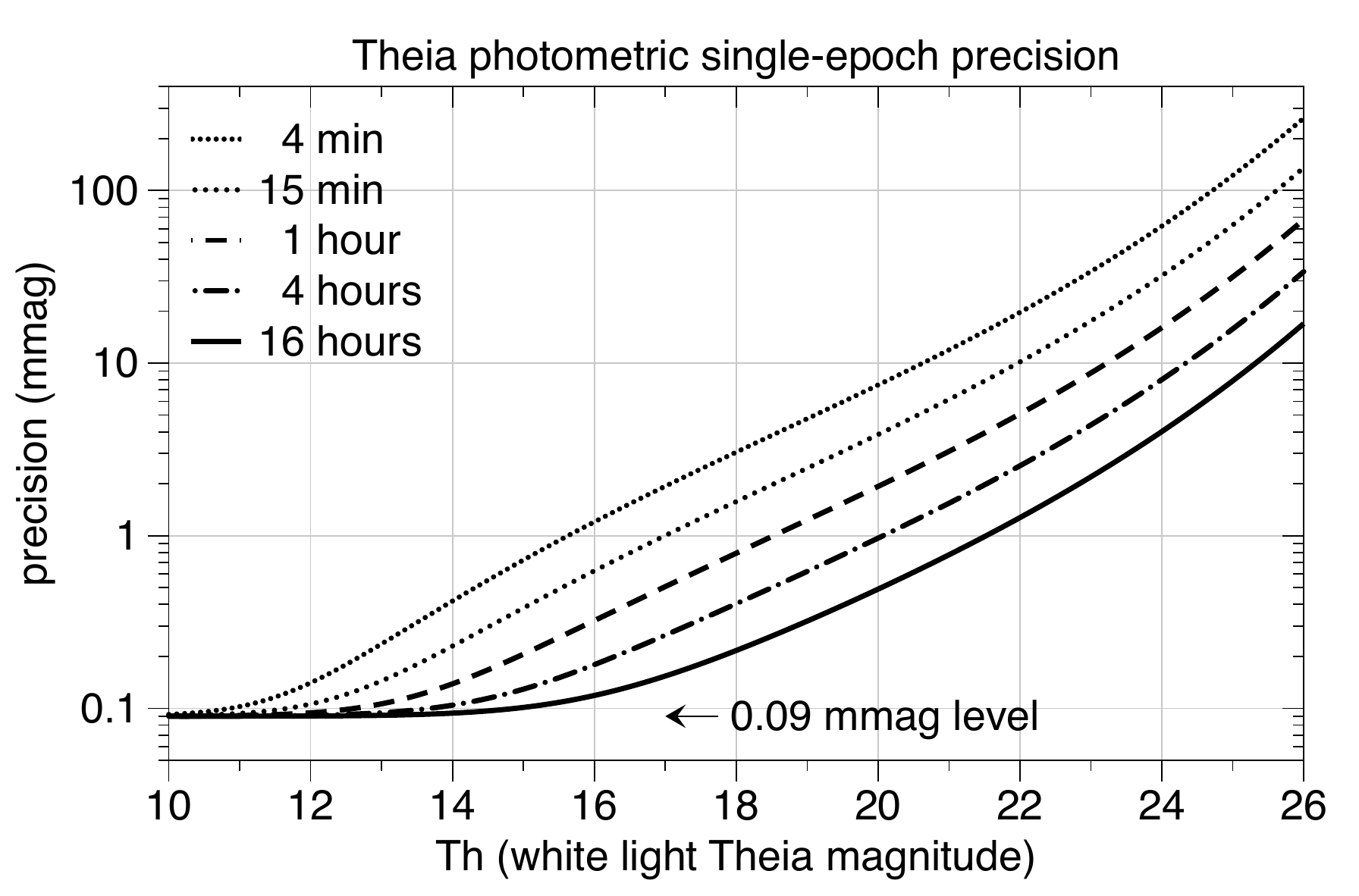}
\caption{Estimated single-epoch photometric performance in the white light \theia band (Th). A systematic noise floor of 0.09~mmag has been assumed following \cite{2016arXiv160800360C}. The $V$ and $R$ magnitudes are +0.0145 and +0.0120 mag above the \theia magnitude for $R-I=0$.}
\label{fig:photPrecision}
\end{figure}

The instrumental effects to consider in the photometric calibration are bias, dark, flat-fielding (PRNU), fringing, background, saturation and non-linearity, contamination, pixel position and characterization of the read-out-noise. All these effects are calibrated through specially designed calibration images from pre-launch measurements and during mission operation from onboard interferometric fringe measurements and the use of the shutter.

The two common methods to derive photometry from CCD images are aperture photometry and PSF fitting. PSF fitting strategy is preferred for \theia, as it is used anyway for astrometry processing, as stated above. Estimates of the relative photometry precision are shown in Fig.~\ref{fig:photPrecision}.

Differential photometry approach can also be used for \theia instead of (or together with) absolute photometry. Constant sources in the field are used to settle the relative photometry of the sources of interest. Finally, in order to standardize the instrumental photometry a set of calibration sources with standard photometry in the field of view are needed. \gaia mission provides standard homogeneous good quality photometry of constant sources in all the sky.

\subsubsection{Software engineering  (Quality assurance, integration, framework, interfaces)}

The proper management of the development of the scientific processing software is crucial to ensure the timely production and quality of the \theia products. This management has to rely on professional software (SW) development procedures and standards and has to ensure an adequate coordination of the geographically disperse contributions to the effort. For this we propose the creation of a "core development" team that will have the responsibility of:

\begin{enumerate}
\item Acting as coordinator of the SW development, defining and enforcing development standards in the \theia consortium. Specifically, the core team will be responsible for the QA of the development, including interface definition, version control, enforcement of ECSS standards.
\item Receiving the contributions from the different sub tasks and integrating them into a common framework. We will maintain a common code repository for collection and control of the code and versioning.
\item Enforcing the definition and implementation of unit tests for the SW. We will set up a continuous integration environment where the unit tests will be run at least daily to ensure repository code integrity and validity.
\item Defining, implementing and running integration tests for the SW. At each SW release the full processing chain will be tested end-to-end to ensure its correctness and integrity. These tests will be defined to cover both technical and scientific validation of the code.
Executing the pipeline. The code will be deployed in the processing system to produce the scientific data from the \theia telemetry. The core team will take responsibility for code deployment, execution control, management and delivery of science-ready products. 
\item The core team will be responsible for monitoring the performance of the instruments by producing, in addition to the final relative corrected position, files with information about the general performance of every data set --- generating warnings when there are failures or the instrument has gone through significant changes.
\end{enumerate}


\section{Mission configuration and profile}
\label{sec:mission}

\subsection{Proposed mission profile}
\label{sec:prop-miss-prof}

\begin{table*}[t]
  \centering
  \caption{Mission main characteristics.}
  \smallskip
  \includegraphics[width=0.7\hsize]{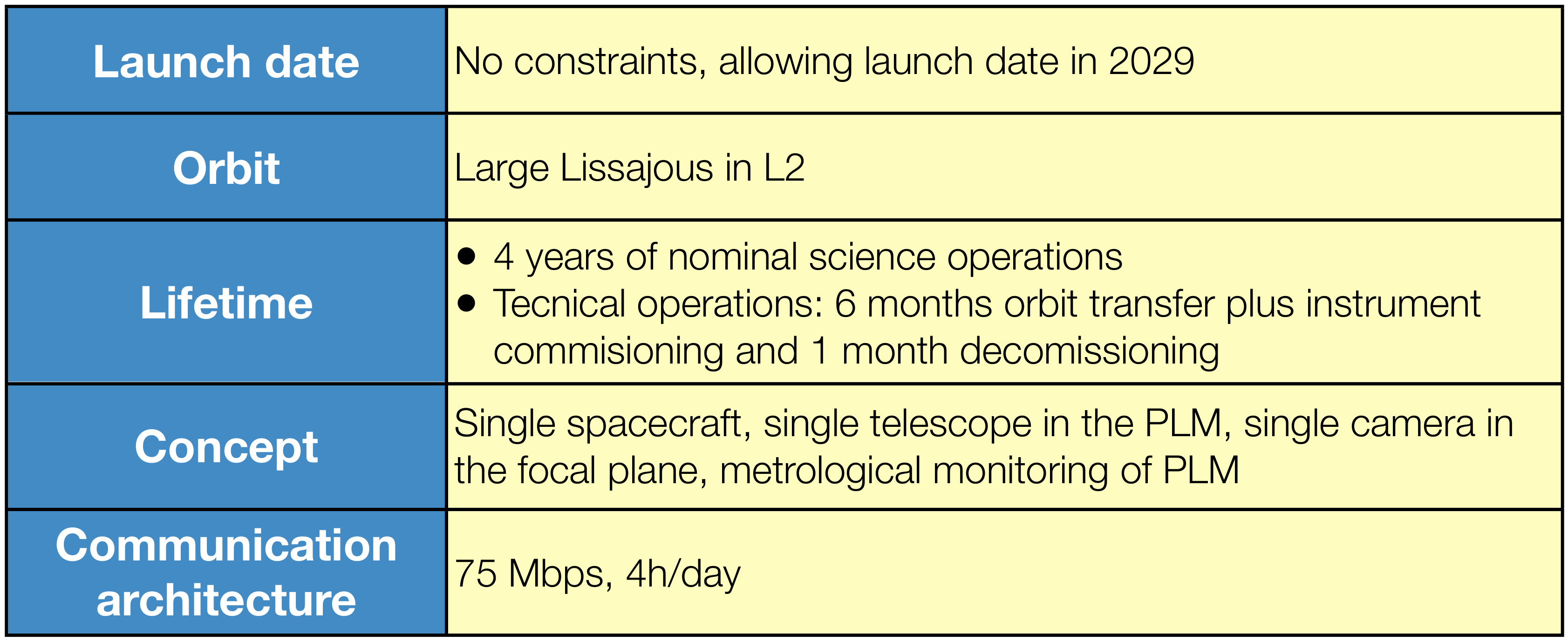}
  \bigskip
  \label{tab:mission-main-characteristics}
\end{table*}

\begin{figure*}[t]
  \centering
  \includegraphics[width=0.75\hsize]{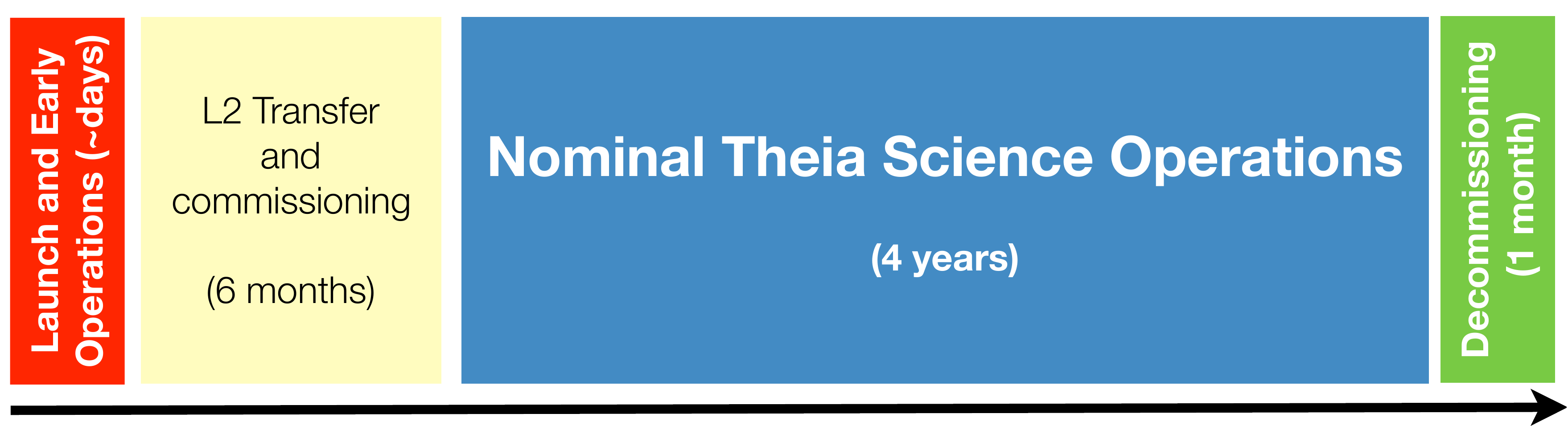}
  \caption{Reference mission timeline.}
  \label{fig:mission.timeline}
\end{figure*}

\subsubsection{Mission Orbit}

\theia is an astrometry mission that needs to point to different directions of the sky. L2 is the selected option for the orbit, since it is very favorable for overall stability because of the absence of gravity gradients, the time available for observation, the environmental conditions characterized by low total ionizing doses. In addition, the thermal conditions over the orbit remains the same, thereby simplifying thermo-elastic design issues. The \theia spacecraft will be directly injected into a large Lissajous or Halo orbit at L2.

To avoid parasitic light from the Sun onto the telescope and the detector, \theia spacecraft have baffles that protect them from Sun light at angle larger than $\pm45^\circ$ in the Sun direction.

\subsubsection{Launcher}
Considering the M5 call programmatics, the targeted orbit and the \theia mass range, the baseline launcher is Ariane 6.2. The launch strategy would consist in a unique burn of upper stage injecting directly the S/C onto a L2-transfer trajectory, avoiding a coasting on a parking orbit. The separation of the satellite would occur after about 1/4 of rotation around the Earth, preventing the Sun illumination to enter into the instrument during ascent and up to the separation.

After the completion of its last burn, the Ariane upper stage would re-orient the S/C into a 3-axes stabilized attitude, with the Sun in the Sunshield normal direction. As for {\textit Herschel} satellite, \theia spacecraft would require a reactive \textit{Safe Mode} control to maintain roughly the Sun in this direction and ensure the instrument protection.

The use of Ariane 6.2 enables a large volume for the spacecraft and its payload. The Ariane fairing allows a maximum diameter of 4500 mm, which conditions: (i) The maximum primary mirror diameter; (ii) the maximum Service module size and thus its internal accommodation capacity,
(iii) the maximum Sun shield and V-groove screens size, linked also with the telescope size and required sky accessibility requirements.

Although not currently specified, the Ariane 6.2 launcher will allow a maximal spacecraft wet mass well in excess of the 2145 kg allowed on Soyuz for a direct transfer to the L2 point.
It shall be noticed that considering Ariane 6.2 performance and the current \theia mass budget, a dual launch is an attractive option that would optimize \theia launch costs if a co-passenger can be identified during the assessment.

\subsubsection{Mission lifetime and timeline} 

\begin{figure*}[t]
  \centering
  \includegraphics[width=1\hsize]{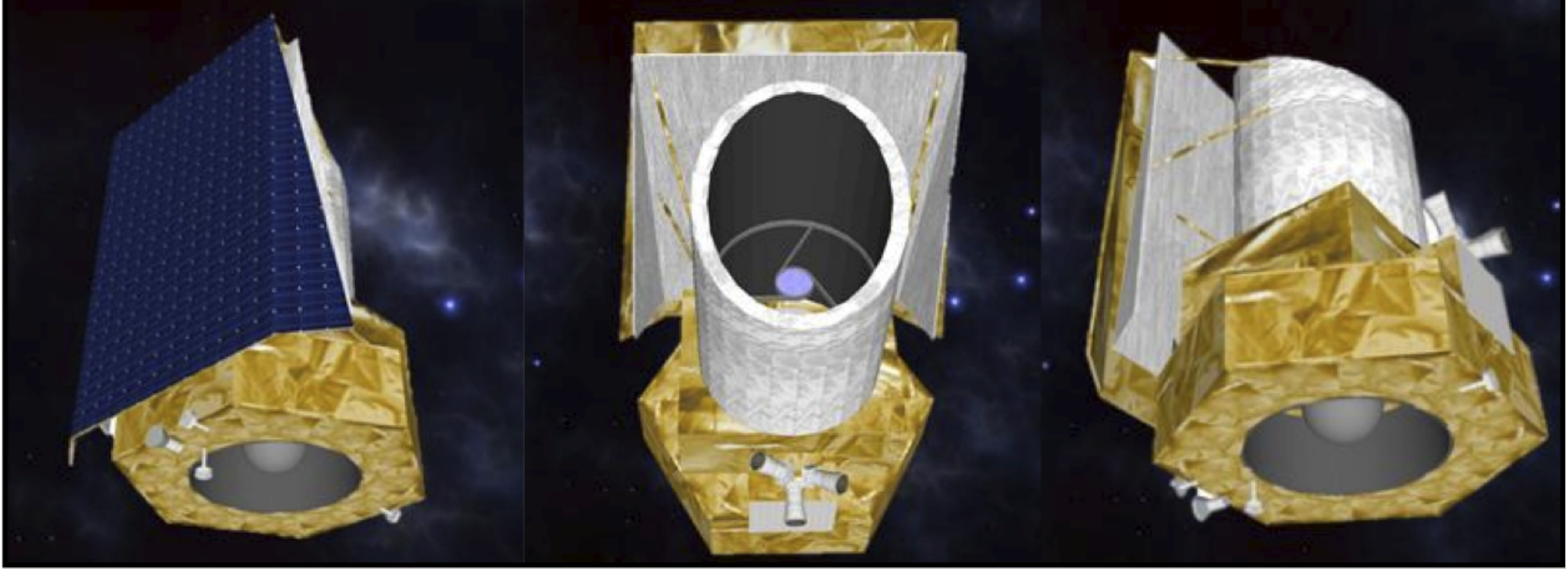}
  \caption{Preliminary \theia satellite rendering (Thales Alenia Space).}
  \label{fig:mission.sc.concept}
\end{figure*}

The time baseline to properly investigate the science program of \theia is 4 years including some time devoted to orbit maintenance. A total of approximately 6 months has been estimated for the orbit transfer including the spacecraft and instrument commissioning. From the total of $\sim 35000$ h dedicated for the scientific program, about 15 min per slew will be dedicated to reconfiguration and station-keeping. The thermal stabilization time is in addition to the slew time.

The primary objective requires 20500 h of nominal time (first column of Table \ref{tab:tech.summary.science}) and about 4000 h will be dedicated to open time observations. The \theia Collaboration is keeping large margins to consider the spacecraft thermal stabilization aspects that can only be correctly known after a detailed thermal modeling of the spacecraft and payload modules and mission scheduling. The mission timeline is flexible and can be optimized together with the target list and the main instrumental characteristics.

\subsection{Spacecraft concepts}
\label{sec:prel-spac-design}

The preliminary \theia analyses performed with the current mission definition allowed to identify a safe and robust mission architecture relying on high TRL technologies, and leaving safe margins and mission growth potential that demonstrates the mission feasibility within the Ariane 6.2 single launch envelope (a dual launch will be an opportunity for cost optimization) and M5 mission cost cap.

\subsubsection{System functional design}
\label{sec:syst-funct-design}

The proposed mission architecture relies on a Korsch three mirror telescope accommodated vertically on top of a platform including all support subsystems.

A high thermal stability of the telescope necessary to ensure its performances is obtained through the use of a Sun Shield on which is accommodated the Solar Array and a vertical V-groove screen.

The micro-arcsecond performance specification places stringent
requirements on the pointing stability. This demands a relative pointing
error (RPE) of $\sim ~20$\,mas ($1\sigma$) in spacecraft $x$ and $y$
direction, and a fraction of arcsec ($1\sigma$) in $z$ (roll) over the
image accumulation time. This can be achieved using chemical (monopropellant hydrazine) propulsion for the transfer corrections, monthly station keeping manoeuvres, and large (180\,deg) slew manoeuvres, and then cold gas micro-propulsion system (MPS) for fine acquisition. This can also be achieved by performing the pointings of the satellite using reaction wheels for large attitude motions between targets and then cold gas. At the approach of the correct pointing, the Satellite will use a MPS for fine relative motion acquisition. During the scientific observation the reaction wheels are stopped to avoid the generation of mechanical noise. Such a synergetic concept was developed for the \euclidx science mode AOCS \citep[e.g.][]{Bacchetta2015}.

\subsubsection{Satellite design description}
\label{sec:satel-design-descr}

The satellite key features are presented in Fig.~\ref{fig:mission.sc.concept}. The design of satellite is mainly based on the \euclidx  service module with a downscaled size to better suit to specific \theia needs and minimize the mass to leave the door open to dual launch opportunities. It uses a central tube and standard 1194mm interface compatible with Ariane planned adapters and an irregular hexagonal shape structure contained inside a 3.5m diameter circle, providing large volumes for platform units and fluid tanks accommodation.
Similarly to \euclid and \textit{Herschel} satellites, \theia Korsch telescope is accommodated on top of the service module in a vertical position leading to a spacecraft height of about 5m. This concept allows to optimize the payload size, Ariane fairing allowing large volumes.

A key driver for the \theia mission is the payload structural stability. Particularly during long observations of 10h or more, the payload shall remain stable within 30\,mK to avoid nanometer distance variations of the primary-secondary mirror. This stringent requirement demands a very stable payload structure with low thermo-elastic deformations. The preliminary analyses led to the selection of a telescope structure largely making use of SiC or Si$_3$N$_4$ ceramic materials. Such materials can be used in a large number of structure components (e.g. truss, brackets, plates).

The payload structural stability requirement imposes a very stable thermal concept for the spacecraft (Fig.~\ref{fig:mission.sc.concept}). 
The preliminary design is making use of a Sun shield supporting also the Solar array. Here too, the concept derives from \euclid and \textit{Herschel}. However, \theia will require additional V-groove vertical screen(s) aiming at minimizing the temperature variations from one target to the other due to different solar impingement on the satellite. In addition, an active control of the payload structure temperature will be added through the use of high precision thermistors and optimally spread heaters. Thermal stability levels more strict than \theia requirements have been demonstrated by \gaiap. Nevertheless, there is still space for some optimization effort related to the V-grooves and the active thermal control system.

Strong heritage does exist on \theia spacecraft avionics and AOCS, particularly coming from \euclid mission having very similar needs in terms of pointing performances. The proposed AOCS configuration would use Star Trackers, FOG gyroscope and Fine Sun Sensors as main sensors, reaction wheels for fast repointing between targets and probably cold gas $\mu$-propulsion (or mini Radio-frequency Ion Thruster, mini-RIT), depending on \euclid final performances. Associated fluid tanks (Nitrogen or Xenon) would be accommodated around the central tube in a symmetrical configuration to keep a centered CoG. This AOCS concept is perfectly compatible with the preliminary pointing performances required by the \theia mission.

The electrical power subsystem would be built a fixed Solar Array installed on the Sun shield as on \euclid with a size compatible with \theia needs and with extreme foreseen Sun depointing angles. A battery has been added to power the S/C during the launch operations and up to separation.

The downlink data rate needs and operational constraints are equivalent to the ones of \euclidx, profiting from its heritage. Similarly, Data Handling units would derive from \euclid ones, with a Control and Data Management Unit (CDMU) including the software and a Mass Memory Unit with several TB storage capacity.  

The required $\Delta V$ for transfer to L2, attitude control, station-keeping maneuvers and end of life disposal will be performed by an hydrazine-based propulsion system operation in blow down. The associated tank will be accommodated inside the central tube.

\subsection{Communications}
\label{sec:communications}

Once in the operational position, data collected by the spacecraft from the detectors or housekeeping data will be sent to ground for analysis and post-processing. The amount of science data produced by the payload module was estimated to a total average of 95\,Gbytes/day (135\,Gbytes/day worst case), including a compression factor of 2.5. The current telecommunication subsystem has been sized to allow a download data rate of $\sim75$\,Mbps with an ESA 35m ground station. Consequently, daily visibility periods of about 4 hours would be necessary, similar to \euclid.

\subsection{Observation scheduling}
\theia scheduling will optimize a number of aspects to guarantee the maximum scientific outcome of the observatory. This will take into account several factors as the absolute value of the Sun Aspect Angle (SAA), the variation of the SAA ($\Delta$SAA) between pointings, the parallactic factor of the sources that require parallax determination, the amount of propeller (cold gas) of the MPSs and the spacecraft data downlink conditions. To verify mission feasibility in terms of the median variation of $\Delta$SAA, a preliminary study of \theia pointings was performed using the method described in \citep{TerenoEtal2015} for \textit{Euclid}.

This study adopted \theia spacecraft parameters and the list of \theia high priority pointings. For this assessment an equal distribution of the observing times among the targets was considered over 6-months periods. Enough margin was left for orbit-keeping manoeuvres and an average slew time of 15 minutes was allocated to each observation. The scheduling was done such that SAA is kept within the S/C limits, resulting in a sequence of observations mostly progressing along ecliptic longitude, while the satellite moves in its orbit around L2. We note that no global mission optimization was adopted at this point.

This study resulted in a scenario in which observations are mostly made orthogonally to the Sun, with a median SAA of $88.4^\circ$. This study also showed that, along the mission, the median $\Delta$SAA is $5.7^\circ$.

The aforementioned result, obtained without a global optimization, indicates that the mission scheduling is feasible. There is margin to improve these results in Phase-A studies. For instance, by optimizing slew time at each longitude step taking into account the target's latitude information, distinguishing between trailing and leading pointings (i.e., forward or backward observations with respect to the orbital movement) -- that will minimize the number of jumps between hemispheres --, and by performing a global optimization in which some targets could be replaced by other targets of the same science case, but that would have less impact on the overall mission.

\section*{Acknowledgments}
We are very grateful to Daniel Coe and the "Concision" team (concision.co.uk) for their help during the preparation of the proposal.


\clearpage
\onecolumn 
\begin{center}
\textbf{\huge Annexes}  
\end{center}
\appendix

\section{Acronyms}

  \begin{multicols}{3}
    \small
    \flushleft
    \textbf{mas}: milli-arcsecond\\ 
    \textbf{\uas}: micro-arcseconds\\
    \textbf{AIV}: Assembly, Integration and Verification\\
    \textbf{AOCS}: Attitude and Orbit Control System\\
    \textbf{BAM}: Basic Angle Monitor\\
    \textbf{BCRS}: Barycentric Celestial Reference System\\
    \textbf{BD}: Brown dwarf\\
    \textbf{BH}: Black Hole\\
    \textbf{CaC}: Cost at Completion\\
    \textbf{CCD}: Charge-Coupled Device\\
    \textbf{CCSDS}: Consultative Committee for Space Data Systems\\
    \textbf{CNES}: Centre National d'Etudes Spatiales\\
    \textbf{CPU}: Central Processing Unit\\
    \textbf{CTA}: Cherenkov Telescope Array\\
    \textbf{CTE}: Charge Transfer Efficiency\\
    \textbf{CTE}: Coefficient of Thermal Expansion\\
    \textbf{CMOS}: Complementary metal–oxide–semiconductor\\
    \textbf{COTS}: Commercial off-the-shelf\\
    \textbf{CDMU}: Control and Data Management Unit\\
    \textbf{CoG}: Centre of Gravity\\
    \textbf{DM}: Dark Matter\\
    \textbf{DPU}: Digital Processing Unit\\
    \textbf{dSph}: dwarf spheroidal galaxy\\
    \textbf{ECSS}: European Cooperation for Space Standards\\
    \textbf{EDAC}: Error Detection and Correction.
    \textbf{ELENA}: Emitted Low-Energy Neutral Atoms\\
    \textbf{ELT}: Extremely Large Telescope\\
   \textbf{EoS}: Equation of State\\ 
   \textbf{EPRAT}: ESA Exoplanet Roadmap Advisory Team\\
    \textbf{ESA}: European Space Agency\\
    \textbf{ESPRESSO}: Echelle SPectrograph for Rocky Exoplanet and Stable Spectroscopic Observation\\
    \textbf{FGS}: Fine Guidance Sensor\\
    \textbf{FOG}: Fiber Optic Gyroscope\\
    \textbf{FOV}: Field Of View\\
    \textbf{FPA}: Focal Plane Array\\
    \textbf{FPA}: Focal Plane Assembly\\
    \textbf{FPAM}: Focal Plane Assembly Module\\
    \textbf{FPGA}: Field-Programmable Gate Array\\
    \textbf{FWHM}: Full Width at Half-Maximum\\
    \textbf{GCRS}: Geocentric Celestial Reference System\\
	\textbf{HIPPARCOS}: High precision parallax collecting satellite\\
    \textbf{H/W}: Hardware\\
	\textbf{HST}: Hubble Space Telescope\\
    \textbf{HVS}: Hyper-Velocity Star\\
    \textbf{HZ}: Habitable Zone\\
    \textbf{IAU}: International Astronomical Union\\
    \textbf{ICD}: Interface control documents\\ 
    \textbf{ICU}: Instrument Control Unit\\
    \textbf{ICRF}: International Celestial Reference Frame\\
    \textbf{IDT}: Initial Data Treatment\\
    \textbf{IMCCE}: Institut de m\'ecanique c\'eleste et de calcul des \'eph\'em\'erides\\
    \textbf{ITAR}: International Traffic in Arms Regulations\\
    \textbf{JPL}: Jet Propulsion Laboratory\\
    \textbf{$\Lambda$CDM} Lambda Cold Dark Matter\\
    \textbf{L-DPU}: Local Digital Processing Unit\\
    \textbf{LSST}: Large Synoptic Survey Telescope\\
    \textbf{LEO}: Low Earth Orbit\\
    \textbf{LEOP}: Launch and Early Operation Phase\\
    \textbf{LGA}: Low Gain Antenna\\
    \textbf{LISA}: Laser Interferometer Space Antenna\\
    \textbf{MACHOs} MAssive Compact Halo Objects\\
    \textbf{MCT}: Micro-pixel Centroid Testbed\\
    \textbf{MGA}: Medium Gain Antenna\\
    \textbf{MOC}: Mission Operations Centre\\
    \textbf{MPS}: Micro-Propulsion System\\
    \textbf{MTG}:  Meteosat Third Generation \\
    \textbf{NASA}: National Aeronautics and Space Administration\\
    \textbf{NEAT}: Nearby Earth Astrometric Telescope\\
    \textbf{NS}: Neutron Star\\
    \textbf{OFAD}: Optical Field Angle Calibration\\
    \textbf{P/L}: Payload\\
    \textbf{PLATO}: PLAnetary Transits and Oscillations of stars\\
    \textbf{PLM}: Payload Module\\
    \textbf{PRNU}: pixel response non uniformity\\
    \textbf{PSF}: Point Spread Function\\
    \textbf{PV}: Peak-to-Valley\\
    \textbf{QA}: Quality Assurance\\
    \textbf{QE}: Quantum Efficiency\\
    \textbf{RF}: Radio Frequency\\
    \textbf{RPE}: Relative Pointing Error\\
    \textbf{RMS}: Root Mean Square\\
    \textbf{ROE}: Read-Out Electronics\\
    \textbf{RV}: Radial Velocities\\
    \textbf{RW}: Reaction Wheels\\
    \textbf{SAA}: Sun Aspect Angle\\
    \textbf{SCU}: System Control Unit\\
    \textbf{SDC}: Science Data Centre\\
    \textbf{SC, S/C}: Spacecraft\\
    \textbf{SDRAM}: Synchronous dynamic random access memory\\
    \textbf{SDSS}: Sloan Digital Sky Survey\\
    \textbf{SERENA}: Search for Exospheric Refilling and Emitted Natural Abundances Experiment\\
    \textbf{SIM}: Space Interferometry Mission\\
    \textbf{SKA}: Square Kilometer Array\\
    \textbf{SM}: Standard Model\\
    \textbf{SNR}: Signal to Noise Ratio\\
    \textbf{SOC}: Science Operations Centre\\
    \textbf{SVM}: Service Module\\    
    \textbf{SW}: Software\\
    \textbf{TAS}: Thales Alenia Space\\
    \textbf{TD}: Time Delay\\
    \textbf{TESS}: Transiting Exoplanet Survey Satellite\\
    \textbf{TMA}: Three Mirror Anastigmat\\
    \textbf{TRL}: Technology Readiness Level\\
    \textbf{TRP}: Technology Research Programme\\
    \textbf{TWTA}: Traveling Wave Tube Amplifiers\\
    \textbf{UCMH}: ultra-compact minihalos\\
    \textbf{VESTA}: Validation Experiment for Solar-system STaring Astrometry\\
    \textbf{VLT}: Very Large Telescope\\
    \textbf{WFS}: Wavefront sensor\\
    \textbf{YAG}: Yttrium Aluminium Garnet\\
  \end{multicols}


\label{sec:references}
\bibliographystyle{aa}
\bibliography{theia}

\end{document}